\documentclass[aps,prd,twocolumn, superscriptaddress,preprintnumbers, nofootinbib,longbibliography,floatfix]{revtex4-2}

\usepackage{graphicx}
\usepackage[caption=false]{subfig}
\usepackage{dcolumn}
\usepackage{bm}
\usepackage{enumitem}

\usepackage[colorlinks=true
,urlcolor=blue
,anchorcolor=blue
,citecolor=blue
,filecolor=blue
,linkcolor=blue
,menucolor=blue
,pagecolor=blue
]{hyperref}
\usepackage[all]{hypcap}
\usepackage{braket}
\usepackage{amsmath,amsfonts}
\usepackage{algorithm, algorithmic, float}
\usepackage{comment}
\usepackage{cleveref}
\usepackage{dsfont}

\usepackage{tabularx}
\usepackage[pass]{geometry}

\DeclareRobustCommand{\Sec}[1]{Sec.~\ref{#1}}

\DeclareRobustCommand{\App}[1]{App.~\ref{#1}}
\DeclareRobustCommand{\Tab}[1]{Table~\ref{#1}}

\DeclareRobustCommand{\Fig}[1]{Fig.~\ref{#1}}

\DeclareRobustCommand{\Eq}[1]{Eq.~(\ref{#1})}
\DeclareRobustCommand{\Eqs}[2]{Eqs.~(\ref{#1}) and (\ref{#2})}
\DeclareRobustCommand{\Ref}[1]{Ref.~\cite{#1}}
\DeclareRobustCommand{\Refs}[1]{Refs.~\cite{#1}}

\usepackage{xcolor}

\newcommand{\ce}{\ensuremath{\textrm{e}}}

\begin{document}

\preprint{MIT-CTP 5137}

\title{Degeneracy Engineering for Classical and Quantum Annealing:\\ A Case Study of Sparse Linear Regression in Collider Physics}

\author{Eric R. Anschuetz}
\email{eans@mit.edu}
\affiliation{Center for Theoretical Physics, Massachusetts Institute of Technology, Cambridge, MA 02139, USA}

\author{Lena Funcke}
\email{lfuncke@mit.edu}
\affiliation{Center for Theoretical Physics, Massachusetts Institute of Technology, Cambridge, MA 02139, USA}
\affiliation{Co-Design Center for Quantum Advantage}
\affiliation{The NSF AI Institute for Artificial Intelligence and Fundamental Interactions}

\author{Patrick T. Komiske}
\email{pkomiske@mit.edu}
\affiliation{Center for Theoretical Physics, Massachusetts Institute of Technology, Cambridge, MA 02139, USA}
\affiliation{The NSF AI Institute for Artificial Intelligence and Fundamental Interactions}

\author{Serhii Kryhin}
\email{serhin@mit.edu}
\affiliation{Center for Theoretical Physics, Massachusetts Institute of Technology, Cambridge, MA 02139, USA}
\affiliation{The NSF AI Institute for Artificial Intelligence and Fundamental Interactions}

\author{Jesse Thaler}
\email{jthaler@mit.edu}
\affiliation{Center for Theoretical Physics, Massachusetts Institute of Technology, Cambridge, MA 02139, USA}
\affiliation{Co-Design Center for Quantum Advantage}
\affiliation{The NSF AI Institute for Artificial Intelligence and Fundamental Interactions}

\begin{abstract}
Classical and quantum annealing are computing paradigms that have been proposed to solve a wide range of optimization problems.
In this paper, we aim to enhance the performance of annealing algorithms by introducing the technique of degeneracy engineering, through which the relative degeneracy of the ground state is increased by modifying a subset of terms in the objective Hamiltonian.
We illustrate this novel approach by applying it to the example of $\ell_0$-norm regularization for sparse linear regression, which is, in general, an $\textsf{NP}$-hard optimization problem.
Specifically, we show how to cast $\ell_0$-norm regularization as a quadratic unconstrained binary optimization (QUBO) problem, suitable for implementation on annealing platforms.
As a case study, we apply this QUBO formulation to energy ﬂow polynomials in high-energy collider physics, finding that degeneracy engineering substantially improves the annealing performance.
Our results motivate the application of degeneracy engineering to a variety of regularized optimization problems.
\end{abstract}

\maketitle

\tableofcontents

\section{Introduction}
\label{sec:intro}

Quantum annealing \cite{Kadowaki1998,farhi2000quantum,kadowaki2002study} is a computing paradigm for solving optimization problems, with applications ranging across computer science problems \cite{farhi2001quantum}, machine learning \cite{lloyd2013quantum},
quantum chemistry \cite{babbush2014adiabatic}, protein folding \cite{perdomo2012finding}, and beyond.
Such optimization problems often require minimizing a cost function, which can be reformulated as finding the ground state of a classical Ising Hamiltonian \cite{Lucas2014}.
Many problems of practical importance, however, have cost functions over exponentially many spin configurations, reminiscent of classical spin glasses \cite{Binder1986, nishimori2001statistical, mezard1987spin}.
These characteristics make it extremely difficult for classical algorithms, including classical annealing, to find the ground state of the classical Ising Hamiltonian \cite{kadowaki2002study}.

Quantum annealing was conceived as an alternative to solve this task, where one elevates the classical Ising Hamiltonian to a quantum spin Hamiltonian to take advantage of tunneling in the optimization landscape~\cite{kadowaki2002study}.
Since the first quantum annealing device became commercially available in 2011 \cite{johnson2011quantum}, a large number of proof-of-principle demonstrations have been performed (see, e.g.,~\Refs{Albash2018,ronnow2014defining,Katzgraber2015,hen2015probing,mott2017solving,PhysRevA.102.062405}).
Quantum annealing still faces several conceptual and hardware challenges, however---in particular the inability to outperform classical annealing algorithms in many applications (see \Ref{Hauke2020} for a review).

In this paper, we introduce the technique of \emph{degeneracy engineering} in order to enhance the performance of classical and quantum annealing.
We show that for some applications, one can bias the spectral landscape toward more optimal solutions, dramatically improving both classical and quantum annealing performance on these problems.
We illustrate this novel concept by applying it to $\ell_0$-norm regularization for sparse linear regression, which is a non-convex optimization problem that is, in general, $\textsf{NP}$-hard~\cite{doi:10.1137/S0097539792240406}.
Specifically, we first show how to cast $\ell_0$-norm regularization as a quadratic unconstrained binary optimization (QUBO) problem, suitable for implementation on (quantum) annealing platforms.
The key insight is to use a \emph{redundant} (qu)bit encoding scheme for the linear fit coefficients, which allows the $\ell_0$-norm penalty term to be written in quadratic form.
The smallest redundant encoding scheme requires only one extra (qu)bit per coefficient.
By using a higher degree of redundancy, though, one is able to increase the \emph{relative degeneracy} of the desired ground-state configuration to the first excited state of the regularizer, which, in practice, yields better annealing performance on the full problem.

Sparse linear regression is a topic of general interest, but here we focus on a case study in high-energy collider physics.
\emph{Energy flow polynomials} (EFPs) are a linear basis of collider observables~\cite{Komiske:2017aww}, which can be used to accomplish a broad range of classification and regression tasks in collider physics.
Most EFP studies to date have used standard linear regression with a subset of $O(1000)$ EFPs~\cite{Komiske:2018vkc,Kasieczka:2019dbj}, but it is likely that many collider tasks could be accomplished to the desired accuracy using only a handful of EFPs.
This is a natural venue to explore sparse linear 
regression, but there are known cases where the two most popular sparse linear regression approaches---ridge regression using $\ell_2$-norm regularization~\cite{hoerl1970ridgeregression} and lasso regression using $\ell_1$-norm regularization~\cite{Tibshirani1996}---yield unsatisfactory results~\cite{Komiske:2017aww,Komiske:2019asc}.
While the $\ell_0$-norm penalty is expected to yield better performance in such cases, it is computationally daunting to implement.
These considerations make this problem an ideal test bed for exploring the performance of degeneracy engineering.

In detailed numerical simulations, we assess the potential gains from quantum annealing by comparing standard simulated annealing~\cite{doi:10.1063/1.1632130} to path integral Monte Carlo (PIMC)~\cite{Barker1979}.
While PIMC is a classical annealing strategy, it is a useful proxy for quantum annealing~\cite{PhysRevLett.117.180402}, and it is exact in the long equilibration time limit.
We compare five different regularization methods, including the standard $\ell_2$-, $\ell_1$-, and $\ell_0$-norm regularizations, as well as two novel heuristics.
Focusing on $\ell_0$-norm regularization, we then compare two different encoding schemes with different degrees of redundancy, thus examining the potential benefits of degeneracy engineering.
Our case study is based on EFP sparse regression tasks with known analytic solutions, so that we have an absolute performance benchmark.
Using our QUBO implementation with the smallest redundant encoding scheme, we find relatively poor regression performance.
Going to a higher degree of redundancy, though, we achieve significantly better performance.
This motivates further studies of degeneracy engineering for other optimization problems beyond sparse linear regression.

The remainder of this paper is organized as follows.
In \Sec{sec:qubo}, we review $\ell_p$-norm-regularized linear regression and its binary encoding on quantum or classical computers, followed by a derivation of $\ell_0$-norm regularization in QUBO form.
In \Sec{sec:DegEng}, we introduce the concept of degeneracy engineering, which improves the annealing performance by increasing the relative degeneracy of the ground-state configuration.
In \Sec{sec:Opt}, we outline different optimization strategies, including a review of classical annealing and PIMC, a proposal of novel heuristics, and considerations for quantum annealing.
In \Sec{sec:efp}, we review EFPs, including a detailed overview of the observables and data sets used in our case study.
In \Sec{sec:NumResults}, we present the numerical results of our case study, comparing the smallest redundant encoding scheme to the scheme with two redundant qubits, comparing the $\ell_0$-norm regularization to its $\ell_1$- and $\ell_2$-norm counterparts, and comparing simulated annealing to PIMC\@.
We conclude in \Sec{sec:conclude}, including a broader discussion of the role of redundant encoding schemes for classical and quantum annealing.

\section{Sparse Linear Regression as a QUBO Problem}\label{sec:qubo}

\subsection{Review of $\ell_p$-Norm Regularization}\label{sec:lp_reg}

For generic regression problems, the goal is to find a function $h:\mathbb{R}^n\to\mathbb{R}$ that approximates the mapping of inputs $\vec{x}$ to outputs $y$ seen in a training data set $\mathcal{S}$.
One way to achieve this is by minimizing the mean squared error (MSE) loss function:
\begin{equation}
\label{eq:MSE}
L_{\rm MSE} = \sum_{s \in \mathcal{S}} \Big(y_s - h(\vec{x}_s) \Big)^2 .
\end{equation}
For linear regression, one chooses a set of $K$ functions $h_a(\vec{x})$ and real fit coefficients $c_a$, such that
\begin{equation}
h(\vec{x}; \{c_a\}) = \sum_{a = 1}^{K} c_a \, h_a(\vec{x}).
\label{eq:functions}
\end{equation}

To avoid overfitting, one is often interested in finding a sparse, approximate minimizer of the MSE.
To achieve this in practice, one introduces a regulator $R$ that penalizes non-zero values of $c_a$:
\begin{equation}
\label{eq:regularized}
L = L_{\rm MSE} + \lambda \, R,
\end{equation}
where $\lambda$ controls the strength of the regularization.
For $\ell_p$-norm regularization, the regularization term is
\begin{equation}
\label{eq:regularized_form}
R^{(p)} =  \sum_{a = 1}^K R_a^{(p)}, \qquad R_a^{(p)} = |c_a|^p,
\end{equation}
where $|\cdot|$ is the absolute value.
When $p = 0$, we define the limit as
\begin{equation}
\lim_{p \to 0} R_a^{(p)}  =
	\begin{cases}   
		0 & c_a = 0 \\
		1 & c_a \not= 0
	\end{cases},
\end{equation}
such that the $\ell_0$-norm penalty depends only on whether $c_a$ is non-zero, independent of its magnitude.
Since $\ell_0$-norm regularized regression is computationally challenging to implement, this problem is well suited for exploring the performance of degeneracy engineering.

\subsection{Redundant Binary Encodings}

To formulate a quadratic representation of the $\ell_0$-norm regularizer, we first consider binary encodings of the real fit coefficients $c_a$.
For an $M$-bit representation, we have
\begin{equation}
\label{eq:binary_rep}
c_a = \sum_{i = 1}^M g_i \, b^{(i)}_a,
\end{equation} 
where $g_i$ are fixed real numbers, and the binary coefficients $b^{(i)}_a$ take values of $0$ or $1$.

For a non-redundant encoding, one typically chooses a standard binary encoding, such as $g_i = 2^i$.
More generally, though, $g_i$ can take any desired fixed value, including a negative value, at the expense of having multiple binary representations for the same real number~\cite{Phatak_1994}. 
As a concrete example, consider a four-bit encoding, where
\begin{equation}
\label{eq:example_redundant_g}
\vec{g} = \{-2,-1,1,2\}.
\end{equation}
For a fixed $a$, there are $2^4 = 16$ possible choices for the values of $b_a^{(i)}$, but only $7$ unique values of $c_a$, namely
\begin{equation}
\label{eq:example_redundant_ca}
c_a \in \{-3,-2,-1,0,1,2,3\}.
\end{equation}
In the context of annealing, these redundant encodings are irrelevant for the ground state if the corresponding values of the loss function are \emph{the same or higher} than for the default encoding.
We will exploit this freedom in implementing $\ell_0$-norm regularization.

\subsection{Quadratic Loss for $\ell_0$-Norm Penalty}
\label{sec:QUBO}

When inserting the binary representation for the fit coefficients $c_a$ in \Eq{eq:binary_rep} into the MSE loss function in \Eq{eq:MSE}, we see that the dependence on the binary coefficients $b^{(i)}_a$ is at most quadratic. Thus, standard linear regression can be cast as a QUBO problem.

A QUBO problem consists of finding a vector
\begin{equation}
x^*=\underset{x\in\mathbb{B}^n}{\arg\min} ~Q(x)
\end{equation}
that is minimal with respect to a quadratic polynomial $Q:\mathbb{B}^n\rightarrow\mathbb{R}$ over binary variables $x_i\in\mathbb{B}$ for $\mathbb{B}=\lbrace 0,1\rbrace$ and $i\in[n]$,
\begin{equation}
Q(x) = \sum_{i=1}^n \sum_{j=1}^i J_{ij} \, x_i \, x_j.
\label{eq:QUBO}
\end{equation}
Here, the coefficients $J_{ij}\in\mathbb{R}$ satisfy $1\leq j\leq i\leq n$, and $\left[n\right]$ is the set of strictly positive integers less than or equal to $n$.

When adding the $\ell_p$-norm regularization in \Eq{eq:regularized_form}, we still have a QUBO form for $p=2$, but not for any other value of $p$.
To understand the role of redundant encodings in this context, it is instructive to first consider the $p = 1$ case.
Because of the absolute value signs in \Eq{eq:regularized_form}, this is not of QUBO form, but it is ``almost QUBO'' since one could remove the absolute value sign if one knew that a given $c_a$ was either always positive or always negative.
Taking inspiration from this observation, consider a redundant encoding of $c_a$ where there are both positive and negative values of $g_i$, such as in \Eq{eq:example_redundant_g}.
In that case, we have the following inequality:
\begin{equation}
\label{eq:p1_inequality}
|c_a| \leq \sum_{i = 1}^M  |g_i| \, b^{(i)}_a.
\end{equation} 
This would be an equality if $b^{(i)}_a$ were only non-zero when $g_i$ was positive, or when $g_i$ was negative, but not both.
From the perspective of minimizing \Eq{eq:regularized}, though, cases with non-zero values of $b^{(i)}_a$ for mixed signs of $g_i$ are irrelevant, as long as there is another encoding of $c_a$ that only uses all positive or all negative values of $g_i$ (and therefore satisfies \Eq{eq:p1_inequality} as an equality).
This is indeed the case for the example in \Eqs{eq:example_redundant_g}{eq:example_redundant_ca}.
Therefore, without changing the solution of the sparse regression problem, we can use a modified $\ell_1$-norm regulator:
\begin{equation}
\label{eq:regularized_form_1}
R_a^{(1-{\rm mod})} = \sum_{i = 1}^M  |g_i| \, b^{(i)}_a,
\end{equation}
which is now of QUBO form.

We can do something similar for the $\ell_0$-norm regulator:
\begin{equation}
\label{eq:regularized_form_0}
R_a^{(0-\text{mod})} = \sum_{i = 1}^M  b^{(i)}_a,
\end{equation}
which is again of QUBO form.
Here, though, for an $N$-bit binary encoding, there are only $N+1$ values of $c_a$ that have the correct regulator, namely all of the individual $g_i$ values (which get a penalty of $1$) and the value $0$ (which gets a penalty of $0$).
Ideally, we would want a large fraction of achievable $c_a$ values to have at least one $b^{(i)}_a$ configuration with the right penalty.
This can be achieved by leveraging a redundant encoding using ancilla bits, as we explain next.

\subsection{Single Ancilla Bit Encoding}
\label{sec:singleABE}

The first example of a redundant encoding involves just a single ancilla bit per fit coefficient.
This ancilla bit $r_a$ plays no role in determining the value of $c_a$, but it appears in the $\ell_0$-norm regulator as follows: 
\begin{equation}
\label{eq:1lb}
R_a^{(0-\text{single})} = r_a + (1 - r_a) \sum_{i = 1}^M  b^{(i)}_a.
\end{equation}
This single ancilla bit encoding (ABE) is shown graphically in \Fig{fig:control_bits_single}, where to match \Eq{eq:positive_negative} below, we have separated out $b^{(i)}_a$ into positive ($p^{(i)}_a$) and negative ($n^{(i)}_a$) fit coefficients.

In the context of annealing, we care about the lowest energy configuration.
Minimizing \Eq{eq:1lb} over the ancilla bit $r_a$, we find that
\begin{equation}
\min_{r_a} R_a^{(0-\text{single})} = 
	\begin{cases}   
		0 & c_a = 0 \\
		1 & c_a \not= 0
	\end{cases},
	\label{eq:reg}
\end{equation}
which is precisely the desired $\ell_0$-norm regulator.

We note that an approximate formulation of $\ell_0$-norm regularization as an optimization problem has recently been proposed in \Ref{desu2021adiabatic}.
This approach, however, is based on the general expression of $k$-local problems as QUBO problems, which requires potentially inefficient gadgetization techniques~\cite{dattani2019quadratization,abel,gabor}.

\begin{figure*}[t]
\subfloat[]{
\includegraphics[width=0.9\columnwidth]{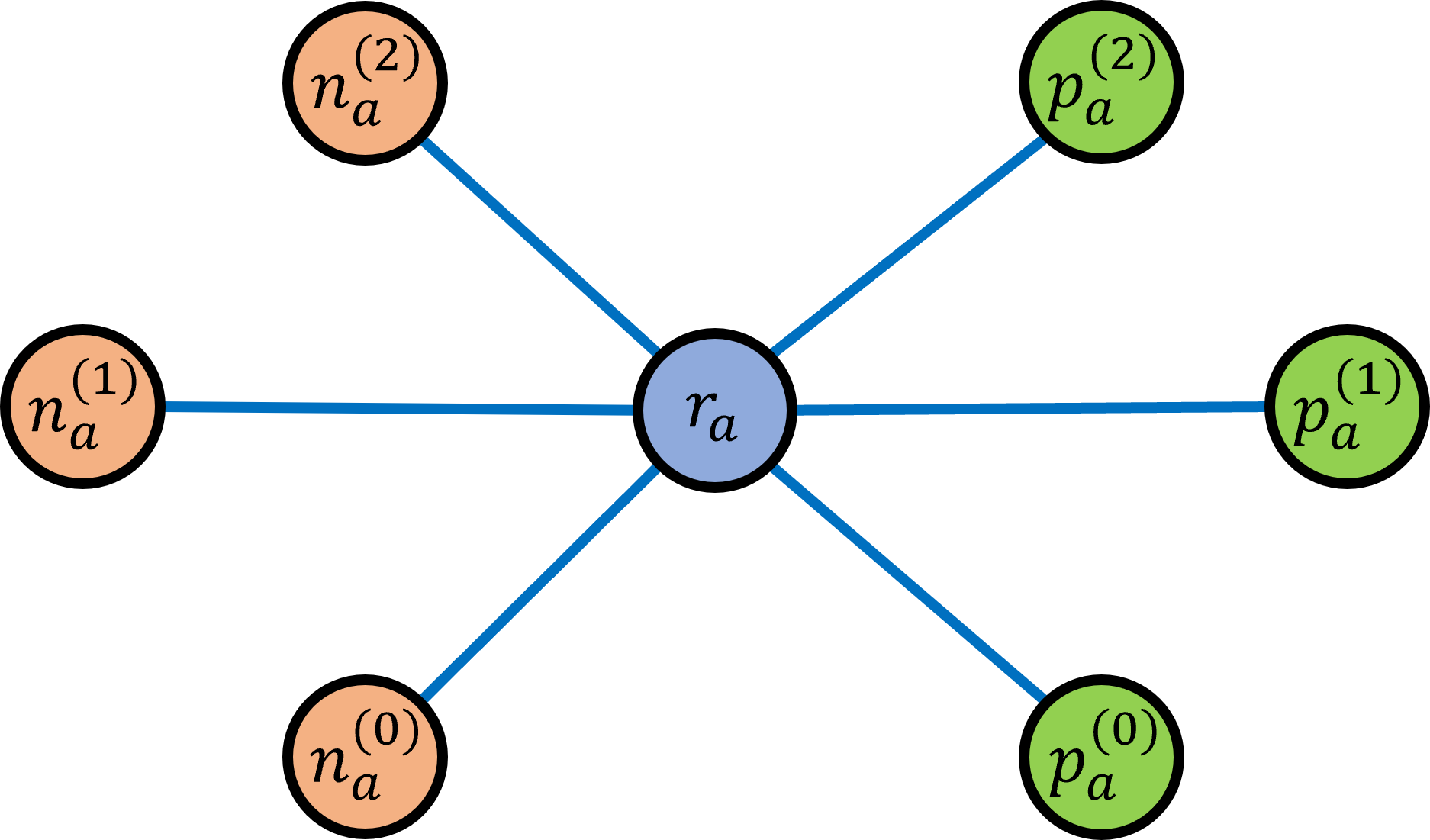}
\label{fig:control_bits_single}
}
$\qquad$
\subfloat[]{
\includegraphics[width=0.9\columnwidth]{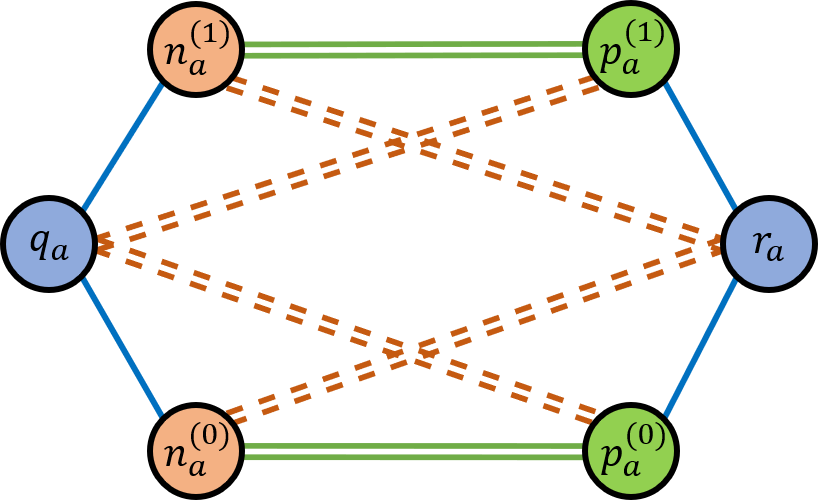}
\label{fig:control_bits_double}
}
\caption{Graphical representation of the $\ell_0$-norm regularizer with (a) single ABE and (b) double ABE.
Circles correspond to a penalty of $+1$ for the ancilla bits $r_a$ and $q_a$ (blue) and positive contributions $p_a^{(i)}$ (green) and negative contributions $n_a^{(i)}$ (orange) to the fit coefficients.
Lines correspond to penalties of $-1$ (single solid blue), $-2$ (double solid green), and $+2$ (double dashed orange).
}
\label{fig:control_bits}
\end{figure*}

\section{Degeneracy Engineering} \label{sec:DegEng}

\subsection{General Principles}

The key idea behind degeneracy engineering is to increase the relative ground-state to excited-state degeneracies of a tractable subset of terms in a given problem Hamiltonian via the addition of ancilla (qu)bits.
More specifically, this technique changes the relative degeneracies (but not the values) associated with this subset of Hamiltonian terms, which in our case is the $\ell_0$-norm regularizer.
Consequently, if one were to optimize the problem Hamiltonian, the success probability of finding the true ground-state energy would be enhanced.
Heuristically, the success probability of finding the true ground-state energy of the \emph{full} Hamiltonian is also enhanced.
Degeneracy engineering is motivated by similar techniques in variational quantum simulation, where it has been shown that a strong over-parametrization of quantum circuits improves the chance of finding a good approximation of the true solution~\cite{Fontana_2021,Kim_2021,anschuetz2022critical,anskiani}.

As we demonstrate in the next subsection, the concept of degeneracy engineering is particularly well suited for Hamiltonians including a penalty term.
While ground-state energies of generic Hamiltonians can be negative, penalty terms employ absolute values and thus vanish under minimization.
This feature makes penalty terms the ideal candidates for degeneracy engineering.
While it is generally hard to engineer multiple \textit{negative} values for generic ground-state energies, one can straightforwardly engineer multiple \textit{zero} values for the ground-state energy of a penalty term.
In particular, this can be achieved by exploiting cancellations of positive and negative contributions to the ground-state energy, as we will exemplify in \Eq{eq:2lb} below.
Thus, degeneracy engineering could provide advantages for any optimization problem containing a penalty term, including penalty terms enforcing physical symmetries.

\subsection{Double Ancilla Bit Encoding}
\label{sec:DoubleEnc}

To illustrate the concept of degeneracy engineering, we apply it to the example of $\ell_0$-norm regularization for sparse linear regression.

The $\ell_0$-norm regulator in \Eq{eq:1lb} has a single minimum, $\min_{r_a} R_a^{(0-\text{single})}=0$, where $r_a=0$ and $b_a^{\left(i\right)}=0$.
However, the regulator also has an exponentially large degeneracy of the first excited state, $\min_{r_a} R_a^{(0-\text{single})}=1$, where $r_a=1$.
Thus, in practice, the optimization using the single ABE is expected to perform poorly.

To mitigate this problem, we want to modify the relative degeneracy of the states under consideration.
Our goal is to match the degeneracy levels of the minimum and the first excited state, without changing the energy values.
To this end, we consider a double ancilla bit encoding (double ABE) of the $\ell_0$-norm loss function.

For concreteness, consider the binary encoding:
\begin{equation}
    g_i = 2^i.
\end{equation}
Next, we introduce a redundant encoding where the fit coefficient zero has multiple representations:
\begin{equation}
\label{eq:positive_negative}
c_a = \sum_{i=0}^M  g_i \big( p_a^{(i)} - n_a^{(i)} \big).
\end{equation}
Here, $p_a^{(i)}$ ($n_a^{(i)}$) are binary coefficients that yield positive (negative) contributions to the fit coefficients.

For the double ABE, we add two ancilla bits ($q_a$ and $r_a$) per fit coefficient:
\begin{align}
\begin{split}
\label{eq:2lb}
R_a^{(0-\text{double})} &= q_a + (1 + 2q_a - r_a) \sum_{i = 1}^M  p^{(i)}_a\\
&\qquad + r_a + (1 + 2r_a - q_a) \sum_{i = 1}^M  n^{(i)}_a \\
&\qquad - 2 \sum_{i = 1}^M p^{(i)}_a \, n^{(i)}_a,
\end{split}
\end{align}
as shown graphically in \Fig{fig:control_bits_double}.
Minimizing \Eq{eq:2lb} over the ancilla bits $r_a$ and $q_a$, we recover the desired $\ell_0$-norm regulator in \Eq{eq:reg}, but with a higher relative ground-state degeneracy; we now describe in more detail why this is so.

\subsection{Comparing the Encodings}

The graphical illustrations in \Fig{fig:control_bits} can help build intuition about the differing behaviors of the single ABE in \Eq{eq:1lb} and the double ABE in \Eq{eq:2lb}.
Here, the ancilla bits $r_a$ and $q_a$ are depicted as blue nodes, the positive contributions $p_a^{(i)}$ to the fit coefficients are shown as green nodes, and the negative contributions $n_a^{(i)}$ are shown as orange nodes.
Turning on any of the nodes is associated with a penalty of $+1$.
Solid blue edges correspond to a pairwise penalty of $-1$, which comes from \Eq{eq:1lb} and from the first two lines of \Eq{eq:2lb}.
Double dashed orange edges correspond to a pairwise penalty of $+2$ from the first two lines of \Eq{eq:2lb}, while double solid green edges correspond to a pairwise penalty of $-2$ from the third line of \Eq{eq:2lb}.

For the single ABE, the only configuration with zero penalty is the one with all nodes turned off, corresponding to $c_a=0$.
The configurations with penalty $+1$ arise from connected graphs, where the connection is enabled by turning on the ancilla bits $r_a$.
Thus, there is only one ground-state configuration with $c_a=0$ and a slew of excited-state configurations for $c_a \not= 0$.

For the double ABE, by contrast, there are a large number of configurations with zero penalty and $c_a=0$, particularly the $2^M$ configurations associated with turning on pairs of nodes connected by double solid green edges.
The configurations with penalty $+1$ and $c_a \not= 0$ arise from connected graphs that do not involve any double solid green edges, of which there are $2^M$.
Thus, there is a balance between the number of $c_a=0$ and $c_a \not= 0$ configurations and therefore an improved loss landscape for our $\ell_0$-norm regularizer.

It is instructive to compare the single and double ABE in the simplest case of $M = 1$, with two binary fit coefficients $p_a$ and $n_a$.
For the single ABE, we have one ancilla bit $r_a$.
There are four different ways to encode $c_a=0$, of which the lowest lying state with $R_a^{(0-\text{single})}=0$ arises from:
\begin{enumerate}[label=(\roman*)]
\item turning off all bits.
\end{enumerate}
There are two different ways to encode $c_a=1$, which are the lowest lying states with $R_a^{(0-\text{single})}=1$:
\begin{enumerate}[label=(\roman*)]
\item turning on just $p_a$; and
\item turning on just $p_a$ and $r_a$.
\end{enumerate}
Thus, the relative degeneracy of the lowest lying $c_a=0$ and $c_a=1$ configurations is 1:2.

For the double ABE, we have two ancilla bits $r_a$ and $q_a$.
There are now eight different ways to encode $c_a = 0$, of which the two lowest lying states with $R_a^{(0-\text{double})}=0$ are:
\begin{enumerate}[label=(\roman*)]
\item turning off all bits, just as for the single ABE; and
\item turning on just $p_a$ and $n_a$.
\end{enumerate}
Similarly, we can encode $c_a=1$ in four different ways, of which the two lowest lying states with $R_a^{(0-\text{double})}= 1$ arise from:
\begin{enumerate}[label=(\roman*)]
\item turning on just $p_a$; and
\item turning on just $p_a$ and $r_a$.
\end{enumerate}
Thus, the relative degeneracy between the lowest lying $c_a=0$ and $c_a=1$ configurations is 1:1.

In this way, we have used the double ABE to successfully engineer a larger ground-state degeneracy without changing the lowest lying energy levels of the system. This general principle of exponentially increasing the ground-state degeneracy of the regularizer can be generalized to $M>1$ in a straightforward fashion, by turning on various combinations of pairs of $(p_a^{(i)},n_a^{(i)})$.

There is some freedom in \Eq{eq:2lb} that could be exploited for practical applications.
We chose a penalty of $+2$ in \Eq{eq:2lb} (i.e.\ the dotted orange edges between $q_a$ and $p^{(i)}_a$ and between $r_a$ and $n^{(i)}_a$) to reduce the degeneracy of the first excited states.
With a penalty of $+1$ instead, one could take a connected configuration with total penalty $+1$ and turn on additional $p^{(i)}_a$ and $n^{(i)}_a$ pairs without additional costs.
As long as it is greater than $+1$, the precise value of this penalty term could be adjusted to optimize the loss landscape.

\subsection{Possible Generalizations}

For concreteness, we perform our case studies using just the two example encodings described above.
There is, however, a whole family of related redundant encodings that might be relevant for practical applications.

As one extreme example, it is possible to avoid highly connected ancilla bits and instead implement tree graph structures, where each node has penalty $+1$ and each edge has penalty $-1$.
In this encoding, there are separate graphs for positive and negative coefficients.
In each graph, the $g_i=1$ node is directly connected to the $g_i=2$ node, instead of being indirectly connected via the ancilla bit.
Then, $g_i=2$ is directly connected to $g_i=4$, which is directly connected to $g_i=8$, and so on.
Meanwhile, $g_i=4$ is connected to an additional $g_i=1$ bit, $g_i=8$ is connected to additional $g_i=1$ and $g_i=2$ bits, and so on.
However, such an encoding not only requires a large overhead of additional bits, but the only configuration with $c_a = 0$ and zero penalty is the one with all nodes turned off.\footnote{We used powers of 2 for simplicity, but there are ways to optimize the coefficients to reduce the size of the required graph.}
Thus, even though such tree graph structures might be advantageous for specific tasks, the double ABE encoding discussed above is, in general, more efficiently implementable.
We leave to future work a study combining these redundant tree graphs with partially connected ancilla bits.

\section{Optimization Strategies}
\label{sec:Opt}

The results in \Sec{sec:NumResults} are based on three different optimization strategies---classical annealing, PIMC, and sparse regularization heuristics---which we describe in this section.
While we do not perform quantum annealing on a quantum computer, we review why PIMC is a useful proxy for studying quantum optimization, and we discuss some general considerations when implementing sparse regression on physical quantum devices.

\subsection{Review of Classical Annealing} \label{sec:ReviewAnnealing}

As a representative measure of the performance of traditional classical optimization algorithms, we perform population annealing~\cite{doi:10.1063/1.1632130}. 
For this, we consider a family of canonical distributions parametrized by the inverse temperature $\beta$,
\begin{equation}
p_\beta(x)=\frac{1}{\mathcal{Z}_\beta}\ce^{-\beta E(x)},
\label{eq:distribution}
\end{equation}
where $E(x)$ is the energy of the state $x$ and $\mathcal{Z}_\beta$ is the partition function. 
As an alternative to the traditional simulated annealing method of optimization, population annealing considers a population of $R_0$ replicas of the state $x$.
This population is initialized randomly (i.e.~infinite temperature), the first annealing step is performed at temperature $1/\beta_0$, and then the system is cooled to some finite temperature $1/\beta_\ell$ by an annealing schedule of $\ell$ steps. 
Unlike simply performing simulated annealing $R_0$ times, however, with each cooling step, replicas are duplicated or deleted based on an estimate of their relative Boltzmann weights.
At each cooling step, the population is reequilibrated according to some Monte Carlo algorithm.
As a representative classical method, we equilibrate using Metropolis-Hastings~\cite{10.1093/biomet/57.1.97}.

\subsection{Path Integral Monte Carlo as a Proxy for Quantum Annealing} \label{sec:ReviewPIMC}

As a representative measure of the performance of quantum optimization algorithms, we consider a proxy for quantum annealing called the PIMC method.
In most stoquastic formulations of quantum annealing, one considers the following parametrized quantum Hamiltonian:
\begin{equation}
    H(s)=(1-s)H_i+sH_f=\Gamma(s)\sum\limits_{i=1}^N\sigma_i^x+J(s) \,\tilde{L},
    \label{eq:QA}
\end{equation}
where $H_i$ is the initial Hamiltonian and $H_f$ is the final Hamiltonian, called the problem Hamiltonian.
The annealing parameter $s=t/t_f\in\left[0,1\right]$ is given by the ratio of the time $t$ and the total annealing time $t_f$, thus linearly increasing from $0$ to $1$.
Here, $\tilde{L}$ is the operator form of the loss function $L$ from \Eq{eq:regularized} encoded in the $\sigma^z$ basis.

To numerically simulate the performance of quantum annealing, we use PIMC.
This method employs the Trotter-Suzuki mapping of the quantum annealing Hamiltonian in \Eq{eq:QA} to a classical energy function with an extra imaginary time dimension, which is discretized into $M$ imaginary time slices~\cite{doi:10.1143/PTP.56.1454}.
This well-known mapping from a $d$-dimensional quantum Ising system to a ($d+1$)-dimensional classical Ising system can be straightforwardly derived using the Trotter breakup formula and spin-$1/2$ algebra; see \App{app:trotter} for details.
We then perform Monte Carlo sampling using the Swendsen-Wang cluster update algorithm~\cite{PhysRevLett.58.86} with the population annealing update heuristic~\cite{doi:10.1063/1.1632130}, forming clusters only in the imaginary time direction on the mapped set of spins~\cite{PhysRevB.68.104409}.
PIMC has been numerically found to accurately simulate quantum annealing in many stoquastic systems~\cite{PhysRevLett.117.180402}.

\subsection{Refined Regression as Novel Heuristics} \label{sec:Heuristics}

To assess the performance of annealing strategies for $\ell_0$-norm regression, we study two novel heuristics:  \emph{refined $\ell_1$-norm regression} and \emph{refined $\ell_0$-norm regression}.

Regression with $\ell_1$-norm regularization is often used as a proxy for regression with $\ell_0$-norm regularization due to the efficiency of the former.
Because the $\ell_1$-norm penalty has constant absolute slope everywhere except the origin, it leads to sparse solutions, just like the $\ell_0$-norm case.
We take this a step further, and consider refined $\ell_1$-norm regression.
In this strategy, coefficients $c_a$ that are set to zero by the initial $\ell_1$-norm regularized regression are clamped to zero.
Then, ordinary linear regression is performed on the remaining coefficients to minimize the MSE loss function of \Eq{eq:MSE}.
The solution found via this heuristic performs at least as well as the originally found solution in terms of sparsity and MSE loss, though not necessarily in terms of the regularized loss.

We use a similar heuristic to post-process the results of our annealing strategies for $\ell_0$-norm regularized linear regression.
In refined $\ell_0$-norm regression, coefficients set to zero by the annealing process are clamped to zero, and ordinary linear regression is performed on the remaining coefficients.
Here, the solution found via this heuristic performs at least as well as the annealed $\ell_0$-regularized solution on all performance measures.
Given the low computational overhead of unregularized linear regression, we implement this refinement step when presenting our baseline annealing results.

\subsection{Considerations for Quantum Annealing}

General adiabatic quantum computation is known to be equivalent to the gate model of quantum computation~\cite{farhi2000quantum}.
Due to experimental considerations, however, most current implementations of quantum annealing platforms use time-dependent \emph{stoquastic} Hamiltonians of the form of Eq.~\eqref{eq:QA}, yielding a model of computation that is not believed to be as powerful as general quantum computation.
Recently, it was shown that even under a restriction to stoquastic Hamiltonians, there exist oracle separations between quantum annealing and classical algorithms for certain classes of problems~\cite{Hastings:2020ubx}.

Outside of these specific classes of problems, however, it has been numerically shown that for many QUBO problems, PIMC---a classical algorithm---performs essentially as well as quantum annealing~\cite{PhysRevLett.117.180402}.
For this reason, we consider PIMC to serve as a good proxy for quantum annealing in our study.

As we will emphasize in \Sec{sec:conclude}, the concept of degeneracy engineering has important implications for both classical and quantum annealing, beyond just QUBO problems.
When solving any optimization problem that employs penalty terms, one can try to engineer multiple zero values for the lowest-energy contribution to the penalty.
Extrapolating from the construction in \Fig{fig:control_bits}, it appears that degeneracy engineering generally requires ancilla qubit(s) that employ a large degree of connectivity to the other qubits on the platform.
Our results therefore stress the importance of good qubit connectivity in quantum annealing platforms.

\section{Case Study with Energy Flow Polynomials}\label{sec:efp}

The results in \Sec{sec:NumResults} are based on a case study in collider physics, where we apply our QUBO formulation of $\ell_0$-norm regularization to EFPs.
In this section, we briefly review the key properties of EFPs and introduce the observable relations and data sets used in our study.

\subsection{Review of Energy Flow Polynomials} \label{sec:EFPs}

EFPs were introduced in \Ref{Komiske:2017aww} to accomplish a wide range of jet analysis tasks in high-energy collider physics.
EFPs form a discrete linear basis for all infrared- and collinear-safe observables, and many common jet observables are exact linear combinations of EFPs.
Many collider tasks can be accomplished using only a handful of EFPs, which makes them an ideal candidate to explore sparse linear regression.

To visualize and calculate the EFPs, \Ref{Komiske:2017aww} established a one-to-one correspondence between EFPs and loopless multigraphs.
For an $M$-particle jet and a multigraph $G$ with $N$ vertices and $d$ edges $(k, l) \in G$, the corresponding functional expression for the EFP reads
\begin{equation}
	\text{EFP}_G = \sum_{i_1 = 0}^{M} \cdot \cdot \cdot \sum_{i_N = 0}^{M} z_1 \cdot \cdot \cdot z_{i_N} \prod_{(k, l) \in G} \theta_{i_k i_l}^\beta,
	\label{eq:EFP}
\end{equation}
where $\beta$ is an angular weighting factor (not to be confused with inverse annealing temperature).
For our numerical studies, we take $\beta = 2$.
In our case study, the energy fraction $z_i$ carried by particle $i$ and the angular distance $\theta_{ij}$ between particles $i$ and $j$ are defined as
\begin{equation}
	z_i = \frac{p_{Ti}}{\sum_j p_{Tj}} \quad \text{and}\quad \theta_{ij} = (\Delta y_{ij}^2 + \Delta \phi_{ij}^2)^{1 / 2}.
	\label{eq:ztheta}
\end{equation}
Here, $p_{Ti}$ is the transverse momentum of particle $i$, and we use the definitions $\Delta y_{ij} = y_i - y_j$ and $\Delta \phi_{ij} = \phi_i - \phi_j$, where $y_i$ and $\phi_i$ are the rapidity and the azimuthal angle of particle $i$.

There are a rich variety of linear relations between different jet observables and EFPs~\cite{Komiske:2017aww,Komiske:2019asc}, a few of which we study in this paper.
Even for fixed $\beta$, the set of all EFPs is an overcomplete basis and therefore needs to be explored using regularized linear regression.
This motivates the application of $\ell_0$-norm regression to study these linear relations.

For our numerical study, we use the \href{https://pkomiske.github.io/EnergyFlow}{\tt EnergyFlow} module, which is based on \Ref{Komiske:2017aww}.
This \textsc{Python} package provides all the necessary tools to compute EFPs on collider events, as well as tools to download, read, and manipulate the data sets described in \Sec{sec:data}.
In our study, we test twelve different linear relations between collider observables and EFPs, which are described in \Sec{sec:observables} and summarized in \Tab{tab:observables}.

\subsection{Testing Relations Between Observables} \label{sec:observables}

{\renewcommand{\arraystretch}{2}

\begin{table*}
	\begin{tabular}{@{\hskip 0.2in} r @{\hskip 0.2in} l @{\hskip 0.2in} l  @{\hskip 0.2in} l @{\hskip 0.2in}} 
	  \hline \hline
	  Label & Name of Observable & Restriction & Multigraph Representation of Linear EFP Relation\\ 
	  \hline\hline\\*[-2em]
	  (a) & Angularity $\alpha = 2$ & None &
	  	$\lambda^{(2)} = \frac{1}{2} \times
		\begin{gathered}
		\includegraphics[scale=.12]{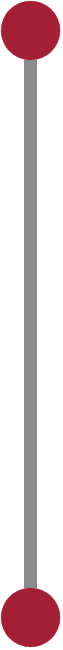}
		\end{gathered}$ \\ 
	  \hline\\*[-2em]
	  (b) & Angularity $\alpha = 4$ & None &  
		$\lambda^{(4)} = \frac{1}{2} \times
		\begin{gathered}
		\includegraphics[scale=.12]{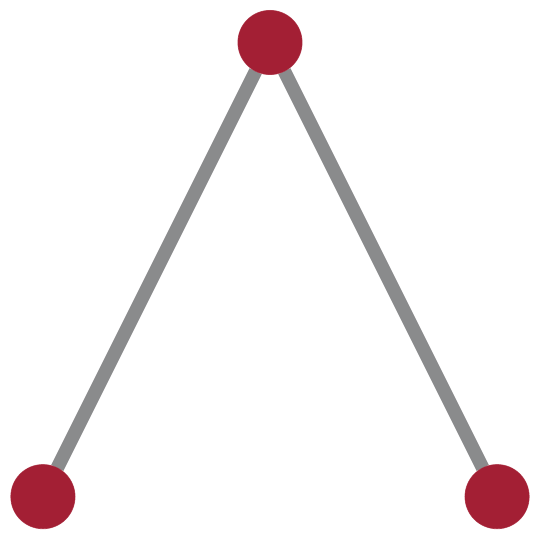}
		\end{gathered}
		- \frac{3}{4} \times
		\begin{gathered}
		\includegraphics[scale=.12]{2_1_1}
		\end{gathered},
		\quad
		\begin{gathered}
		\includegraphics[scale=.12]{2_1_1}
		\end{gathered}$\\ 
	  \hline\\*[-2em]
	  (c) & Angularity $\alpha = 6$ & None & 
		$\lambda^{(6)} = \frac{1}{2} \times
		\begin{gathered}
		\includegraphics[scale=.12]{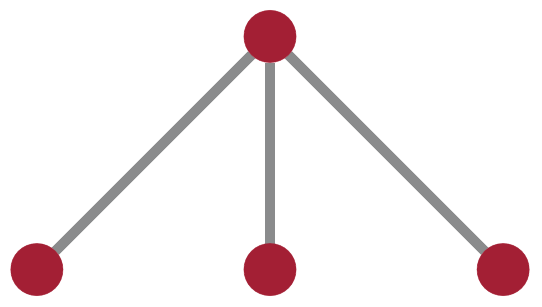}
		\end{gathered}
		-\frac{3}{2} \times
		\begin{gathered}
		\includegraphics[scale=.12]{3_2_1}
		\end{gathered}
		\quad
		\begin{gathered}
		\includegraphics[scale=.12]{2_1_1}
		\end{gathered}
		+ \frac{5}{8} \times
		\begin{gathered}
		\includegraphics[scale=.12]{2_1_1}
		\end{gathered}
		\quad
		\begin{gathered}
		\includegraphics[scale=.12]{2_1_1}
		\end{gathered}
		\quad
		\begin{gathered}
		\includegraphics[scale=.12]{2_1_1}
		\end{gathered}$
	 \\
	  \hline\\*[-2em]
	  (d) & Determinant $C$ & None &  
		$\det C = \frac{1}{4} \times
		\begin{gathered}
		\includegraphics[scale=.12]{3_2_1}
		\end{gathered}
		- \frac{1}{2} \times
		\begin{gathered}
		\includegraphics[scale=.12]{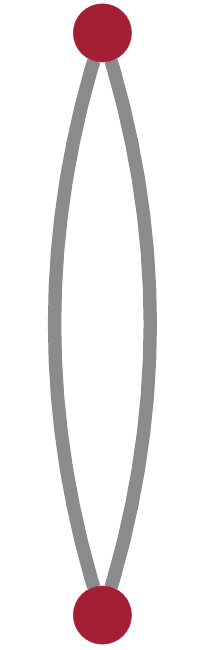}
		\end{gathered}$
		\\
	  \hline\\*[-2em]
	  (e) & Triple Dumbbell & $M \leq 2$ & 
		$\begin{gathered}
		\includegraphics[scale=.12]{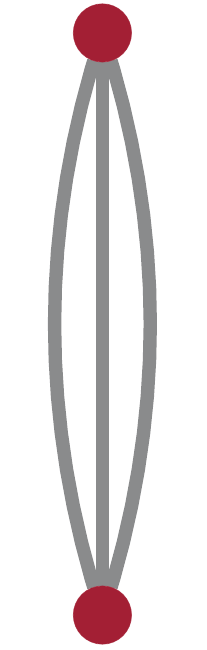}
		\end{gathered}
		= 2 \times
		\begin{gathered}
		\includegraphics[scale=.12]{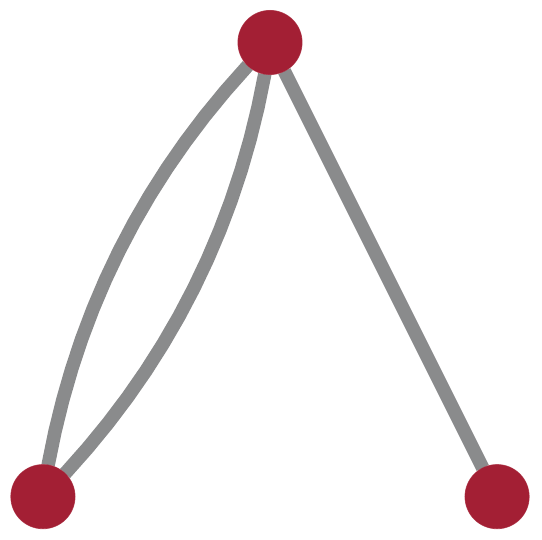}
		\end{gathered}$
		 \\
	  \hline\\*[-2em]
	  (f) & Triple Dumbbell (Approx.) & None & 
		$\begin{gathered}
		\includegraphics[scale=.12]{2_3_1}
		\end{gathered}
		\approx 2 \times
		\begin{gathered}
		\includegraphics[scale=.12]{3_3_2}
		\end{gathered}$
	\\
	  \hline\\*[-2em]
	  (g) & Lollipop & $M \leq 2$ &  
		$\begin{gathered}
		\includegraphics[scale=.12]{3_3_2}
		\end{gathered}
		= 
		\begin{gathered}
		\includegraphics[scale=.12]{4_3_1}
		\end{gathered}
		+ 
		\begin{gathered}
		\includegraphics[scale=.12]{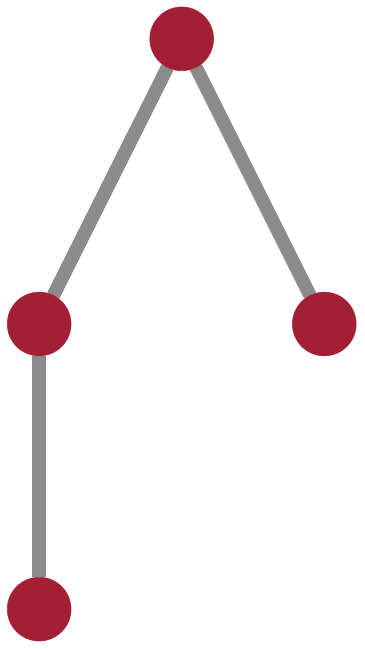}
		\end{gathered}$
		\\
	  \hline\\*[-2em]
	  (h) & Lollipop (Approx.) & None & 
		$\begin{gathered}
		\includegraphics[scale=.12]{3_3_2}
		\end{gathered}
		\approx 
		\begin{gathered}
		\includegraphics[scale=.12]{4_3_1}
		\end{gathered}
		+ 
		\begin{gathered}
		\includegraphics[scale=.12]{4_3_2}
		\end{gathered}$
		\\
	  \hline\\*[-2em]
	  (i) & Five Dots & $M \leq 3$ &  
		$\begin{gathered}
		\includegraphics[scale=.12]{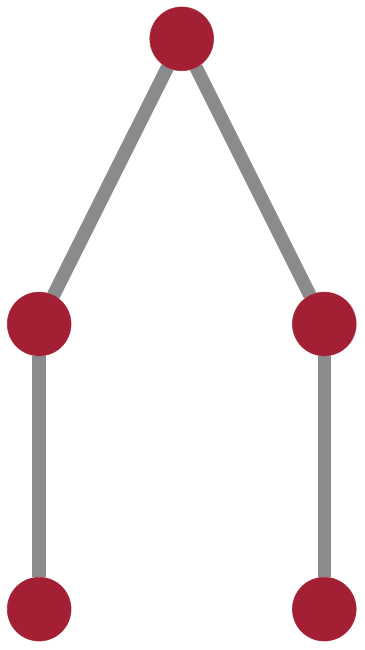}
		\end{gathered}
		=
		\begin{gathered}
		\includegraphics[scale=.12]{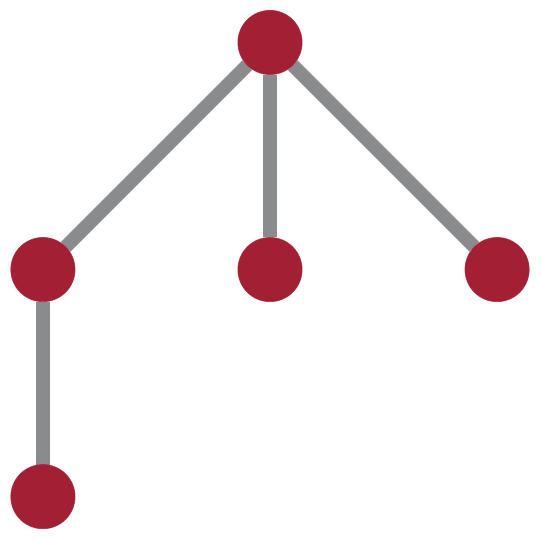}
		\end{gathered}
		+ \frac{1}{2} \times
		\begin{gathered}
		\includegraphics[scale=.12]{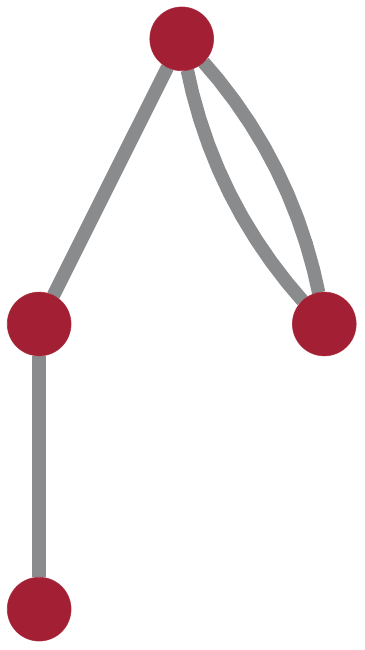}
		\end{gathered}
		- \frac{1}{2} \times
		\begin{gathered}
		\includegraphics[scale=.12]{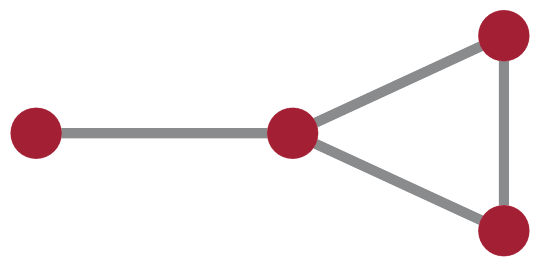}
		\end{gathered}$
		\\
	  \hline\\*[-2em]
	  (j) & Five Dots (Approx.) & None & 
$\begin{gathered}
\includegraphics[scale=.12]{5_4_3}
\end{gathered}
\approx
\begin{gathered}
\includegraphics[scale=.12]{5_4_2}
\end{gathered}
+ \frac{1}{2} \times
\begin{gathered}
\includegraphics[scale=.12]{4_4_5}
\end{gathered}
- \frac{1}{2} \times
\begin{gathered}
\includegraphics[scale=.12]{4_4_4}
\end{gathered}$
\\
	  \hline\\*[-2em]
	  (k) & Planar Event & $n\leq 2$ &  
{$\!\begin{aligned} 
               \begin{gathered}
\includegraphics[scale=.12]{4_3_2}
\end{gathered}
&=
\frac{1}{2} \times
\begin{gathered}
\includegraphics[scale=.12]{5_4_3}
\end{gathered}
+ \frac{1}{2} \times
\begin{gathered}
\includegraphics[scale=.12]{2_2_1}
\end{gathered}
\quad
\begin{gathered}
\includegraphics[scale=.12]{2_1_1}
\end{gathered}
+ \frac{1}{3} \times
\begin{gathered}
\includegraphics[scale=.12]{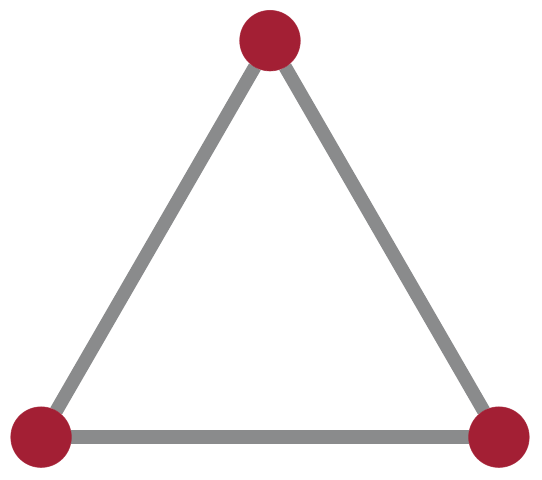}
\end{gathered}
\\
&-\frac{1}{6} \times
\begin{gathered}
\includegraphics[scale=.12]{3_3_1}
\end{gathered}
\quad
\begin{gathered}
\includegraphics[scale=.12]{2_1_1}
\end{gathered}
-\frac{1}{4} \times
\begin{gathered}
\includegraphics[scale=.12]{3_2_1}
\end{gathered}
\quad
\begin{gathered}
\includegraphics[scale=.12]{2_2_1}
\end{gathered}
 \end{aligned}$}
		\\
	  \hline\\*[-2em]
	  (l) &  Planar Event (Approx.) & None &  
{$\!\begin{aligned} 
	\begin{gathered}
	\includegraphics[scale=.12]{4_3_2}
	\end{gathered}
	&\approx
	\frac{1}{2} \times
	\begin{gathered}
	\includegraphics[scale=.12]{5_4_3}
	\end{gathered}
	+ \frac{1}{2} \times
	\begin{gathered}
	\includegraphics[scale=.12]{2_2_1}
	\end{gathered}
	\quad
	\begin{gathered}
	\includegraphics[scale=.12]{2_1_1}
	\end{gathered}
	+ \frac{1}{3} \times
	\begin{gathered}
	\includegraphics[scale=.12]{3_3_1}
	\end{gathered}
	\\
	&-\frac{1}{6} \times
	\begin{gathered}
	\includegraphics[scale=.12]{3_3_1}
	\end{gathered}
	\quad
	\begin{gathered}
	\includegraphics[scale=.12]{2_1_1}
	\end{gathered}
	-\frac{1}{4} \times
	\begin{gathered}
	\includegraphics[scale=.12]{3_2_1}
	\end{gathered}
	\quad
	\begin{gathered}
	\includegraphics[scale=.12]{2_2_1}
	\end{gathered}
	\end{aligned}$}
\\
\hline
\hline
	\end{tabular}
\caption{Labels and names of the twelve observable relations used in our EFP case study.
The third column indicates possible restrictions on their range of applicability, where $M$ is the number of particles in the jet and $n$ is the number of spatial dimensions.
The fourth column gives the corresponding multigraph representations of the linear EFP relations and represents \Eqs{eq:MSE}{eq:functions}, where $y_s$ corresponds to the observable on the left-hand side, $h_a$ corresponds to the EFPs on the right-hand side, and $c_a$ corresponds to the coefficients to be determined.
}
\label{tab:observables}
\end{table*}
}

Many common jet observables, including the jet mass, energy correlation functions~\cite{Larkoski:2013eya}, and angularities~\cite{Ellis:2010rwa,Larkoski:2014pca}, are exact finite linear combinations of EFPs.
This makes them useful targets for our annealing studies since there is a ground truth definition of successful regularized regression.
We consider twelve different linear relations between collider observables and EFPs, which have been extensively studied in \Refs{Komiske:2017aww,Komiske:2019asc}.
These twelve relations, summarized in \Tab{tab:observables}, will serve as benchmarks for testing our QUBO formulation of $\ell_0$-norm regression. In \Tab{tab:observables}, the fourth column represents \Eqs{eq:MSE}{eq:functions}, where $y_s$ corresponds to the observable on the left-hand side, $h_a$ corresponds to the EFPs on the right-hand side, and $c_a$ corresponds to the coefficients to be determined, which optimally match the numbers given in the table.

The first set of observables is given by the infrared- and collinear-safe jet angularities~\cite{Ellis:2010rwa,Larkoski:2014pca} defined as
\begin{align}
\lambda^{(\alpha)} = \sum_{i=1}^Mz_i\theta_i^\alpha,
\label{eq:angthetais}
\end{align}
where $\alpha>0$ is an angular exponent and $\theta_{i}$ denotes the distance of particle $i$ to the $p_T$-weighted centroid axis $(y_J,\phi_J)$ of the jet located at
\begin{equation}
y_J=\sum_{j=1}^M z_j y_j,\quad\phi_J=\sum_{j=1}^M z_j\phi_j.
\label{eq:jetaxis}
\end{equation}
Using Eq.~\eqref{eq:jetaxis}, the angularities in Eq.~\eqref{eq:angthetais} can be expressed in terms of pairwise distances as
\begin{align}
\lambda^{(\alpha)} =\sum_{i_1=1}^M z_{i_1} \left(\sum_{i_2 = 1}^Mz_{i_2}\theta_{i_1i_2}^2 - \frac12 \sum_{i_2 = 1}^M\sum_{i_3 = 1}^M z_{i_2}z_{i_3} \theta_{i_2i_3}^2\right)^{\alpha/2},
\label{eq:angthetaijs}
\end{align}
where $\theta_{ij} = (\Delta y_{ij}^2 + \Delta \phi_{ij}^2)^{1/2}$.

For even $\alpha$, the parenthetical in \Eq{eq:angthetaijs} can be expanded and identified to be a linear combination of EFPs with $N = \alpha$ and $d=\alpha$~\cite{Gur-Ari:2011cjr}.
Focusing on the cases $\alpha \in \{2, 4, 6\}$ and using the multigraph representation of \Eq{eq:EFP}, we can write down the following linear relations for the jet angularities:
\begin{align}
\lambda^{(2)} &= \frac{1}{2} \times
\begin{gathered}
\includegraphics[scale=.15]{2_1_1}
\end{gathered}, \label{eq:lambda2}
\\
\lambda^{(4)} &= \frac{1}{2} \times
\begin{gathered}
\includegraphics[scale=.15]{3_2_1}
\end{gathered}
- \frac{3}{4} \times
\begin{gathered}
\includegraphics[scale=.15]{2_1_1}
\end{gathered}
\quad
\begin{gathered}
\includegraphics[scale=.15]{2_1_1}
\end{gathered},
\\
\lambda^{(6)} &= \frac{1}{2} \times
\begin{gathered}
\includegraphics[scale=.15]{4_3_1}
\end{gathered}
-\frac{3}{2} \times
\begin{gathered}
\includegraphics[scale=.15]{3_2_1}
\end{gathered}
\quad
\begin{gathered}
\includegraphics[scale=.15]{2_1_1}
\end{gathered}
+ \frac{5}{8} \times
\begin{gathered}
\includegraphics[scale=.15]{2_1_1}
\end{gathered}
\quad
\begin{gathered}
\includegraphics[scale=.15]{2_1_1}
\end{gathered}
\quad
\begin{gathered}
\includegraphics[scale=.15]{2_1_1}
\end{gathered}.
\end{align}
In these three multigraph representations, each edge $(k,l)$ corresponds to a term $\theta_{i_k i_l}$ in \Eq{eq:angthetaijs}, and each vertex $j$ corresponds to a summation $\sum_{i_j=1}^M z_{i_j}$.

Next, we consider a jet observable based on the two-dimensional geometric moment tensor of the energy distribution in the $(y,\phi)$-plane~\cite{Gur-Ari:2011cjr,Gallicchio:2012ez}:
\begin{equation}
	C = \sum_{i = 1}^M z_i
	\begin{bmatrix}
		(y_i - y_J)^2 & (\phi_i - \phi_J)(y_i - y_J) \\
		(\phi_i - \phi_J)(y_i - y_J) & (\phi_i - \phi_J)^2
	\end{bmatrix},
	\label{eq:C}
\end{equation}
where the distances are measured with respect to the $p_T$-weighted centroid axis $(y_J,\phi_J)$ of the jet in \Eq{eq:jetaxis}.
Both the trace and the determinant of this matrix can be expressed as a linear combination of EFPs.
The trace is related to the $\alpha = 2$ angularity, while the determinant satisfies~\cite{Komiske:2017aww}
\begin{equation}
\det C = \frac{1}{4} \times
\begin{gathered}
\includegraphics[scale=.15]{3_2_1}
\end{gathered}
- \frac{1}{2} \times
\begin{gathered}
\includegraphics[scale=.15]{2_2_1}
\end{gathered}.
\label{eq:detC}
\end{equation}
In \Ref{Komiske:2019asc}, a variety of relations were derived from cutting the graph nodes.
These relations only hold for a limited number of particles, and they can be derived from the fact that anti-symmetrizing $L$ indices of a tensor in $M$ dimensions yields zero for $L > M$.
A useful organizational scheme for the EFPs is by the number of edges $d$ in the associated multigraph.
We consider two linear relations at $d=3$, called ``Triple Dumbbell'' and ``Lollipop,'' which are valid only for events containing $M \leq 2$ particles~\cite{Komiske:2019asc}:
\begin{align}
M \leq 2: \quad
\begin{gathered}
\includegraphics[scale=.15]{2_3_1}
\end{gathered}
& = 2 \times
\begin{gathered}
\includegraphics[scale=.15]{3_3_2}
\end{gathered},\label{eq:dumbbell_relation}
\\
M \leq 2: \quad
\begin{gathered}
\includegraphics[scale=.15]{3_3_2}
\end{gathered}
&= 
\begin{gathered}
\includegraphics[scale=.15]{4_3_1}
\end{gathered}
+ 
\begin{gathered}
\includegraphics[scale=.15]{4_3_2}
\end{gathered}.
\label{eq:lollipop_relation}
\end{align}
We consider one example at $d=4$, called ``Five Dots,'' for events containing $M \leq 3$ particles~\cite{Komiske:2019asc}:
\begin{equation}
M \leq 2: \quad
\begin{gathered}
\includegraphics[scale=.15]{5_4_3}
\end{gathered}
=
\begin{gathered}
\includegraphics[scale=.15]{5_4_2}
\end{gathered}
+ \frac{1}{2} \times
\begin{gathered}
\includegraphics[scale=.15]{4_4_5}
\end{gathered}
- \frac{1}{2} \times
\begin{gathered}
\includegraphics[scale=.15]{4_4_4}
\end{gathered}.
\end{equation}

As the last example, we consider a linear relation called ``Planar Event,'' which is subject to a spatial constraint on the event.
In particular, this relation is only applicable to planar events with two (or fewer) spatial degrees of freedom~\cite{Komiske:2019asc}:
\begin{align}
\begin{split}
n \leq 2: \quad
\begin{gathered}
\includegraphics[scale=.12]{4_3_2}
\end{gathered}
&=
\frac{1}{2} \times
\begin{gathered}
\includegraphics[scale=.12]{5_4_3}
\end{gathered}
+ \frac{1}{2} \times
\begin{gathered}
\includegraphics[scale=.12]{2_2_1}
\end{gathered}
\quad
\begin{gathered}
\includegraphics[scale=.12]{2_1_1}
\end{gathered}
+ \frac{1}{3} \times
\begin{gathered}
\includegraphics[scale=.12]{3_3_1}
\end{gathered}
\\
&-\frac{1}{6} \times
\begin{gathered}
\includegraphics[scale=.12]{3_3_1}
\end{gathered}
\quad
\begin{gathered}
\includegraphics[scale=.12]{2_1_1}
\end{gathered}
-\frac{1}{4} \times
\begin{gathered}
\includegraphics[scale=.12]{3_2_1}
\end{gathered}
\quad
\begin{gathered}
\includegraphics[scale=.12]{2_2_1}
\end{gathered}.
\end{split}
\end{align}

A summary of these linear relations is given in \Tab{tab:observables}, along with the restrictions that constrain their range of applicability.
Additionally, we list ``approximate'' linear relations, where we consider exact linear relations outside of their range of applicability.
This allows us to test the performance of sparse linear regression in regimes where we expect to find good, but not perfect, solutions.
In total, we have four exact relations that always hold, four exact relations that hold only with restrictions, and four approximate relations, which yields twelve linear relations that are tested in our numerical study.

\subsection{Data Sets}
\label{sec:data}

For our numerical study, we use a data set from the CMS Open Data Portal \cite{OpenData2014, OpenData2016} in the MOD HDF5 Format~\cite{CMSOpenData}, which was created for jet-based studies.
These dijet events are generated in \textsc{Pythia} 6.4.25 \cite{Sjostrand:2006za}, and we do not consider any detector simulation effects.
In our study, we use 100,000 shower-generated events with $p_T \in [475, 525]$ GeV and absolute values of the rapidity $|y| < 1.9$.
Even though the event samples are weighted, for simplicity we treat the events as having equal weights.%
\footnote{Event weights could be straightforwardly incorporated by generalizing the MSE loss function.}

For most of the observables in \Sec{sec:observables}, we can use generic events to test the given functional relations.
In specific cases, however, we need to constrain the data to incorporate the specific conditions listed in \Tab{tab:observables}.
For example, some of the linear relations are only applicable to planar events or to events with a specific number of particles.
To generate planar events, we constrain the particle motion to two spatial dimensions, which is accomplished by setting the azimuthal angles of all particles to zero, $\phi_i = 0$.
To generate events with a fixed particle number, we consider events with larger particle numbers and sequentially delete random particles until reaching the required number.
In this process, we preserve the total transverse momentum $p_T$ of the jet by rescaling the transverse momenta of the remaining particles.

As mentioned above, we only apply this preprocessing when testing the ``exact'' linear relations that are subject to constraints.
When testing the ``approximate'' versions of these linear relations, we leave the data unmodified.

\section{Numerical Results} \label{sec:NumResults}

We now present the results of our numerical study, in which we apply sparse regression to test the twelve linear EFP relations in \Tab{tab:observables}.
Since we have two different annealing encoding schemes, five different optimization strategies, and twelve observable relations to test, we only present selected results to highlight the main points of our study; we present some additional results in App.~\ref{app:additional_plots}.

First, we demonstrate the advantage of degeneracy engineering by comparing the baseline encoding from \Sec{sec:singleABE} to the degeneracy-engineered encoding in \Sec{sec:DoubleEnc}. 
Second, we demonstrate the advantage of the refinement approach introduced in \Sec{sec:Heuristics}, showing that refined $\ell_0$-norm regression performs better than its unrefined version.
Third, we demonstrate the advantage of $\ell_0$-norm regularization, by showing that it yields a better sparsity/performance trade-off than $\ell_1$- or $\ell_2$-norm regularization.
Finally, we assess the potential gains from quantum computing by comparing classical annealing to PIMC, finding no dramatic difference in performance.

\subsection{Advantage of Degeneracy Engineering}
\label{sec:ResultsDegEng}

\begin{figure*}
\centering
\subfloat[]{
	\includegraphics[width=0.27\textwidth]{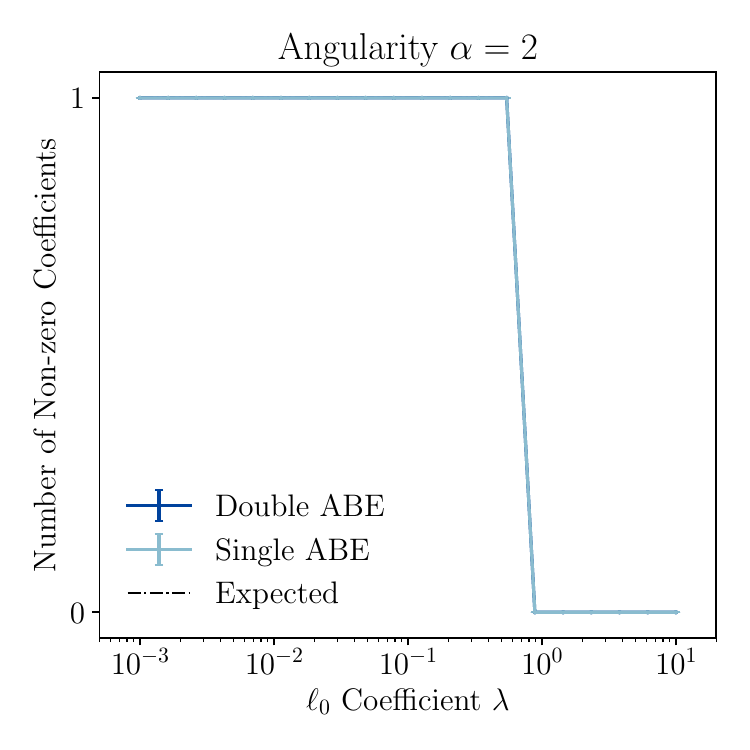}
}
\subfloat[]{
	\includegraphics[width=0.27\textwidth]{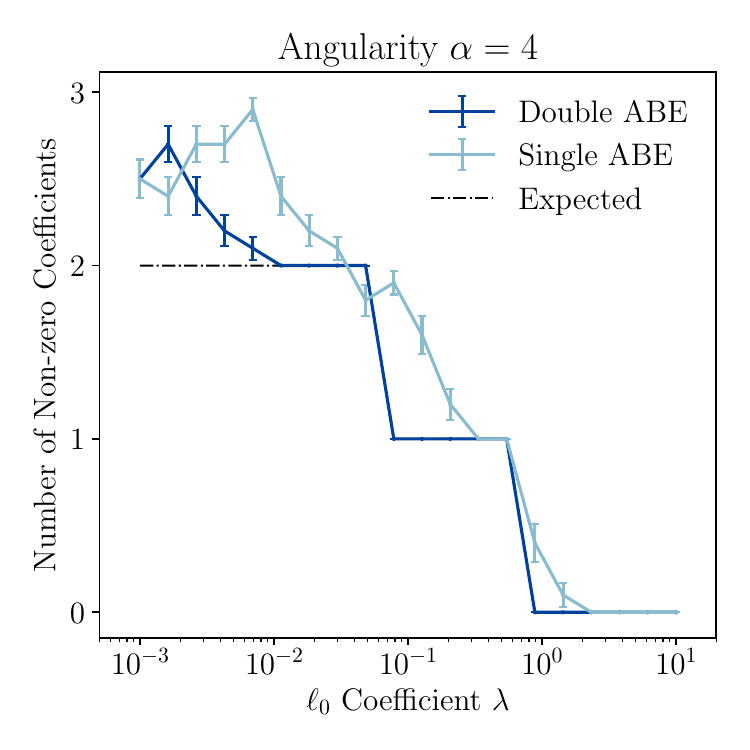}
}
\subfloat[]{
	\includegraphics[width=0.27\textwidth]{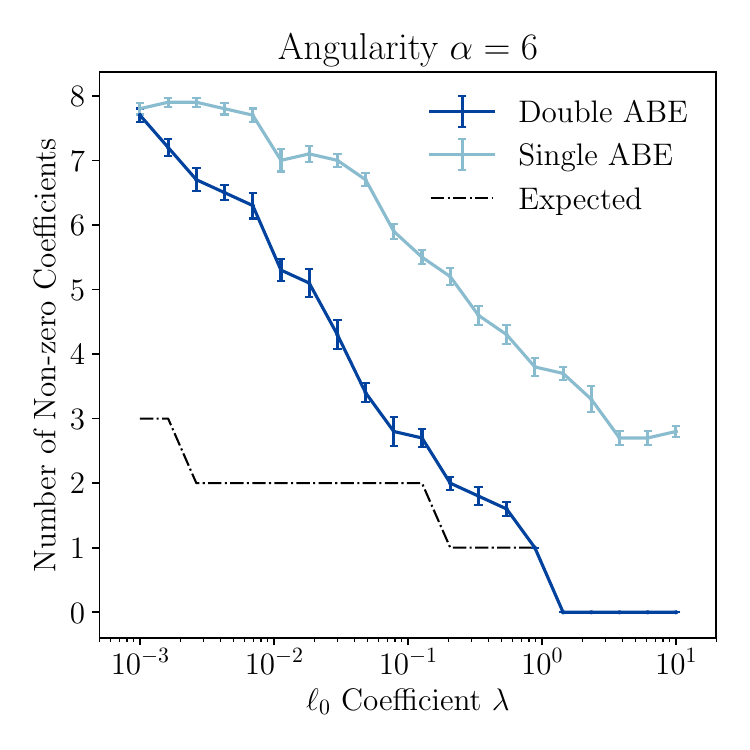}
}\\
\subfloat[]{
	\includegraphics[width=0.27\textwidth]{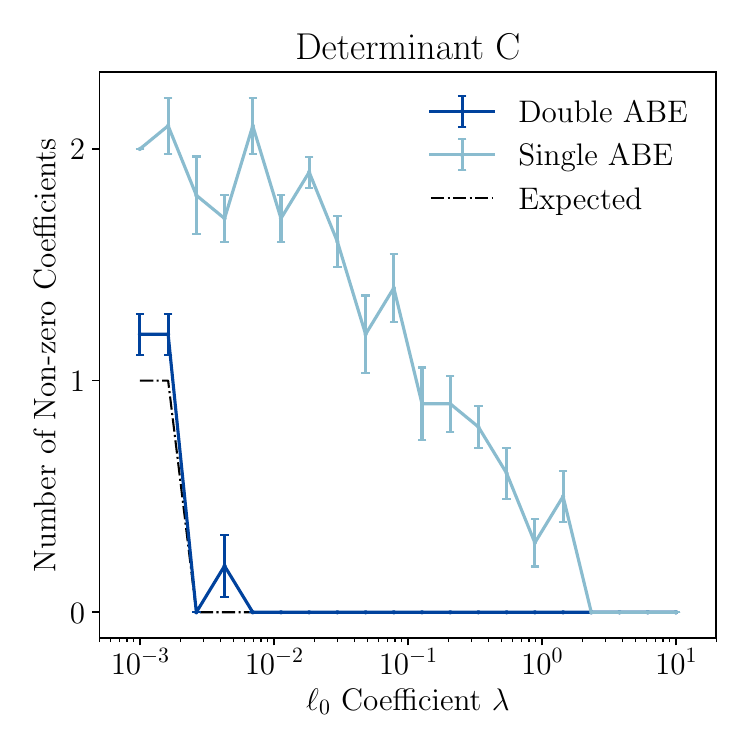}
}
\subfloat[]{
	\includegraphics[width=0.27\textwidth]{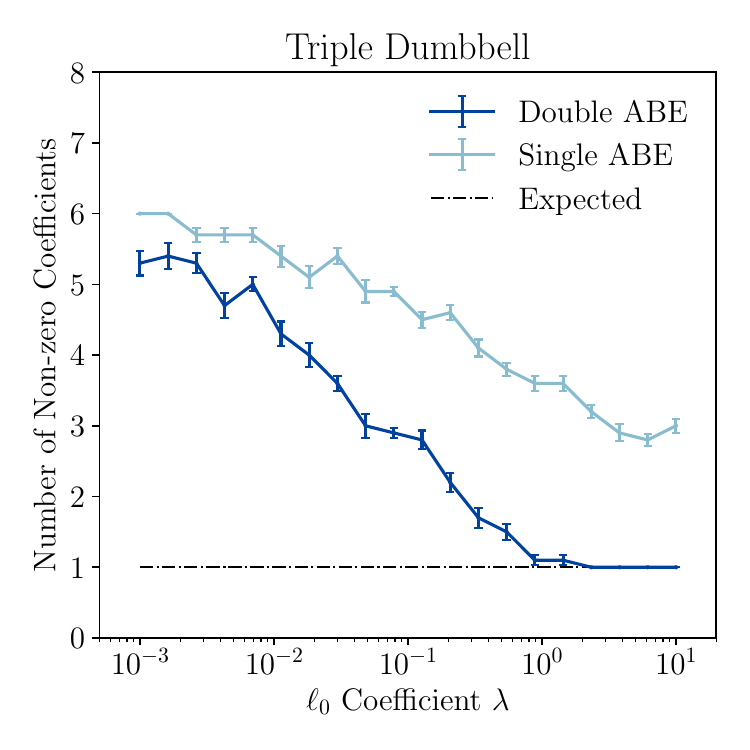}
}
\subfloat[]{
	\includegraphics[width=0.27\textwidth]{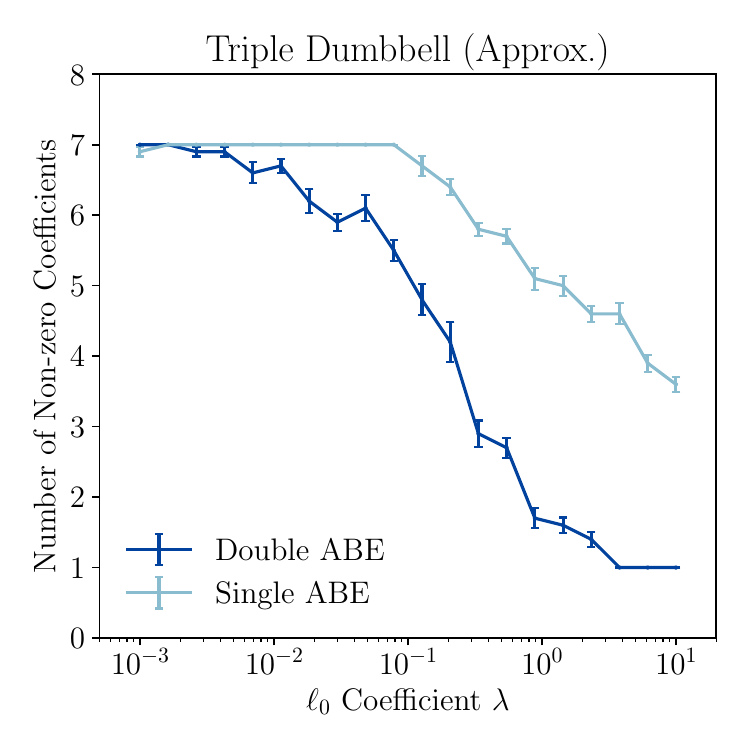}
}\\
\subfloat[]{
	\includegraphics[width=0.27\textwidth]{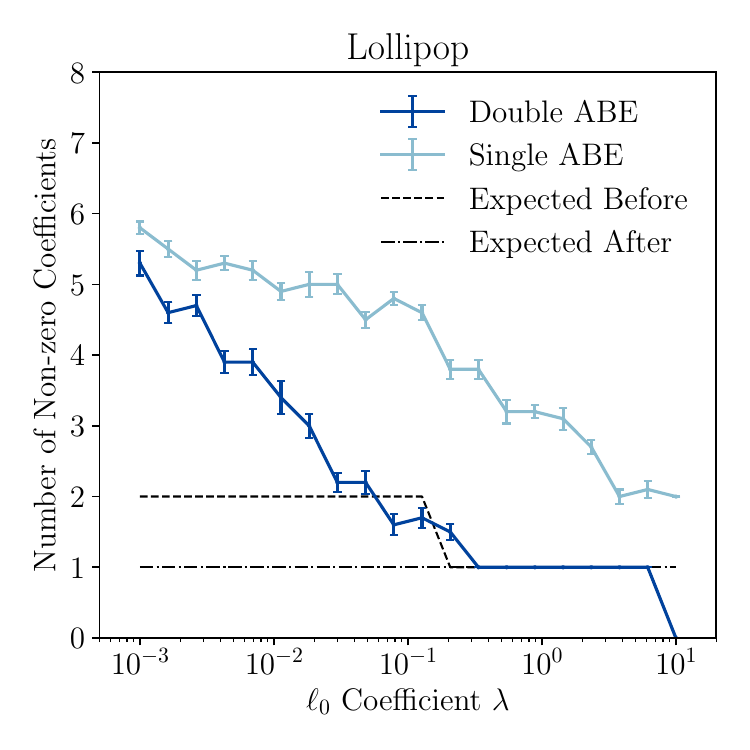}
	\label{fig:3g_Lolipop}
	}
\subfloat[]{
	\includegraphics[width=0.27\textwidth]{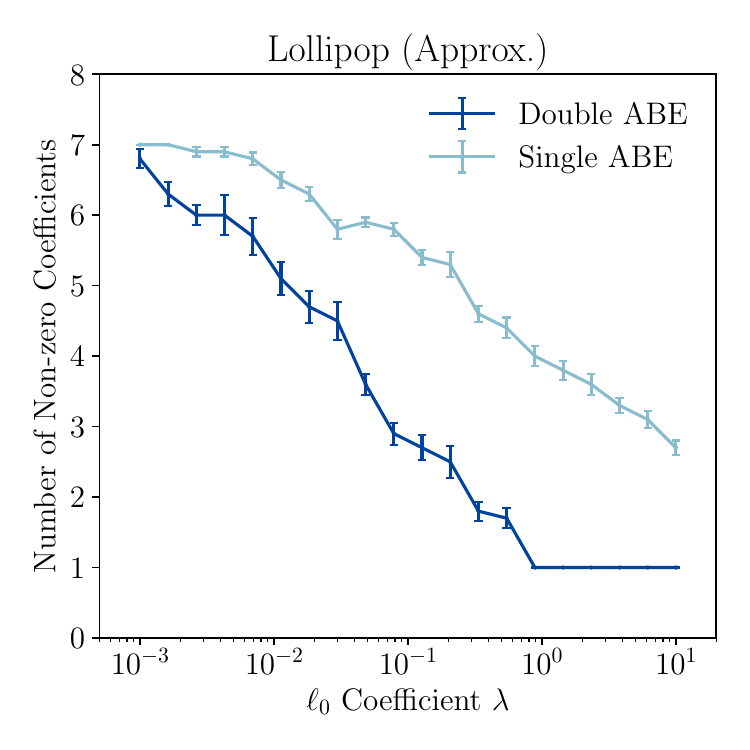}
}
\subfloat[]{
	\includegraphics[width=0.27\textwidth]{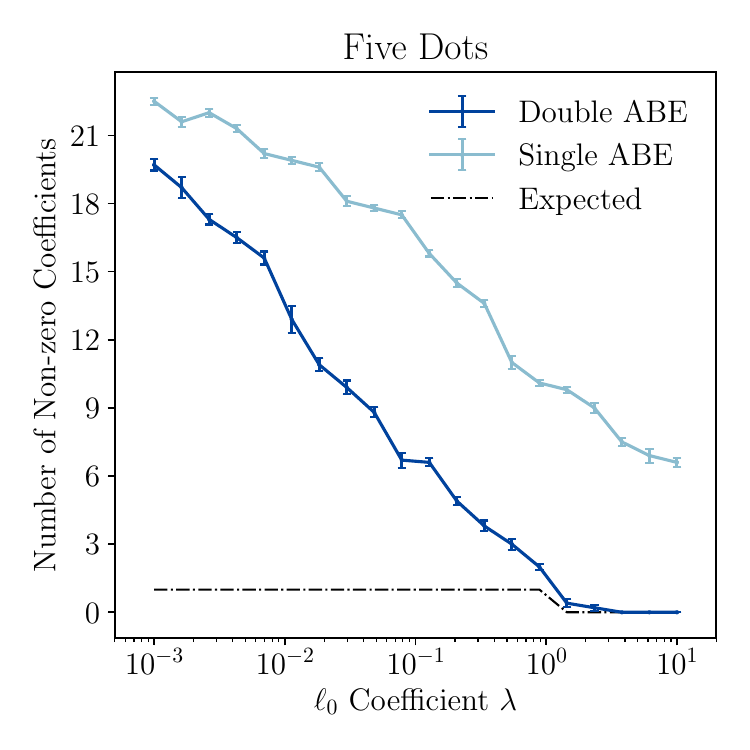}
}\\
\subfloat[]{
	\includegraphics[width=0.27\textwidth]{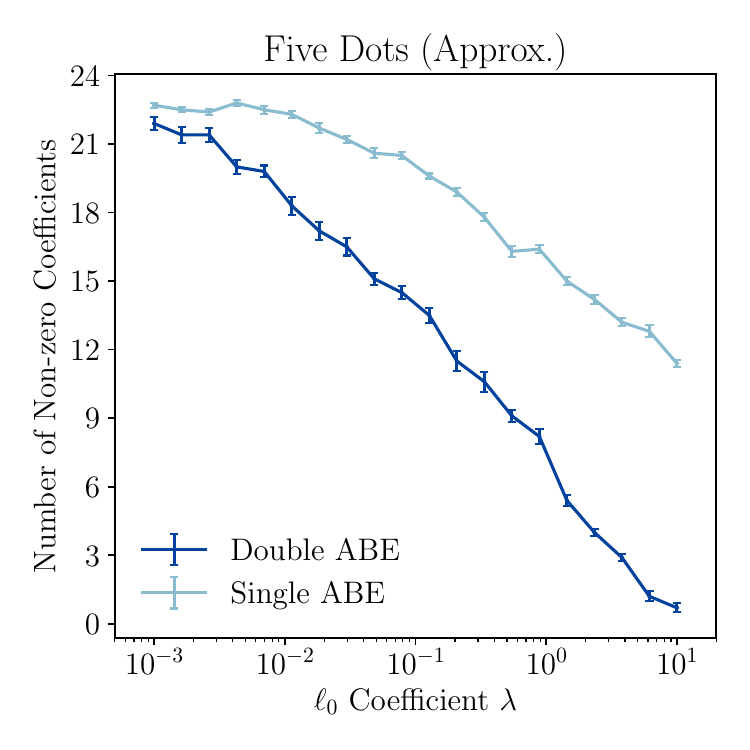}
}
\subfloat[]{
	\includegraphics[width=0.27\textwidth]{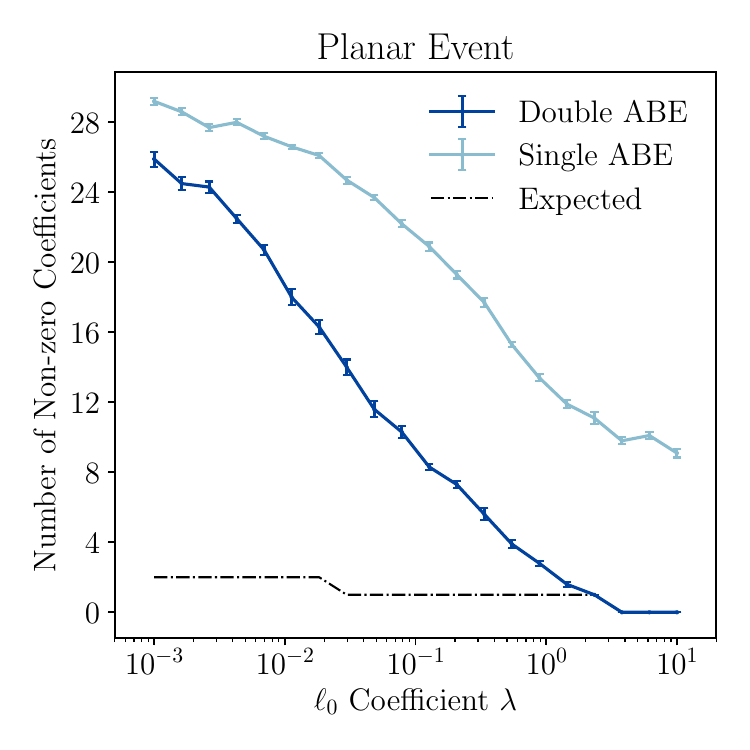}
}
\subfloat[]{
	\includegraphics[width=0.27\textwidth]{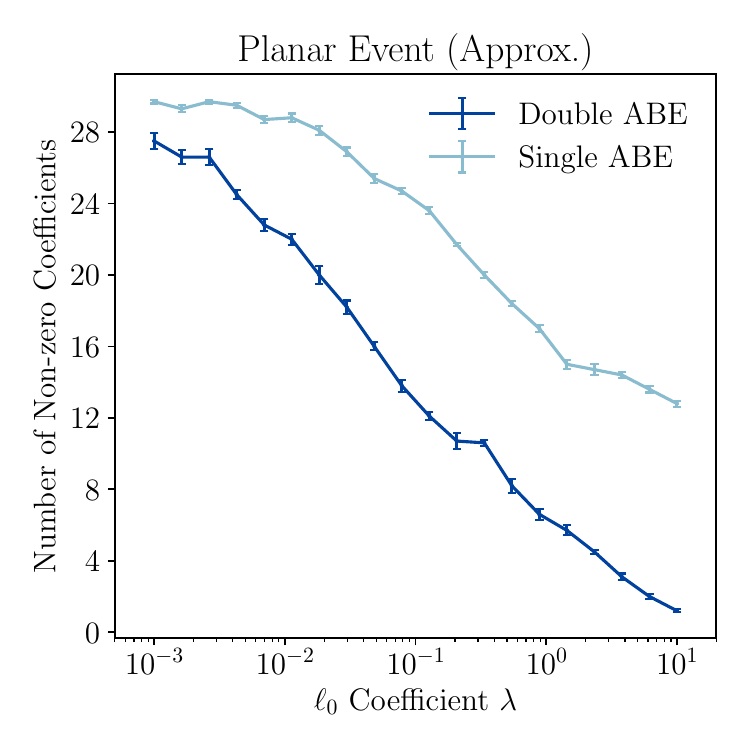}
}
	\caption{
	Number of non-zero fit coefficients as a function of the $\ell_0$-norm coefficient $\lambda$, comparing single ABE (light blue) with double ABE (dark blue) on classical annealing.
	The twelve observable relations and their (a)--(l) labels are given in \Tab{tab:observables}.}
	\label{fig:double_vs_single_nnz}
\end{figure*}

To evaluate the performance of degeneracy engineering from \Sec{sec:DegEng}, we compare the performance of the single ABE in \Eq{eq:1lb} versus the double ABE in \Eq{eq:2lb}.
For both encodings, we use the same classical annealing algorithm with the same training parameters for each observable.
As described in \Sec{sec:ReviewAnnealing}, this optimization algorithm is based on classical population annealing with a geometric annealing schedule, with inverse temperature at annealing step $i$ given by
\begin{equation}
  \beta_i=\beta_0\left(\frac{\beta_\ell}{\beta_0}\right)^{\frac{i}{\ell}}.
  \label{eq:geo_sch}
\end{equation}
The population is initialized at the temperature $\beta_0=1/T_0 = 10$ and then cooled to the temperature $\beta_\ell=1/T_l = 10^{10}$ by an annealing schedule of $\ell=2^{14}$ steps.
For the $\ell_0$-norm coefficient $\lambda$, we study a range that spans four orders of magnitude, $\lambda \in [10^{-3},10]$.

In \Fig{fig:double_vs_single_nnz}, we show the number of non-zero fit coefficients as a function of $\lambda$, comparing the single ABE (light blue) to the double ABE (dark blue).
The twelve plots in this figure correspond to the twelve different relations in \Tab{tab:observables}.
The results are averaged over ten independent runs, with the standard deviation shown as error bars.
For all observables, we find that the degeneracy-engineered version with double ABE performs either equally well or better in terms of the number of identified non-zero fit coefficients.
In \Fig{fig:double_vs_single_loss} of \App{app:additional_plots}, we plot the loss function versus $\lambda$ as an alternative way to highlight the improved behavior of the double ABE.

For all non-approximate relations in \Tab{tab:observables}, we can analytically compute the best-case theoretical expectation by considering all possible combinations of non-zero coefficients given by a particular analytical relation.
In \Fig{fig:double_vs_single_nnz}, this analytical result is displayed as the black-dashed ``expected'' line. 
Interestingly, for the Lollipop observable in \Fig{fig:3g_Lolipop}, our original theoretical expectation from \Eq{eq:lollipop_relation} was outperformed by the double ABE algorithm.
Indeed, the corresponding black-dashed line (``Expected Before'') has more non-zero coefficients than we found numerically.
This inspired us to find a different analytic relation for the Lollipop observable, which yielded an improved black-dotted line (``Expected After'') with a smaller loss function:
\begin{equation}
M \leq 2: \quad
\begin{gathered}
\includegraphics[scale=.12]{3_3_2}
\end{gathered}
= \frac{1}{2} \times
\begin{gathered}
\includegraphics[scale=.12]{2_3_1}
\end{gathered}.
\label{eq:new_relation}
\end{equation}
Amusingly, this is just the reversed Triple Dumbbell relation from \Eq{eq:dumbbell_relation}.

\subsection{Advantage of Refinement}
\label{sec:l0_norm_refinement}

\begin{figure*}
\centering
\subfloat[]{
	\includegraphics[width=0.27\textwidth]{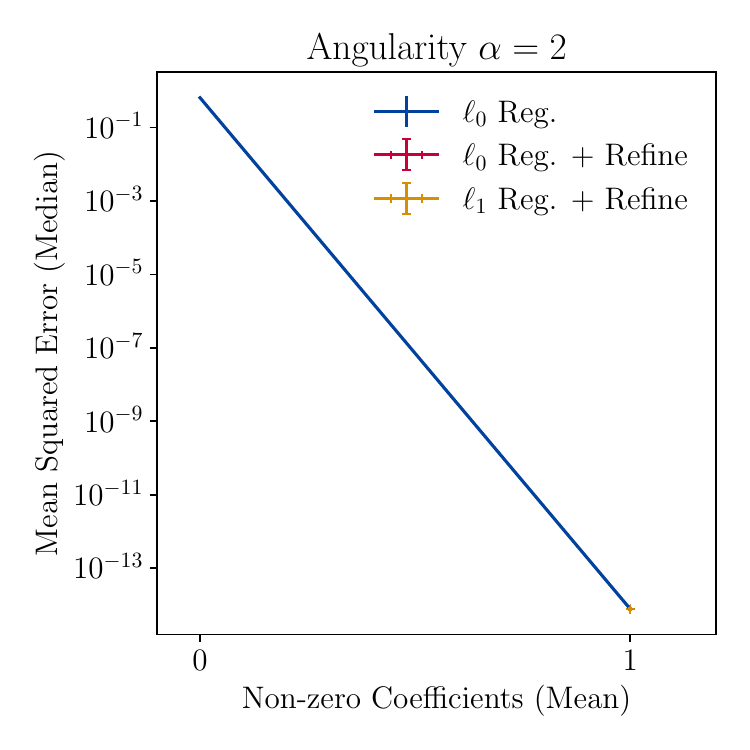}
}
\subfloat[]{
	\includegraphics[width=0.27\textwidth]{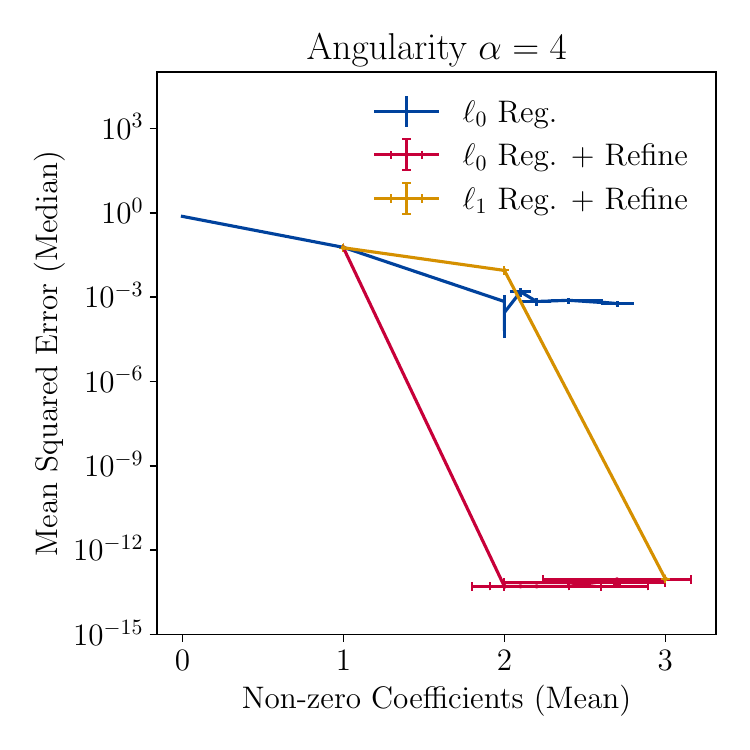}
}
\subfloat[]{
	\includegraphics[width=0.27\textwidth]{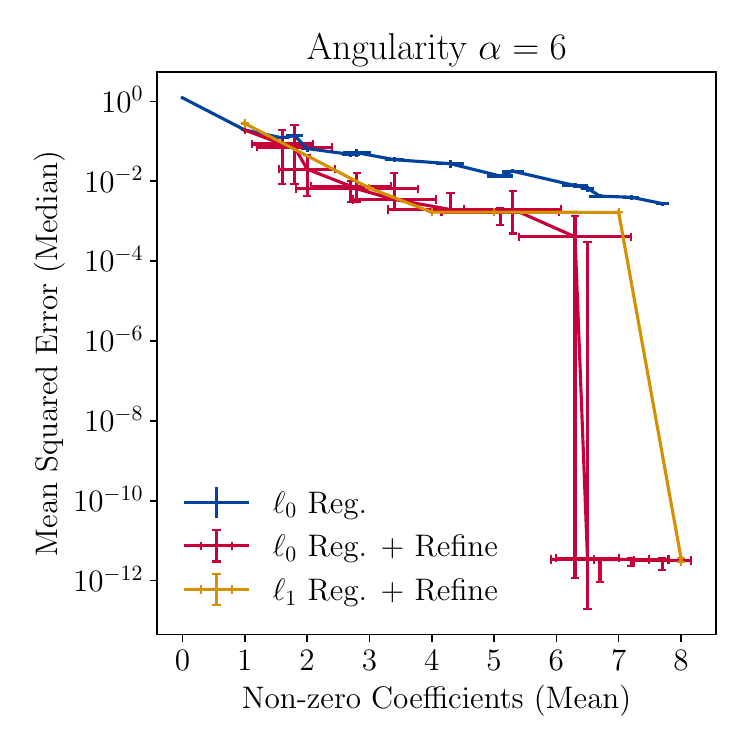}
}\\
\subfloat[]{
	\includegraphics[width=0.27\textwidth]{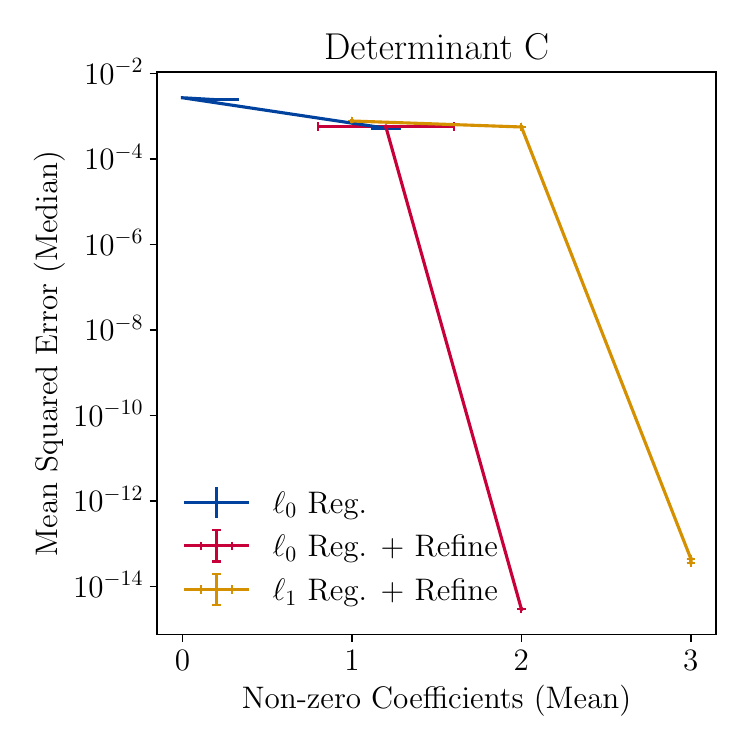}
}
\subfloat[]{
	\includegraphics[width=0.27\textwidth]{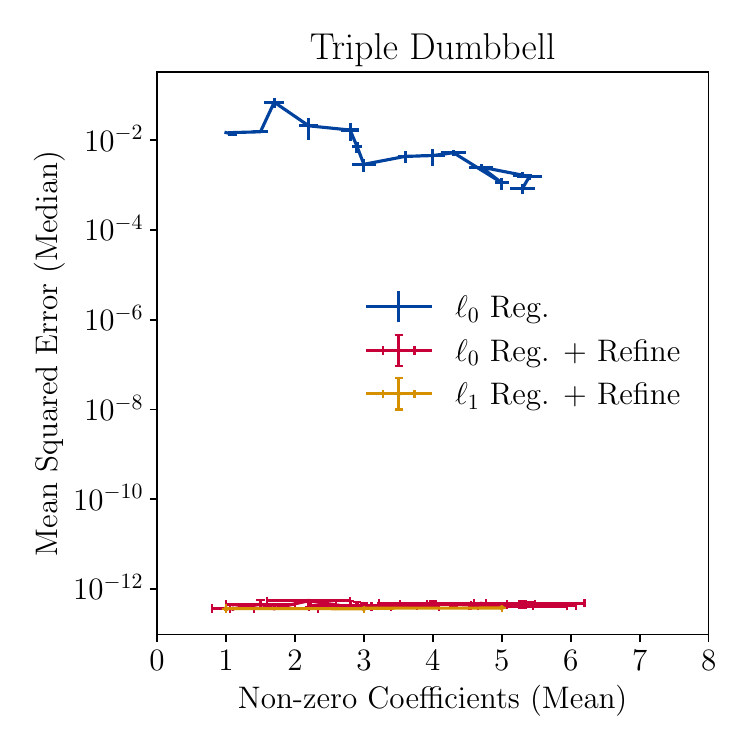}
}
\subfloat[]{
	\includegraphics[width=0.27\textwidth]{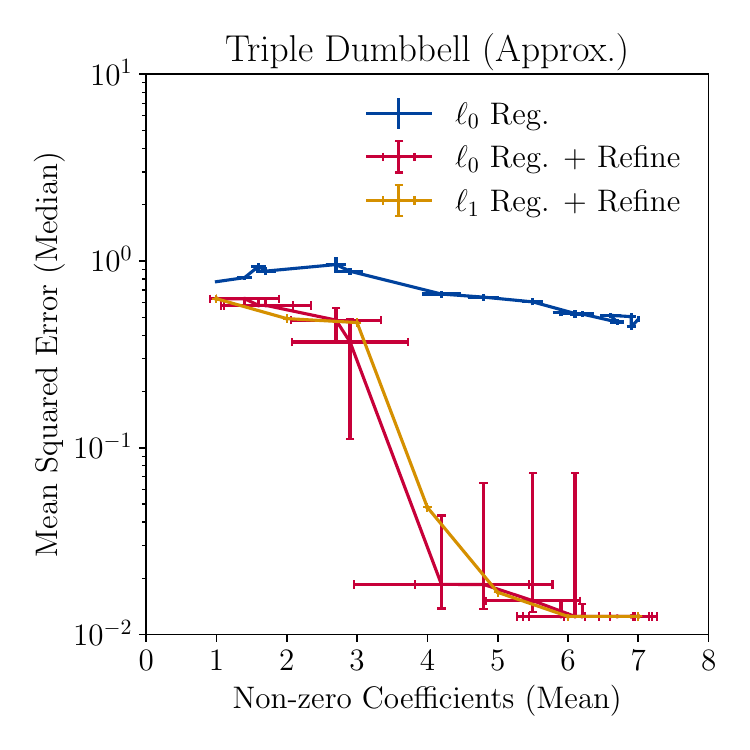}
}\\
\subfloat[]{
	\includegraphics[width=0.27\textwidth]{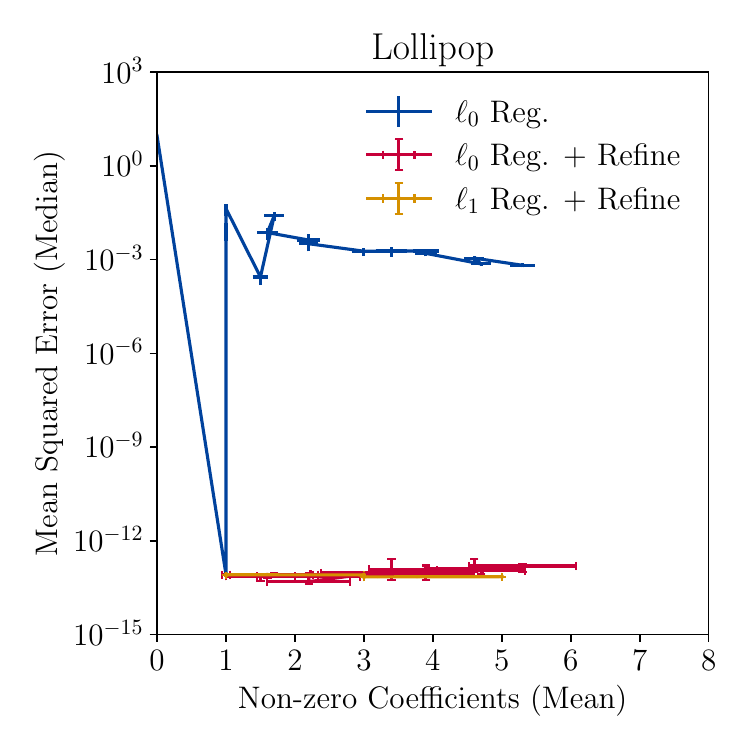}
}
\subfloat[]{
	\includegraphics[width=0.27\textwidth]{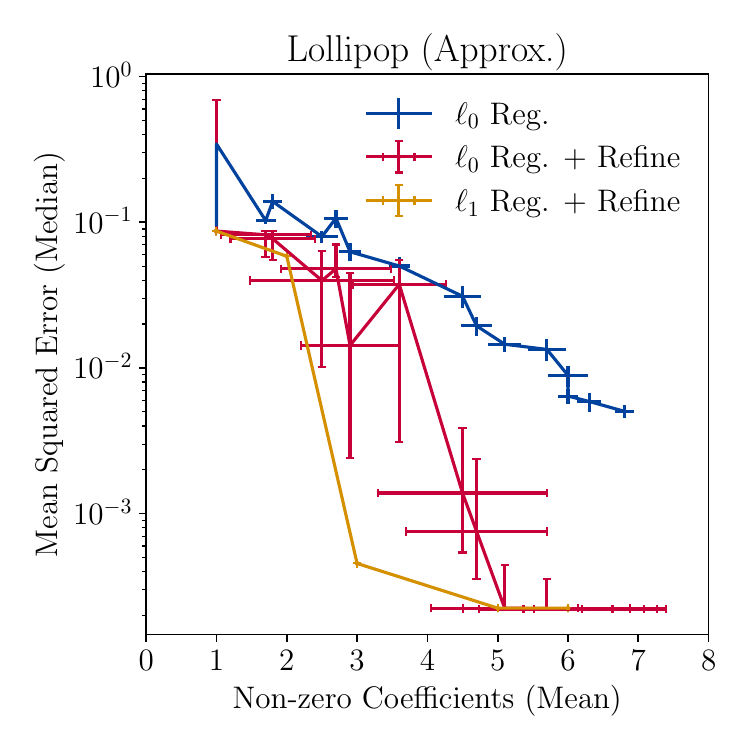}
}
\subfloat[]{
	\includegraphics[width=0.27\textwidth]{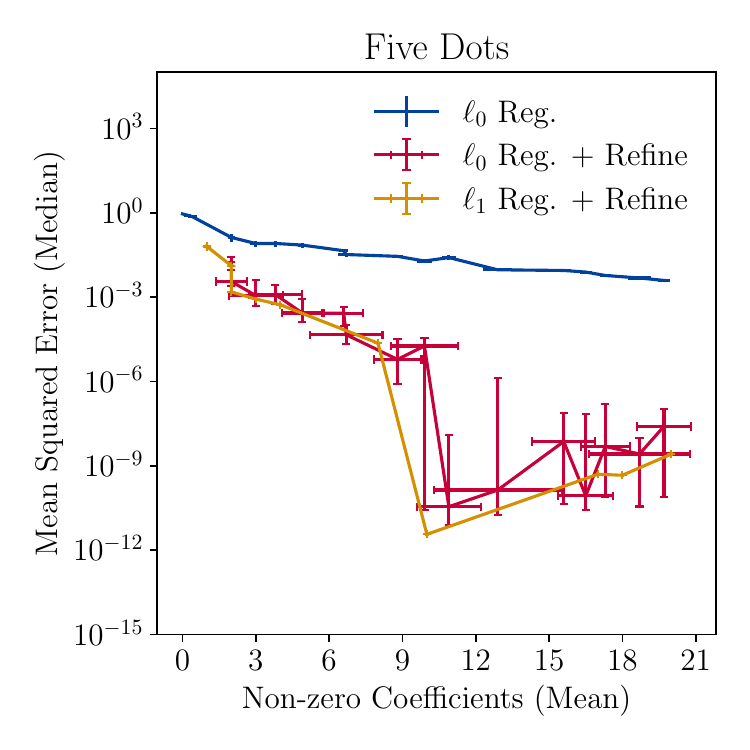}
}\\
\subfloat[]{
	\includegraphics[width=0.27\textwidth]{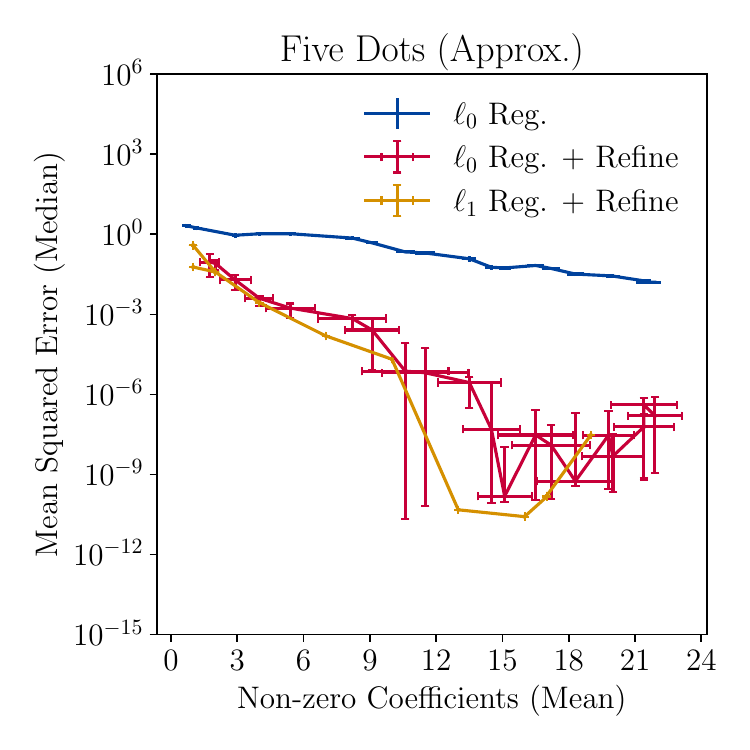}
}
\subfloat[]{
	\includegraphics[width=0.27\textwidth]{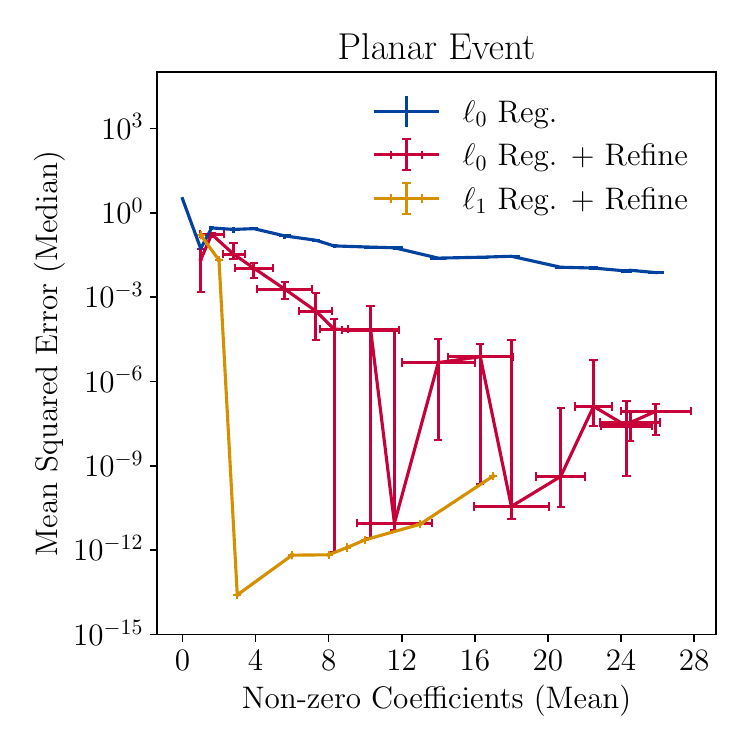}
}
\subfloat[]{
	\includegraphics[width=0.27\textwidth]{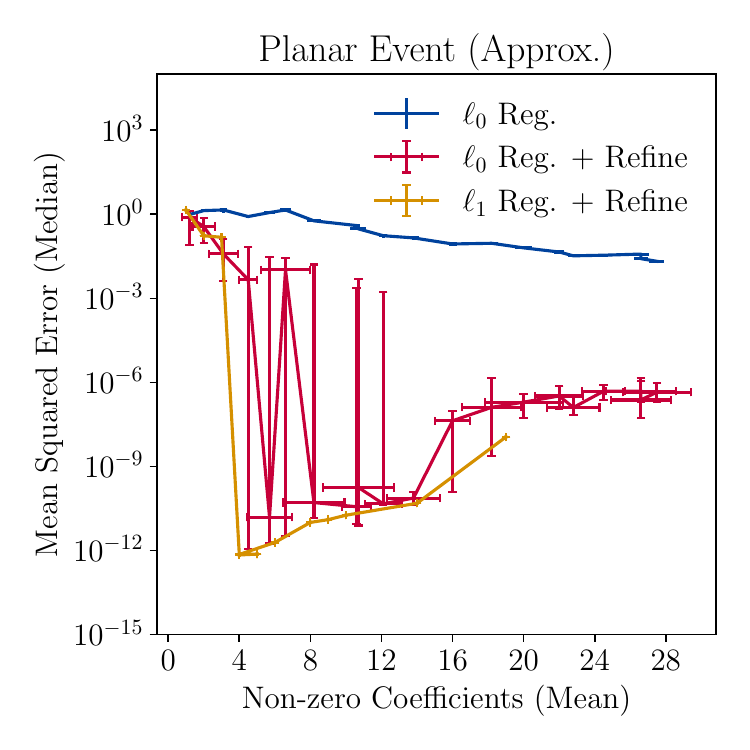}
}
	\caption{
	Median MSE loss function in \Eq{eq:MSE} as a function of the mean number of non-zero coefficients, comparing $\ell_0$-norm regression (blue) with refined $\ell_0$-norm (red) and $\ell_1$-norm (orange) regression.  The same twelve observables from \Tab{tab:observables} with (a)--(l) labels are shown.}
	\label{fig:L0_vs_L0ref_vs_L1ref}
\end{figure*}

\begin{figure*}
\centering
\subfloat[]{
	\includegraphics[width=0.27\textwidth]{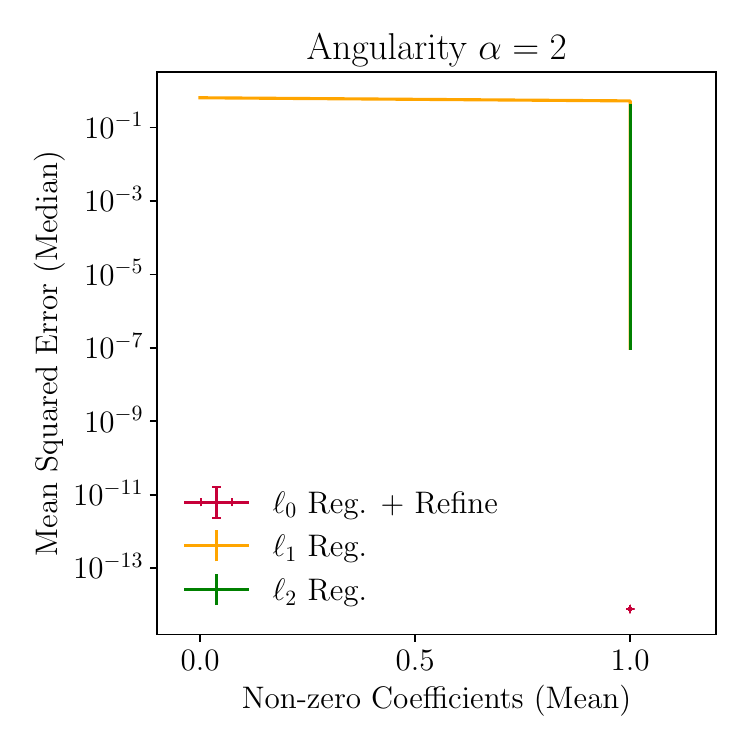}
}
\subfloat[]{
	\includegraphics[width=0.27\textwidth]{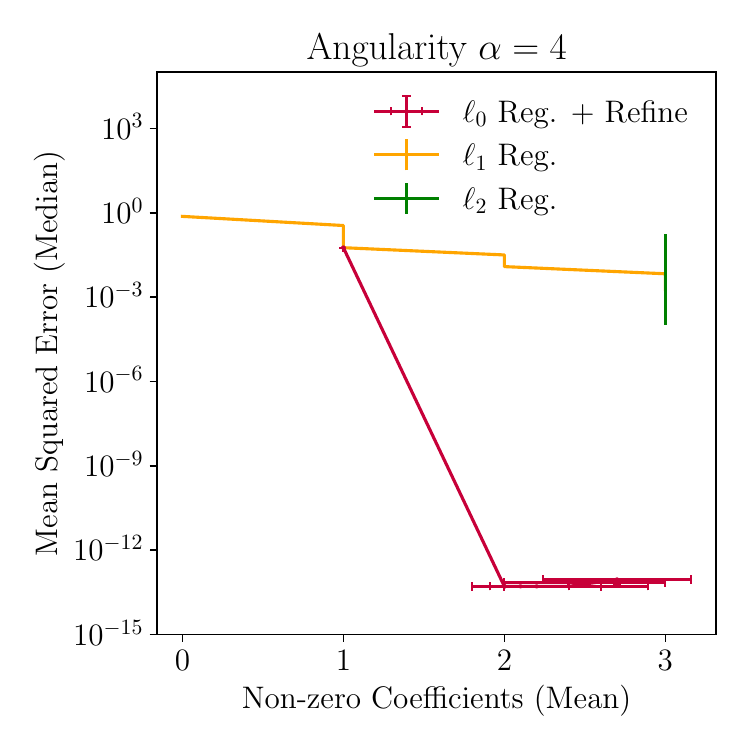}
}
\subfloat[]{
	\includegraphics[width=0.27\textwidth]{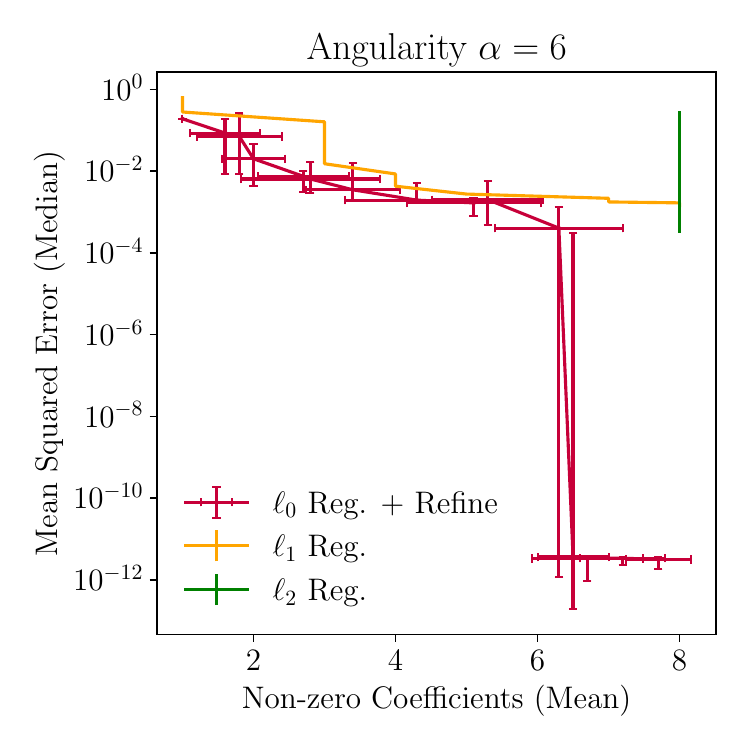}
}\\
\subfloat[]{
	\includegraphics[width=0.27\textwidth]{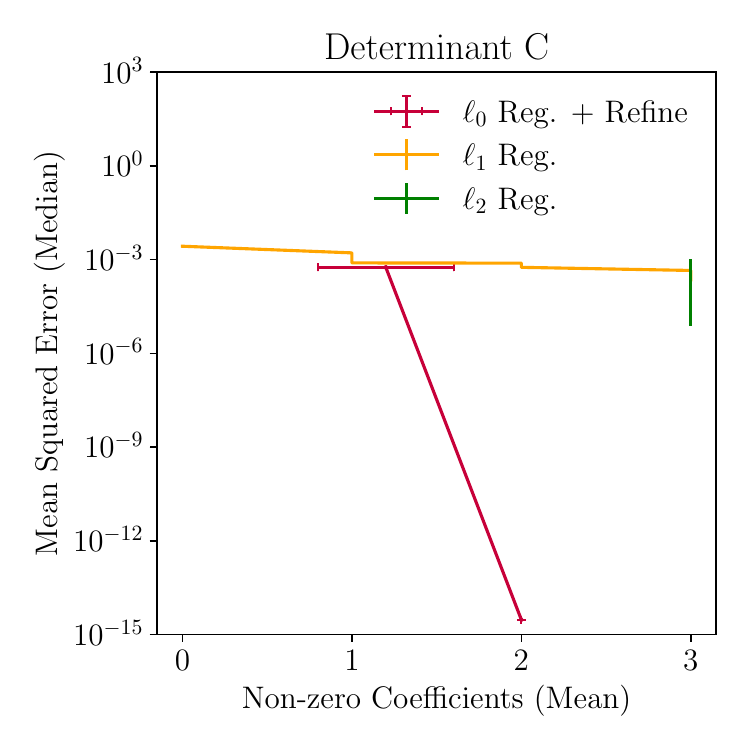}
}
\subfloat[]{
	\includegraphics[width=0.27\textwidth]{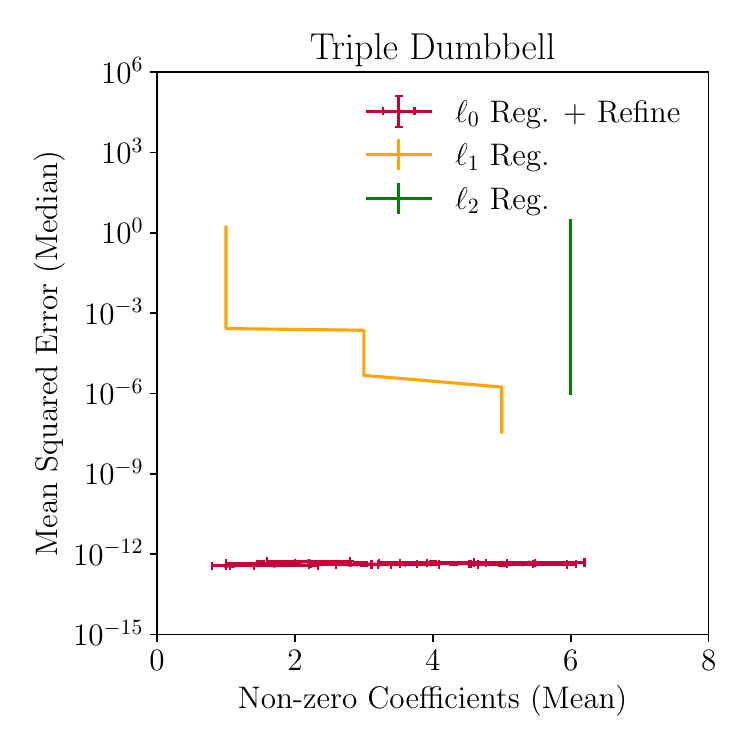}
}
\subfloat[]{
	\includegraphics[width=0.27\textwidth]{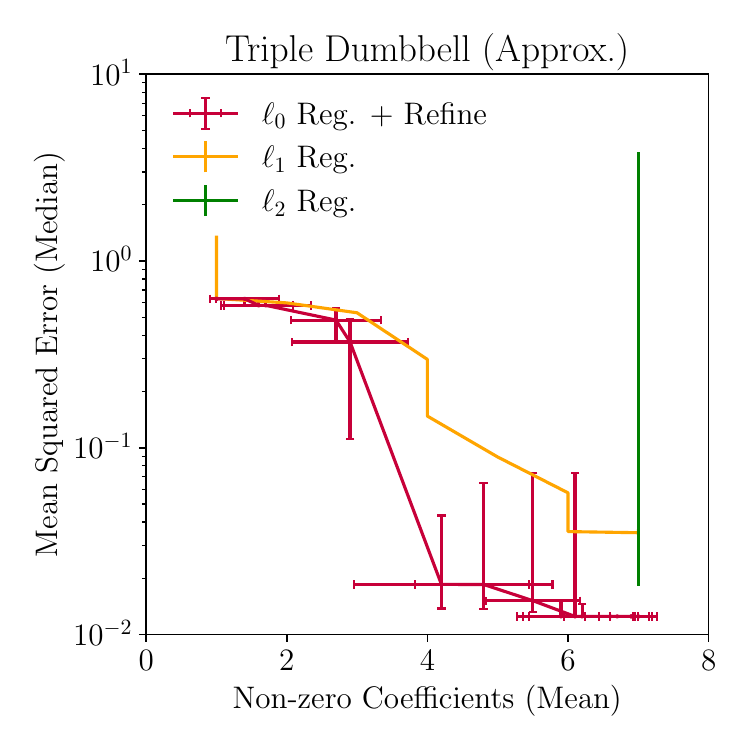}
}\\
\subfloat[]{
	\includegraphics[width=0.27\textwidth]{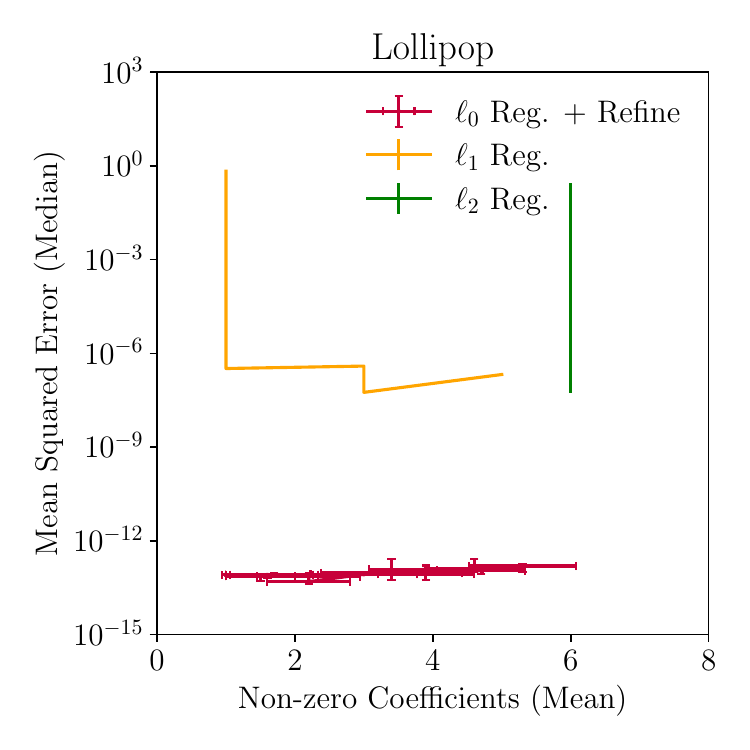}
	\label{fig:L1_vs_L2_vs_L0ref_g_lollipop}
}
\subfloat[]{
	\includegraphics[width=0.27\textwidth]{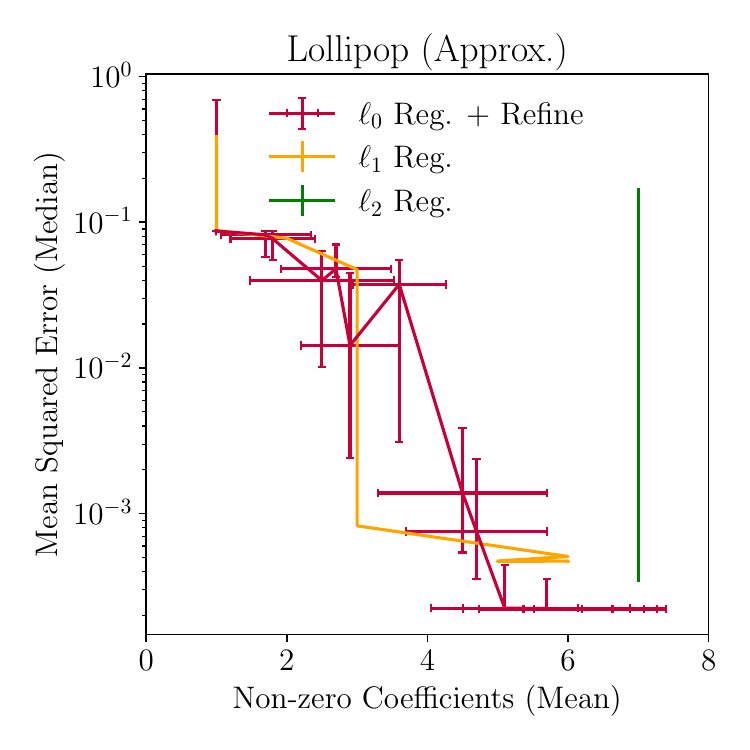}
}
\subfloat[]{
	\includegraphics[width=0.27\textwidth]{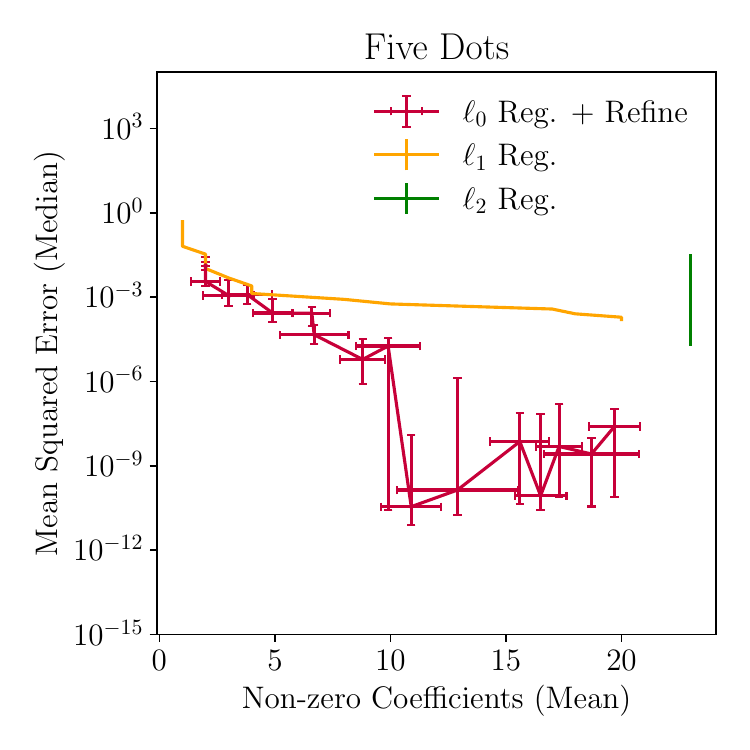}
}\\
\subfloat[]{
	\includegraphics[width=0.27\textwidth]{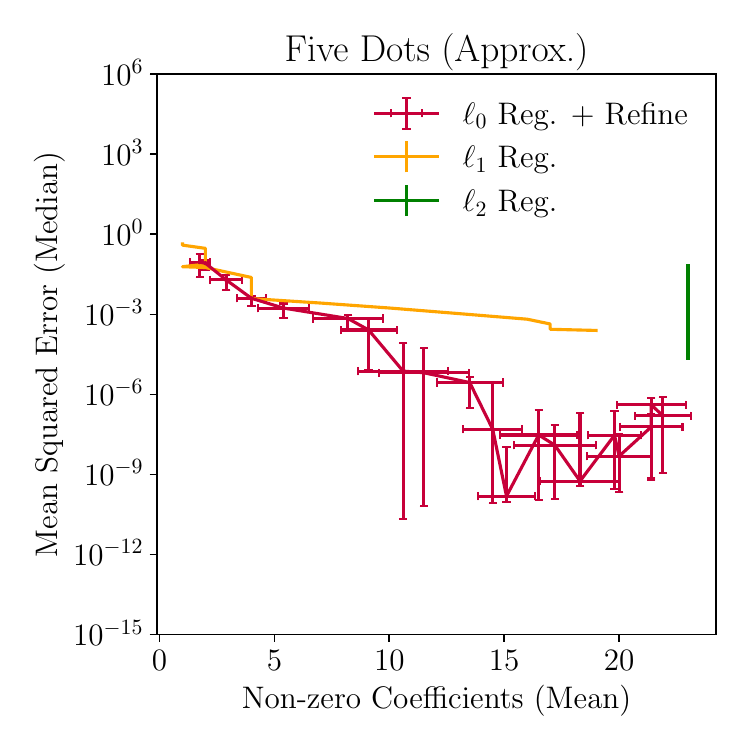}
}
\subfloat[]{
	\includegraphics[width=0.27\textwidth]{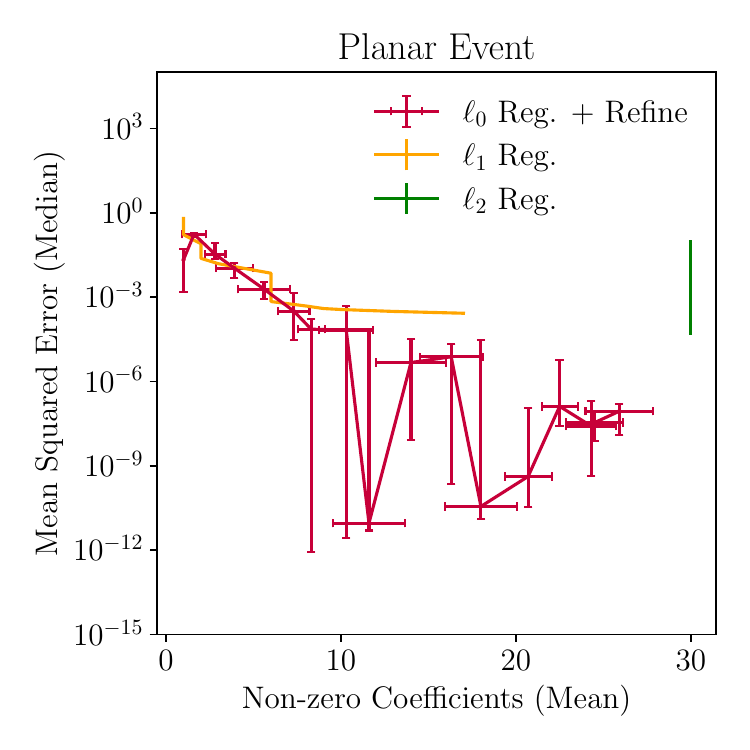}
}
\subfloat[]{
	\includegraphics[width=0.27\textwidth]{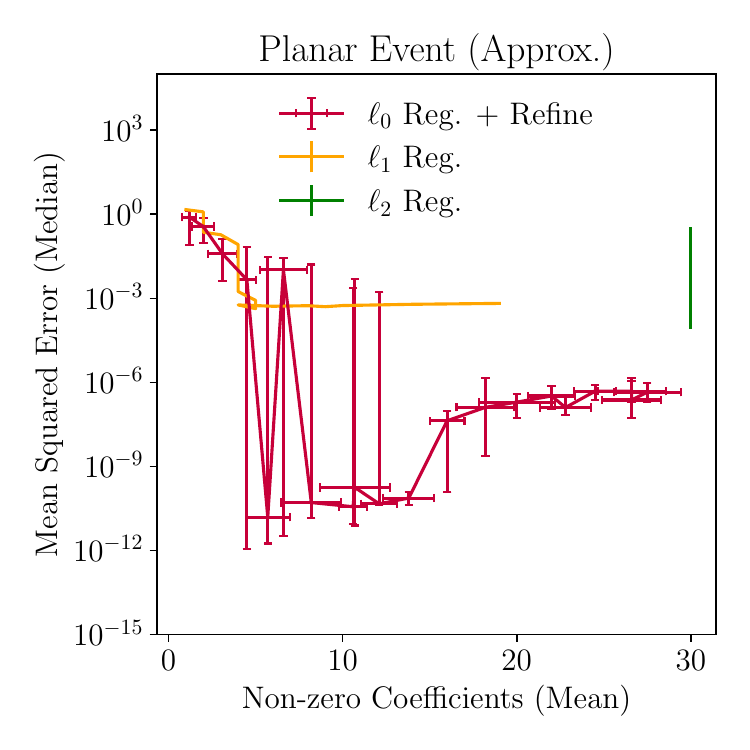}
}
	\caption{Same as \Fig{fig:L0_vs_L0ref_vs_L1ref}, but now comparing refined $\ell_0$-norm regression (red) to standard $\ell_1$-norm (orange) and $\ell_2$-norm (green) regression. The (a)--(l) are defined in \Tab{tab:observables}.}
	\label{fig:L1_vs_L2_vs_L0ref}
\end{figure*}

We now evaluate the refinement approach given by the two novel heuristics introduced in \Sec{sec:Heuristics}.
For this, we use the degeneracy-engineered classical annealing with double ABE, employing the same annealing parameters, annealing schedule, and observables as in \Sec{sec:ResultsDegEng}.

When studying the performance of the two novel heuristics, we have to account for the fact that $\ell_0$-norm and $\ell_1$-norm regression have different loss functions; see \Eq{eq:regularized}.
This requires us to choose an alternative presentation compared to \Fig{fig:double_vs_single_nnz} since the meaning of $\lambda$ differs.
We choose to plot the median of the unregularized MSE loss function in \Eq{eq:MSE} as a function of the mean number of non-zero fit coefficients, since both of these quantities have meaning for any regularization scheme.
To compute the error bars for the MSE, we take the 25\% and 75\% quantiles from ten distinct runs.
To compute the mean number and the corresponding error bars of the non-zero fit coefficients, we average over these ten distinct runs.

In \Fig{fig:L0_vs_L0ref_vs_L1ref}, we compare the standard $\ell_0$-norm regression (blue) to the two novel heuristics:  refined $\ell_0$-norm regression (red) and refined $\ell_1$-norm regression (orange). 
As explained in \Sec{sec:Heuristics}, we use unregularized regression to refine the non-zero coefficient values while clamping coefficients that were set to zero in the original regularized regression.
Refinement improves the MSE performance of standard $\ell_0$- and $\ell_1$-norm regression with only moderate computational overhead.

The large fluctuations in the median MSE in \Fig{fig:L0_vs_L0ref_vs_L1ref} are due to the fact that even after refinement, single bit flips in the solution can yield large changes in the model.
This makes it somewhat difficult to interpret these plots, but we can draw two general lessons.
First, there is a tradeoff between lowering the number of relevant non-zero fit coefficients---implicitly via making $\lambda$ larger---and increasing the MSE.
As the number of non-zero coefficients decreases, the accuracy of the regression solution worsens as expected.
Second, the refined $\ell_0$-norm regression and the refined $\ell_1$-norm regression perform similarly well for all twelve observables we studied.
For a fixed number of non-zero coefficients, both refined heuristics yield substantially lower MSE compared to unrefined $\ell_0$-norm regression.

\subsection{Advantage of $\ell_0$-Norm Regularization}
\label{sec:l0_norm_reg_num}

The key premise of our analysis is that $\ell_0$-norm regularization should yield sparser solutions to EFP regression problems than $\ell_1$- and $\ell_2$-norm regularization.
To test this, we compare the refined version of $\ell_0$-norm regularization to the standard versions of $\ell_1$- and $\ell_2$-norm regression.
As in \Sec{sec:l0_norm_refinement}, we plot the median MSE loss as a function of the mean number of non-zero fit coefficients.

Results are shown in \Fig{fig:L1_vs_L2_vs_L0ref}, for refined $\ell_0$-norm regression (red), $\ell_1$-norm regression (orange), and $\ell_2$-norm regression (green).
Because $\ell_2$-norm regression does not yield a sparse solution for any value of $\lambda$, the green line is vertical on these plots.
For specific observables, including the Lollipop observable in \Fig{fig:L1_vs_L2_vs_L0ref_g_lollipop}, only refined $\ell_0$-norm regression manages to consistently find the exact solution, independently of the number of non-zero coefficients.
This can be seen from comparing the very small MSE values for the $\ell_0$-norm case to the large MSE values obtained for $\ell_1$- and $\ell_2$-norm regression.
Thus, the heuristic of refined $\ell_0$-norm regression manages to minimize the number of non-zero coefficients as effectively as $\ell_1$-norm regression, while also finding an exact solution.
For all observables, only refined $\ell_0$-norm regression manages to consistently find a nearly exact solution (as measured by MSE), when the number of non-zero coefficients is large.

\subsection{Challenges for Quantum Annealing}
\label{sec:ResultsPIMC}

As our final numerical study, we assess the potential gains from quantum computing by comparing classical annealing to PIMC.
Recall from \Sec{sec:ReviewPIMC} that PIMC serves as a proxy for quantum annealing.
We use the same observables as \Sec{sec:ResultsDegEng}, but a different annealing schedule to attempt an apples-to-apples comparison.
For classical annealing, the distributions are initialized at the inverse temperature $\beta_0=1/T_0 = 10$ and then cooled to the inverse temperature $\beta_\ell=1/T_\ell = 10^{8}$ by a geometric annealing schedule of $\ell=2048$ steps.
For PIMC, $J(s)$  increases geometrically from $10$ to $10^8$, while $\Gamma(s)$ decreases geometrically from $\Gamma=1$ to $\Gamma=0$, once again over $\ell=2048$ annealing steps.
We use double ABE for both methods.

\begin{figure*}
\centering
\subfloat[]{
	\includegraphics[width=0.27\textwidth]{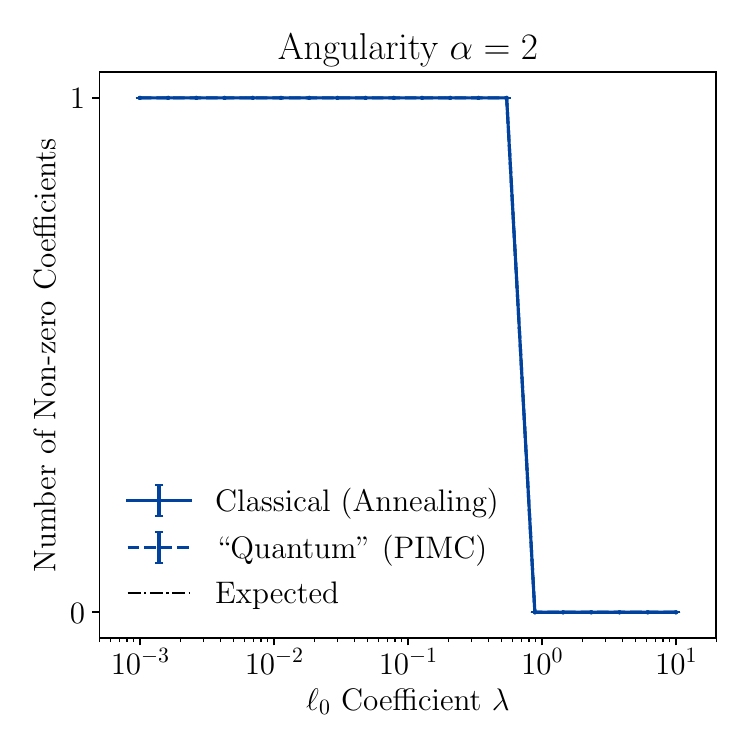}
}
\subfloat[]{
	\includegraphics[width=0.27\textwidth]{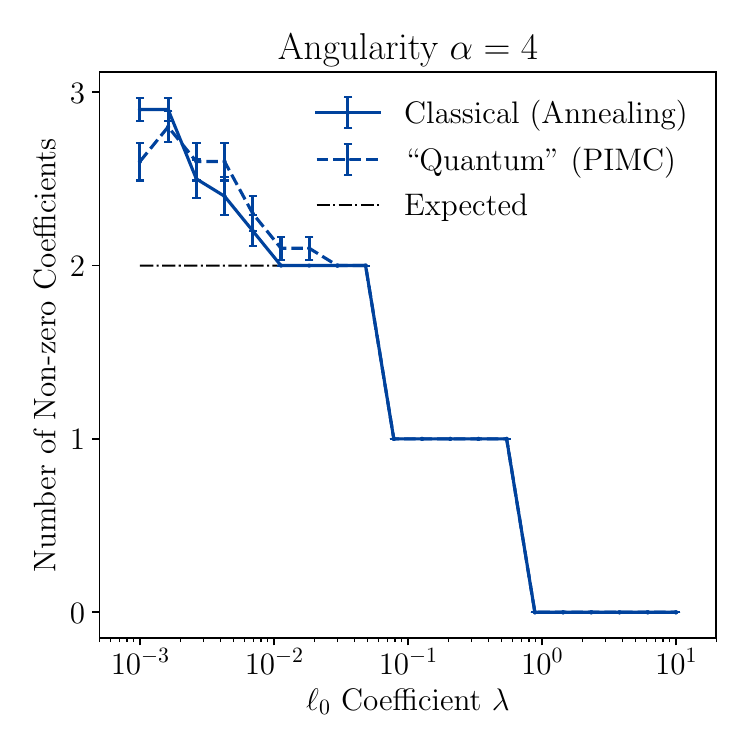}
}
\subfloat[]{
	\includegraphics[width=0.27\textwidth]{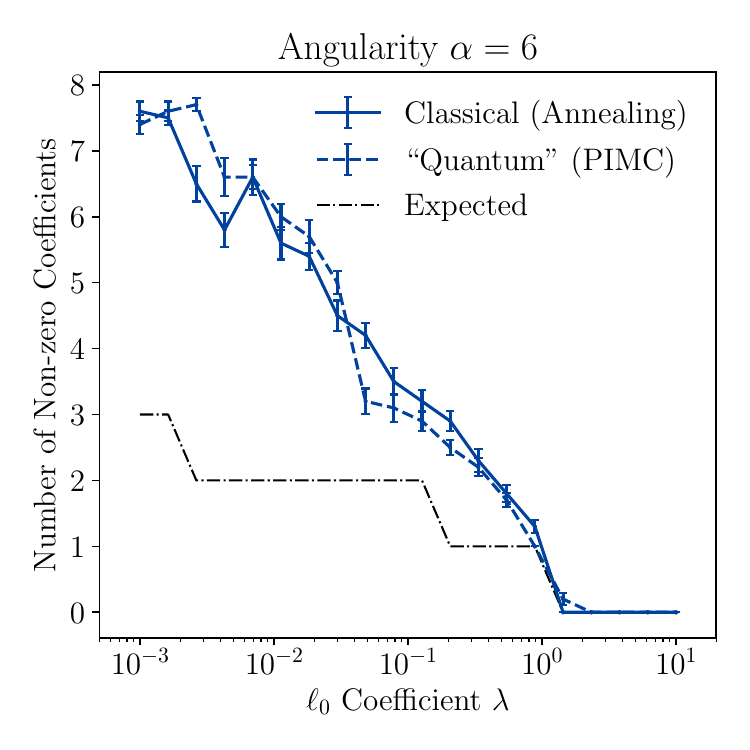}
}\\
\subfloat[]{
	\includegraphics[width=0.27\textwidth]{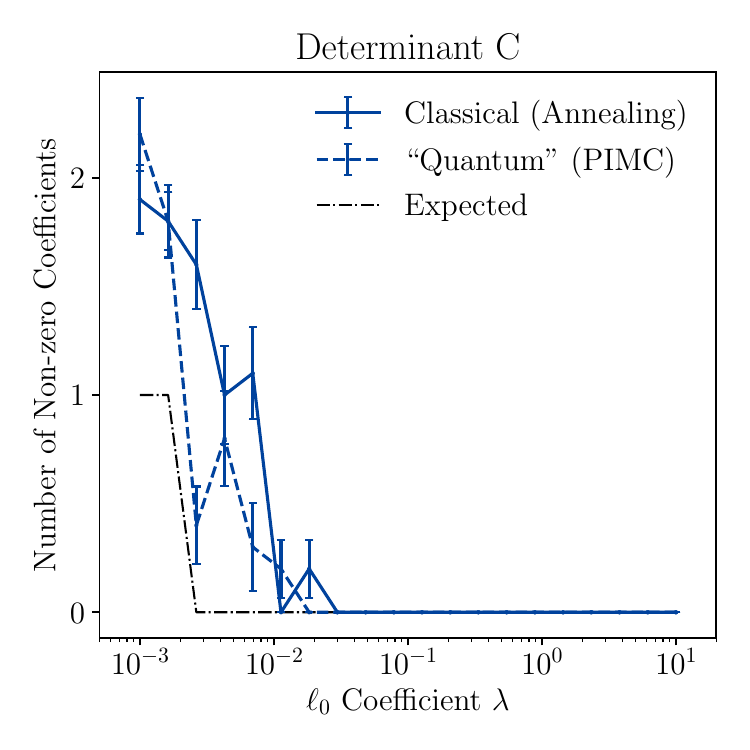}
}
\subfloat[]{
	\includegraphics[width=0.27\textwidth]{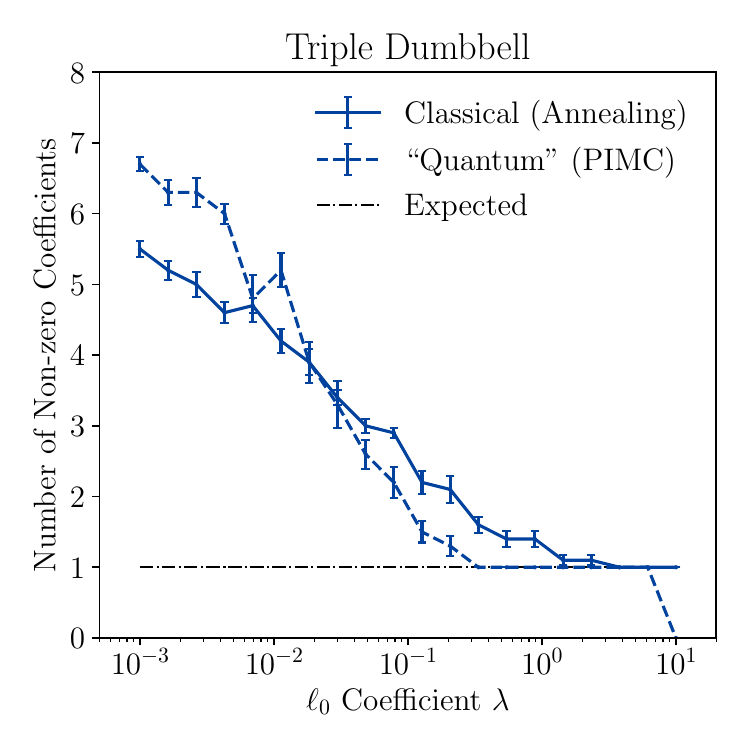}
}
\subfloat[]{
	\includegraphics[width=0.27\textwidth]{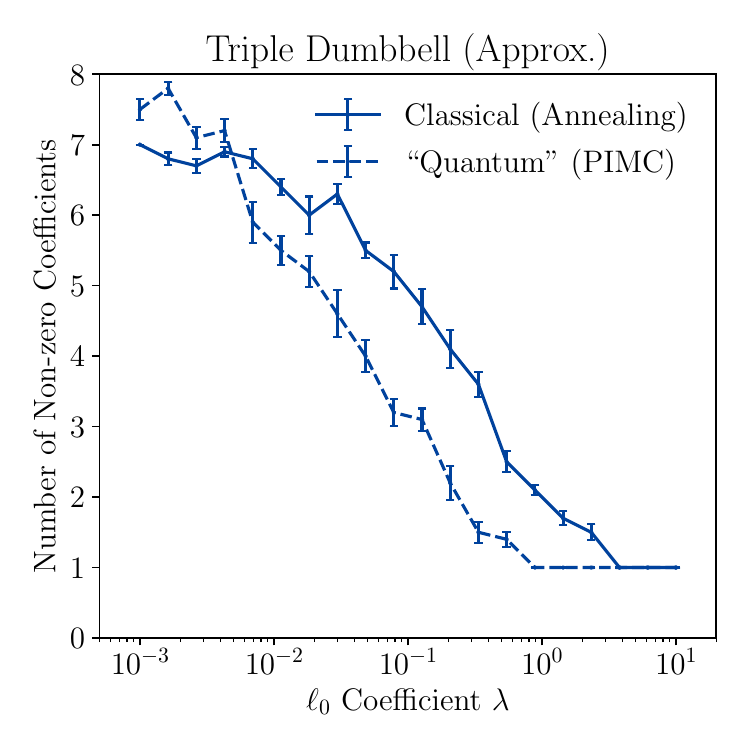}
}\\
\subfloat[]{
	\includegraphics[width=0.27\textwidth]{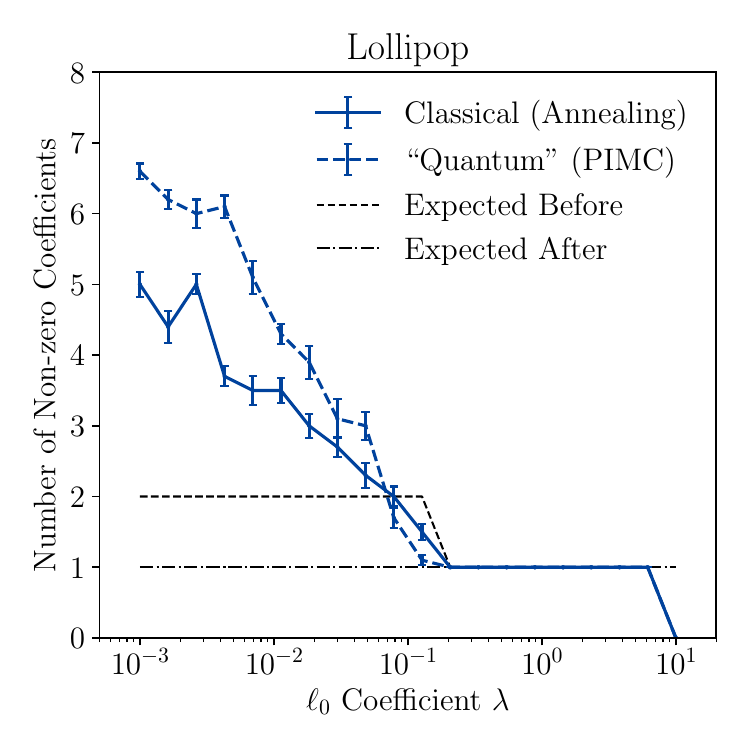}
\label{fig:lollipop2}}
\subfloat[]{
	\includegraphics[width=0.27\textwidth]{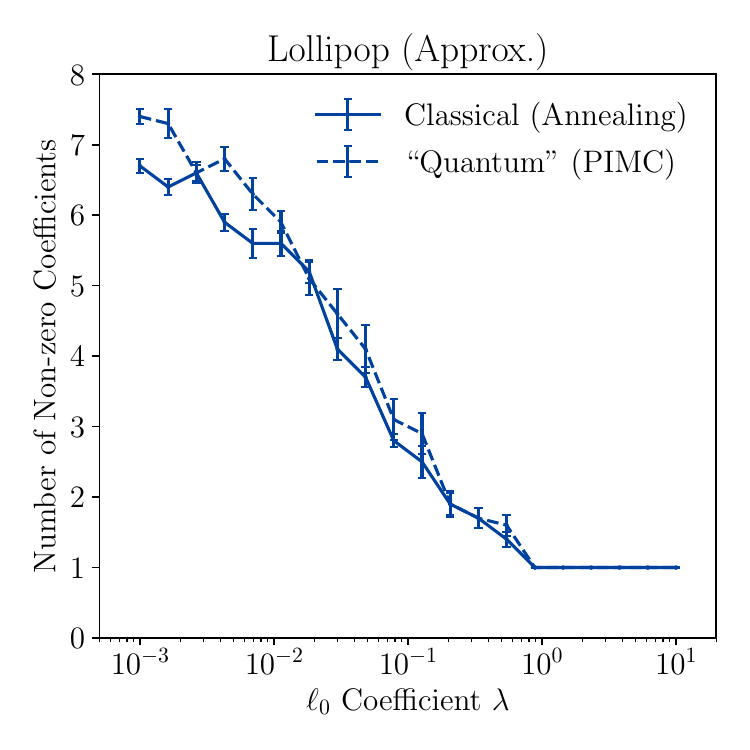}
}
\subfloat[]{
	\includegraphics[width=0.27\textwidth]{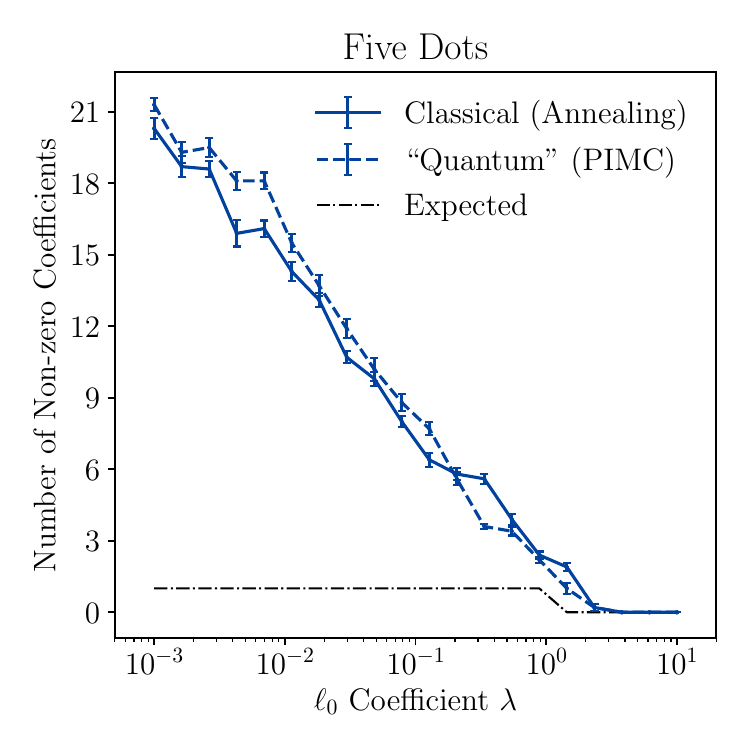}
}\\
\subfloat[]{
	\includegraphics[width=0.27\textwidth]{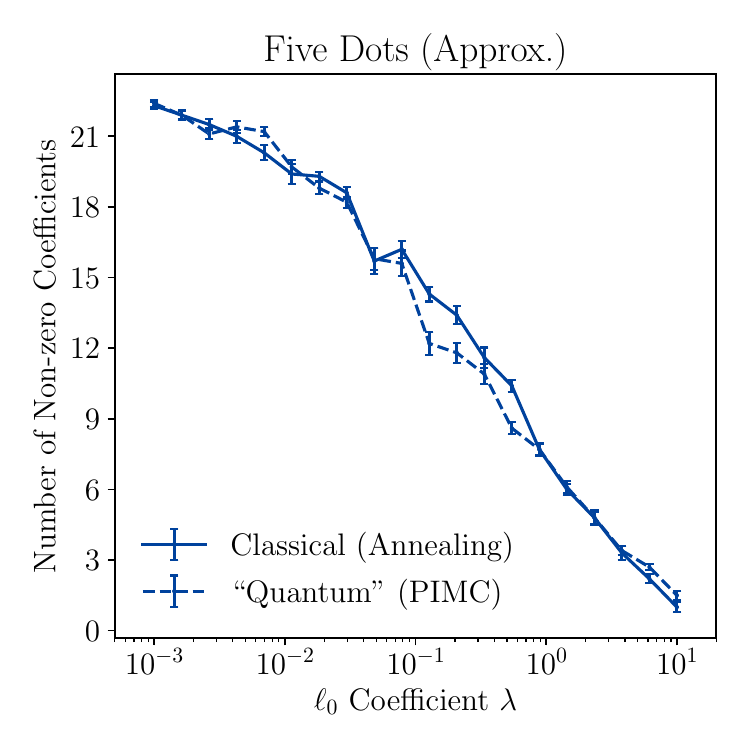}
}
\subfloat[]{
	\includegraphics[width=0.27\textwidth]{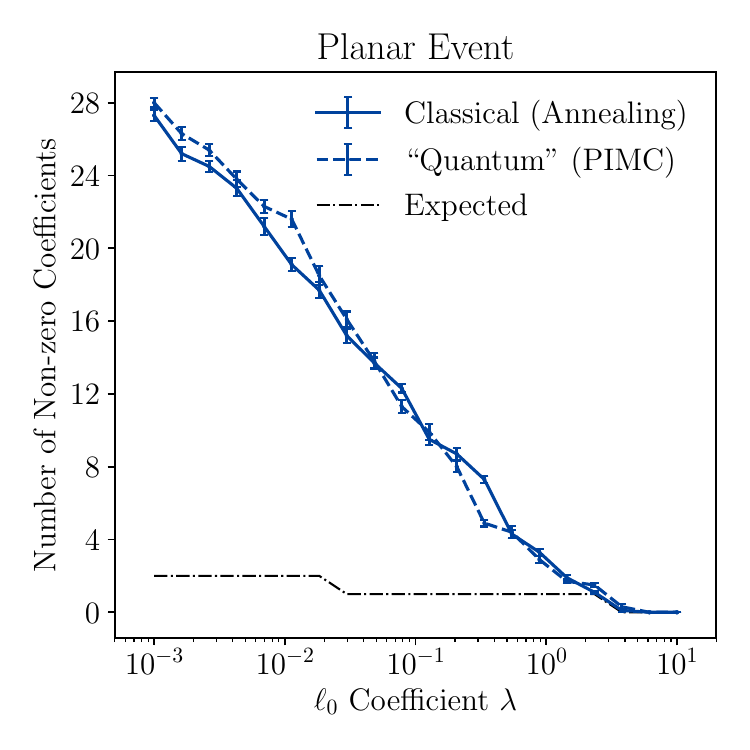}
}
\subfloat[]{
	\includegraphics[width=0.27\textwidth]{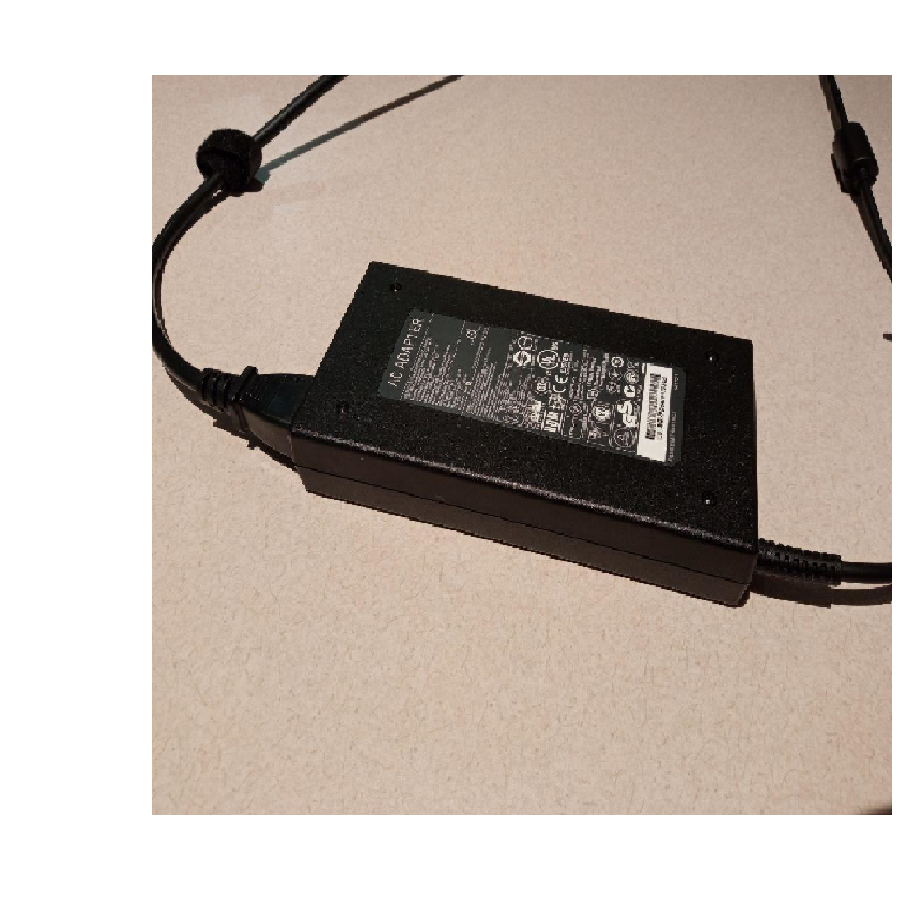}
	\label{fig:charger2}
}
	\caption{
	Same as \Fig{fig:double_vs_single_nnz}, but comparing classical annealing (solid blue) to PIMC (dashed blue) as a proxy for quantum annealing, using the double ABE.
	As discussed in footnote \ref{footnote:burnt}, \Fig{fig:charger2} has been replaced by a burnt charger. The (a)--(l) are defined in \Tab{tab:observables}.}
	\label{fig:classic_vs_pimc_nnz}
\end{figure*}

In \Fig{fig:classic_vs_pimc_nnz}, we plot the number of non-zero coefficients as a function of the $\ell_0$-norm coefficient $\lambda$. 
We compare the performance of classical annealing (solid blue) with PIMC (dashed blue). 
As in \Fig{fig:double_vs_single_nnz}, the error bars are computed by averaging the results over ten distinct runs, and we plot the best-case analytical expectation (black dashed) for all non-approximate relations.
In \Fig{fig:classic_vs_pimc_loss} of \App{app:additional_plots}, we plot the loss as a function of $\lambda$ as an alternative way to assess the potential gains from quantum computing.

For the Lollipop observable in \Fig{fig:lollipop2}, we again observe that our original theoretical expectation was outperformed by both classical annealing and PIMC.
PIMC is actually able to do a better job in the vicinity of $\lambda \simeq 0.1$, though classical annealing does slightly better at smaller $\lambda$. 
For all observables, we find that the performance of classical annealing and PIMC are similar, both with respect to the number of non-zero coefficients and with respect to the loss function; see \Fig{fig:classic_vs_pimc_loss}.
The main difference between classical annealing and PIMC is the significantly higher (classical) computation cost of the latter.%
\footnote{\label{footnote:burnt}When trying to perform PIMC on the twelfth observable ``Planar Event (Approx.)'', we burnt out a laptop power supply, as illustrated in \Fig{fig:charger2}.  We decided against tempting fate to test this observable on a high-performance computer.}

These results demonstrate the robustness of the regression performance with respect to changing the annealing method.
On the other hand, these findings suggest that true quantum annealing may not yield performance gains for this particular optimization problem.

\section{Conclusions}
\label{sec:conclude}

In this paper, we introduced the technique of degeneracy engineering, which is a strategy to improve the performance of both classical and quantum annealing algorithms by increasing the relative degeneracy of the ground state by manipulating a subset of terms in the problem Hamiltonian.
We applied this new concept to the $\textsf{NP}$-hard problem of $\ell_0$-norm regularization for sparse linear regression, focusing on a case study in high-energy collider physics.

The key theoretical insights of this paper are twofold. 
First, we discovered an efficient representation of $\ell_0$-norm regularization as a QUBO problem, which opens up the possibility to study this problem with quantum annealing without relying on potentially inefficient gadgetization~\cite{dattani2019quadratization}.
Second, we found that the \emph{relative degeneracy} of the ground state of the $\ell_0$ regularizer can be increased by increasing the degree of \textit{redundancy} in the qubit encoding scheme for the linear fit coefficients.
In practice, our numerical simulations suggest that this changes the spectrum of the total problem Hamiltonian to a spectrum that is more amenable to annealing strategies.

In detailed numerical experiments, we demonstrated the advantages of using $\ell_0$-norm regularization for sparse linear regression and of employing degeneracy engineering.
In a case study on energy flow polynomials in collider physics, we compared five different regularization methods, including the standard $\ell_2$-, $\ell_1$-, and $\ell_0$-norm regularization, as well as two novel heuristics that refine $\ell_0$-norm regularization.
We found an advantage of $\ell_0$-norm regularization compared to $\ell_1$- and $\ell_2$-norm regularization, with the best performance obtained using the two refinement heuristics.
We also compared standard simulated annealing to path integral Monte Carlo as a proxy for quantum annealing, where we found similar performances for both approaches.
Most importantly, we compared different encoding schemes with different degrees of redundancy, finding significantly better performance from the degeneracy-engineered QUBO implementation with a higher degree of redundancy.

What are the prospects, limitations, and requirements of degeneracy engineering?
The concept of degeneracy engineering potentially has wide-range applicability, in particular, for Hamiltonians containing a penalty term that is easy to study analytically.
Penalty terms are ubiquitous in optimization problems and beyond, ranging from $\ell_0$-norm regularization terms to penalty terms enforcing physical constraints or symmetries (see \Ref{Gray:2021bkw} for an example of penalty terms in particle track reconstruction).
While ground-state energies of generic Hamiltonians can be negative, penalty terms employ absolute values and thus vanish under minimization.
This feature makes penalty terms the ideal candidates for degeneracy engineering, as one can potentially engineer multiple zero values for the ground-state energy of a penalty term.
As we exemplified in \Eq{eq:2lb}, this can be achieved by exploiting cancellations of positive and negative contributions to the ground-state energy.
Quantum annealing platforms could substantially benefit from this concept, but would require a large degree of connectivity for the ancilla qubit(s), as illustrated in \Fig{fig:control_bits}.

Our results motivate studies of degeneracy engineering for optimization problems beyond sparse linear regression.
We also expect degeneracy engineering to be applicable to optimization methods beyond classical and quantum annealing, including variational quantum simulations on digital quantum computers~\cite{Peruzzo2014, McClean:2016} and classical optimization methods like tensor-network methods~\cite{Orus:2018dya}.
If a task is aimed at finding the ground-state energy of a Hamiltonian, then it will likely benefit from engineering more ground-states configurations.

\section*{Acknowledgements} 

E.R.A. is supported by the National Science Foundation Graduate Research Fellowship Program under Grant No.\ 4000063445, and a Lester Wolfe Fellowship and the Henry W.\ Kendall Fellowship Fund from M.I.T.
L.F. and J.T. are supported by the U.S.\ Department of Energy (DOE), Office of Science, National Quantum Information Science Research Centers, Co-design Center for Quantum Advantage (C$^2$QA) under Contract No.\ DE-SC0012704 and by the DOE QuantISED program through the theory consortium “Intersections of QIS and Theoretical Particle Physics” at Fermilab (FNAL 20-17).
L.F. is additionally supported by the U.S.\ DOE Office of Nuclear Physics under Grant Contracts No.\ DE-SC0011090 and No.\ DE-SC0021006.
This work was supported by the U.S.\ DOE Office of High Energy Physics under Grant Contract No.\ DE-SC0012567 and by the National Science Foundation under Cooperative Agreement No.\ PHY-2019786 (The NSF AI Institute for Artificial Intelligence and Fundamental Interactions, \url{http://iaifi.org/}).


\appendix

\section{Technical Details of Path Integral Monte Carlo} 
\label{app:trotter}

In this appendix, we review some technical details~\cite{PhysRevB.66.094203} of deriving the path-integral representation of the Ising model used to simulate quantum annealing.
We start with the transverse Ising Hamiltonian in \Eq{eq:QA},
\begin{equation}
    H=\sum\limits_{\langle ij\rangle} J_{ij}\sigma_i^z\sigma_j^z+\Gamma\sum\limits_{i=1}\sigma_i^x,
    \label{eq:TIHamiltonian}
\end{equation}
where $J_{ij}$ are couplings between nearest-neighbor sites and $\Gamma$ is the transverse field. The latter does not commute with the classical Ising term and therefore turns the Ising model from classical to quantum.

To derive the path-integral representation of the quantum Hamiltonian in \Eq{eq:TIHamiltonian}, we first split this Hamiltonian into its kinetic energy term $K$ and its potential energy term $U$ given by
\begin{equation}
K=\Gamma\sum\limits_{i=1}\sigma_i^x,\quad U=\sum\limits_{\langle ij\rangle} J_{ij}\sigma_i^z\sigma_j^z,
\end{equation}
such that $H=K+U$ and $[K,U]\neq 0$.

Then, we write down the partition function $Z$ at the temperature $T=1/\beta$ as
\begin{align}
\begin{split}
Z&={\rm Tr} e^{-\beta H}\\
&={\rm Tr} \left(e^{-\beta (K+U)/P}\right)^P\\
&= \sum_{s^1}\ldots\sum_{s^P}\langle s^1| e^{-\beta (K+U)/P} | s^2\rangle\\
& \times \langle s^2| e^{-\beta (K+U)/P} \ldots | s^P\rangle \langle s^P| e^{-\beta (K+U)/P} | s^1\rangle ,
\end{split}
\label{eq:Z}
\end{align}
where we inserted the identity operator $\mathds{1}=\sum_{s^m} |s^m\rangle\langle s^m|$ in the last equality and denoted $s^m=\{ s_i^m\}$ as a configuration of all spins in the $m$th Trotter slice.

Next, we turn the exact expression for the partition function in \Eq{eq:Z} into an approximate expression,
\begin{align}
\begin{split}
Z&\approx Z_P= \sum_{s^1}\ldots\sum_{s^P}\langle s^1| e^{-\beta K/P}e^{-\beta U/P} | s^2\rangle\\
& \times \langle s^2| e^{-\beta K/P}e^{-\beta U/P} \ldots | s^P\rangle \langle s^P| e^{-\beta K/P}e^{-\beta U/P} | s^1\rangle,
\end{split}
\label{eq:ZP}
\end{align}
by using the Trotter breakup formula,
\begin{equation}
e^{-\beta (K+U)/P}\approx e^{-\beta K/P}e^{-\beta U/P},
\end{equation}
which neglects non-zero commutators of $K$ and $U$. The expression for $Z_P$ in \Eq{eq:ZP} approximates the original partition function $Z$ in \Eq{eq:Z} with an error that is proportional to $(\Delta t)^2$, where $\Delta t= \beta/P$ is the so-called Trotter breakup time.

As a next step, we observe that the potential energy $U$ is diagonal in the chosen spin basis. Thus, the only non-trivial term in \Eq{eq:ZP} is the average of the kinetic term $K$ between two Trotter slices,
\begin{align}
\begin{split}
&\langle s^m| e^{-\beta K/P}e^{-\beta U/P} | s^{m+1}\rangle\\
&= \langle s^m| e^{-\beta K/P} | s^{m+1}\rangle e^{-\beta U(s^{m+1})/P}.
\end{split}
\label{eq:nontrivial}
\end{align}
The kinetic part of this equation contains a sum over the spin sites in the exponential, which can be expressed as a product of expectation values,
\begin{align}
\begin{split}
\langle s^m| e^{-\beta K/P} | s^{m+1}\rangle  &= \langle s^m| \exp\left(-\frac{\beta \Gamma}{P} \sum_{i=1}^N\sigma_i^x\right) | s^{m+1}\rangle\\
&=\prod_{i=1}^N \langle s^m| \exp\left(-\frac{\beta \Gamma}{P} \sigma_i^x\right) | s^{m+1}\rangle,
\end{split}
\label{eq:spinprod}
\end{align}
because spin operators at different sites $k$ and $k+1$ commute. Here, $N$ is the number of lattice sites.

The most crucial step of the derivation, which turns the model from quantum into classical, is the following. In the case of spin-1/2, one can show that
\begin{align}
\begin{split}
\langle \uparrow|e^{\alpha\sigma_x}|\uparrow\rangle &= \langle \downarrow|e^{\alpha\sigma_x}|\downarrow\rangle = \cosh(\alpha),\\
\langle \uparrow|e^{\alpha\sigma_x}|\downarrow\rangle &= \langle \downarrow|e^{\alpha\sigma_x}|\uparrow\rangle = \sinh(\alpha),
\end{split}
\end{align}
which implies that one can rewrite the transversal-field (quantum) term as an Ising-like (classical) interaction between different spins $s$ and $s'$ with $ss'=\pm 1$,
\begin{align}
\begin{split}
\langle s|e^{\alpha\sigma_x}|s'\rangle &= \sqrt{(1/2)\sinh(2\alpha)} e^{-(1/2)\ln \tanh(\alpha) ss'}\\
&\equiv C e^{B ss'}.
\end{split}
\label{eq:spins}
\end{align}
Combining Eqs.~\eqref{eq:nontrivial}, \eqref{eq:spinprod}, and \eqref{eq:spins}, we find
\begin{align}
\begin{split}
&\langle s^m| e^{-\beta K/P}e^{-\beta U/P} | s^{m+1}\rangle\\
 &= C^N \exp\left(\frac{J_\perp}{PT} \sum_{i}s_i^{m}s_i^{m+1}\right) \exp\left(\frac{1}{PT} \sum_{\langle ij\rangle}J_{ij} s_i^{m}s_j^{m}\right),
\end{split}
\label{eq:spinfinal}
\end{align}
where we have defined
\begin{align}
\begin{split}
J_\perp &= \frac{PT}{2} \ln \tanh\left(\frac{\Gamma}{PT}\right)>0,\\
C^2 &= \frac{1}{2}\sinh\left( \frac{2\Gamma}{PT}\right).
\end{split}
\label{eq:JC}
\end{align}
Thus, the $J_\perp$ term in \Eq{eq:spinfinal} yields a ferromagnetic Ising-like coupling between the spins $s_i^m$ and $s_i^{m+1}$, which are nearest neighbors along the Trotter dimension.

Finally, we can express the partition function of the $d$-dimensional quantum system in \Eq{eq:ZP} as a partition function of a ($d+1$)-dimensional classical system,
\begin{equation}
Z\approx Z_P = C^{NP}\sum_{s^1}\ldots\sum_{s^P}e^{-H_{d+1}/PT},
\end{equation}
where the ($d+1$)-dimensional classical Hamiltonian is given by 
\begin{equation}
H_{d+1}=-\sum_{m=1}^P\left(\sum_{\langle ij\rangle}J(s)s_i^ms_j^m+J^T\sum_is_i^ms_i^{m+1} \right).
\end{equation}
Here, $s^m=\{s^m_i\}$ denotes a configuration of all the spins in the $m$th Trotter slice, where $M+1$ is identified with $m$ and $J^T$ is the uniform coupling along the extra (imaginary time) direction.

\section{Additional Plots}
\label{app:additional_plots}

In this appendix, we present additional plots to complement the discussion in \Sec{sec:NumResults}.

\begin{figure*}
\centering
\subfloat[]{
	\includegraphics[width=0.27\textwidth]{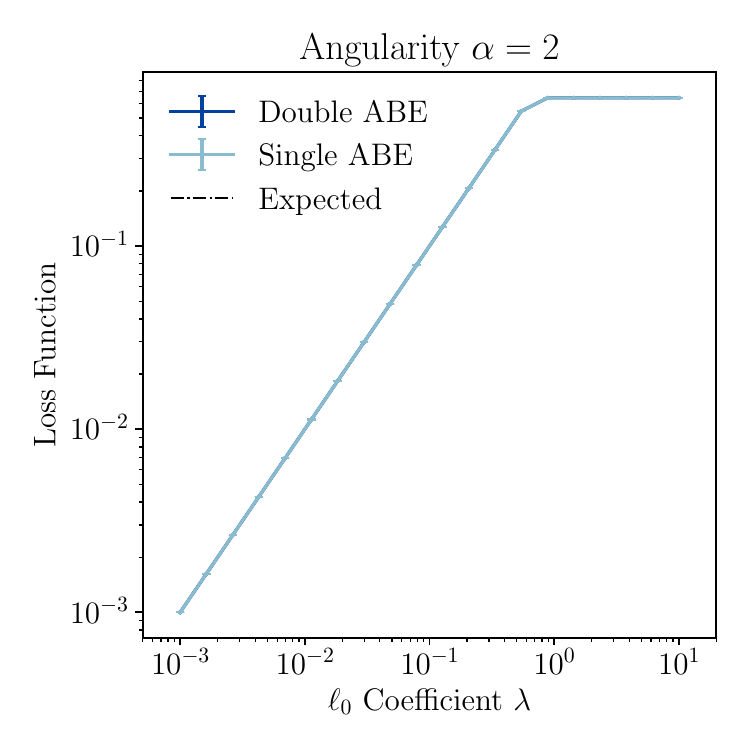}
	\label{fig:2a_angularity2}
}
\subfloat[]{
	\includegraphics[width=0.27\textwidth]{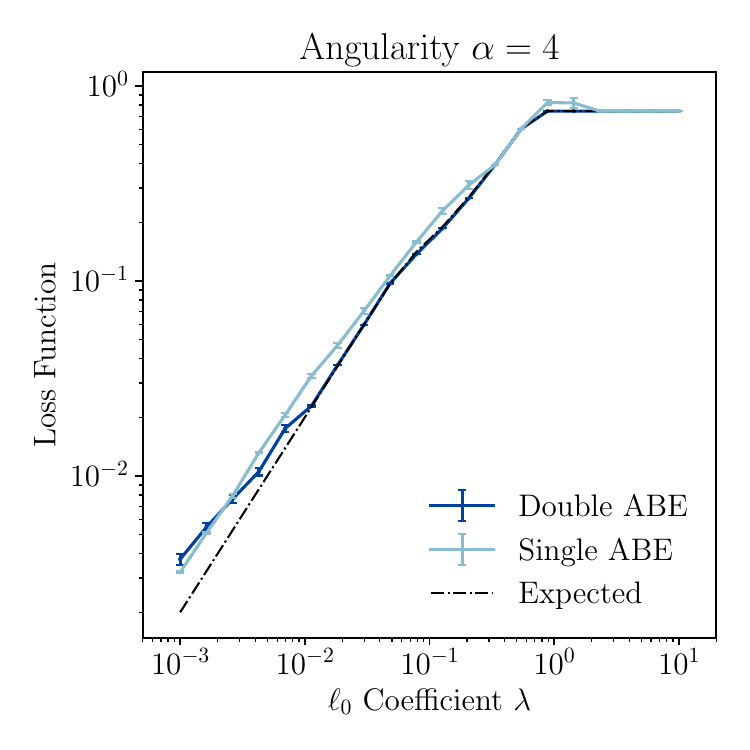}
	\label{fig:2b_angularity4}
}
\subfloat[]{
	\includegraphics[width=0.27\textwidth]{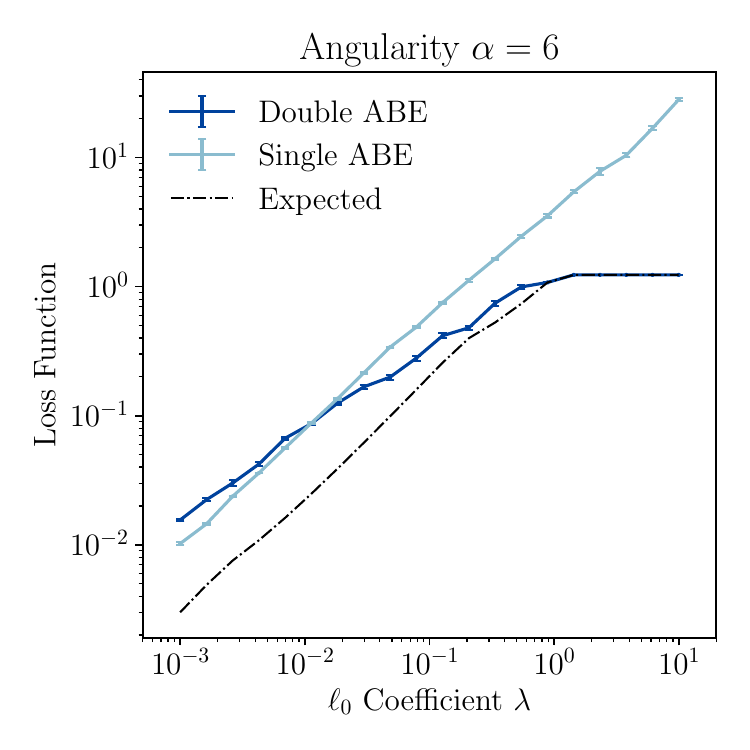}
	\label{fig:2c_angularity6}
}\\
\subfloat[]{
	\includegraphics[width=0.27\textwidth]{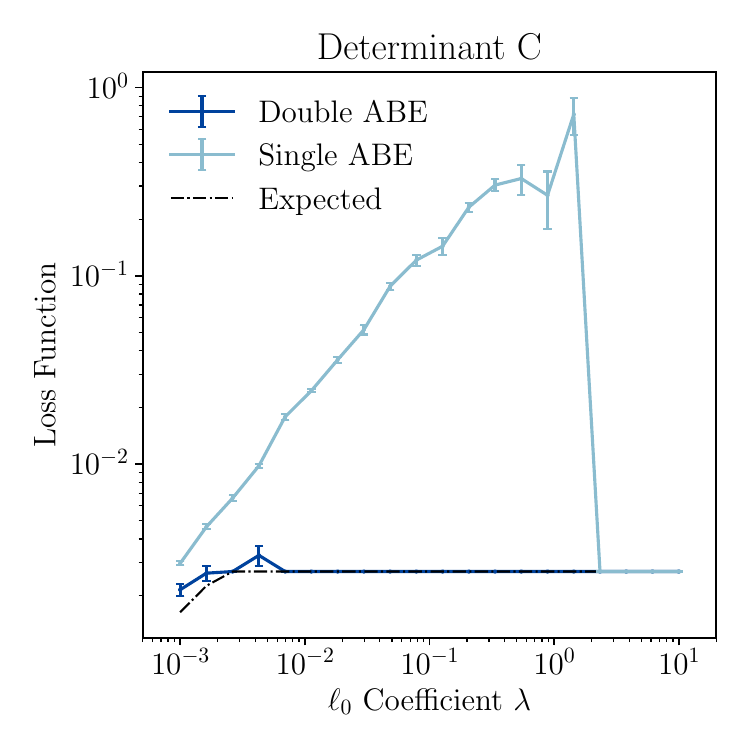}
	\label{fig:2d_detC}
}
\subfloat[]{
	\includegraphics[width=0.27\textwidth]{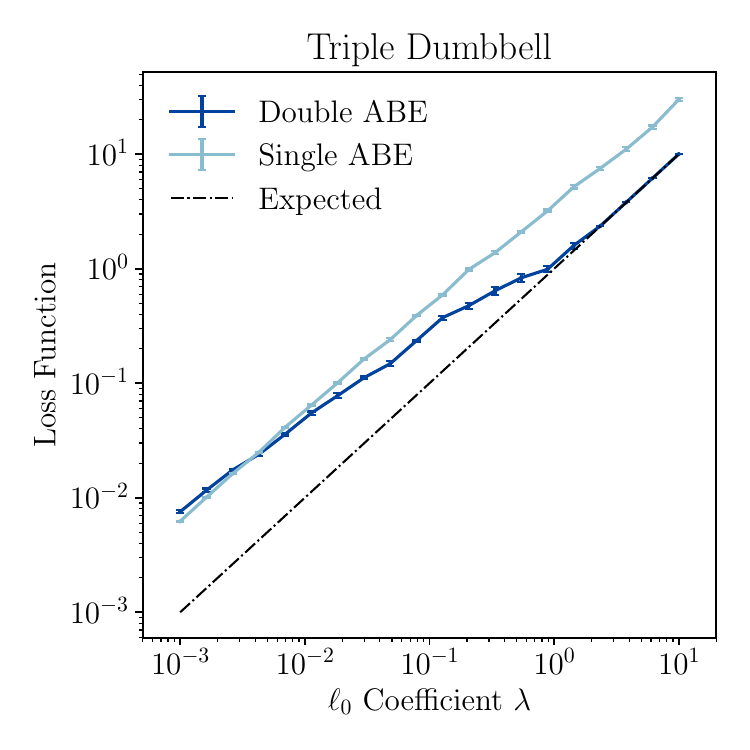}
	\label{fig:2e_TripDum}
}
\subfloat[]{
	\includegraphics[width=0.27\textwidth]{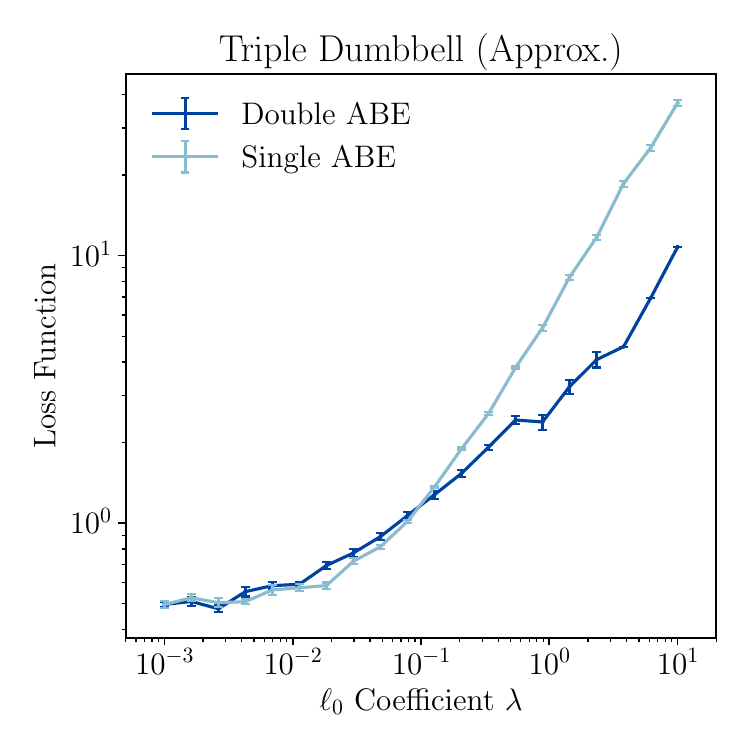}
	\label{fig:2f_TripDumApprox}
}\\
\subfloat[]{
	\includegraphics[width=0.27\textwidth]{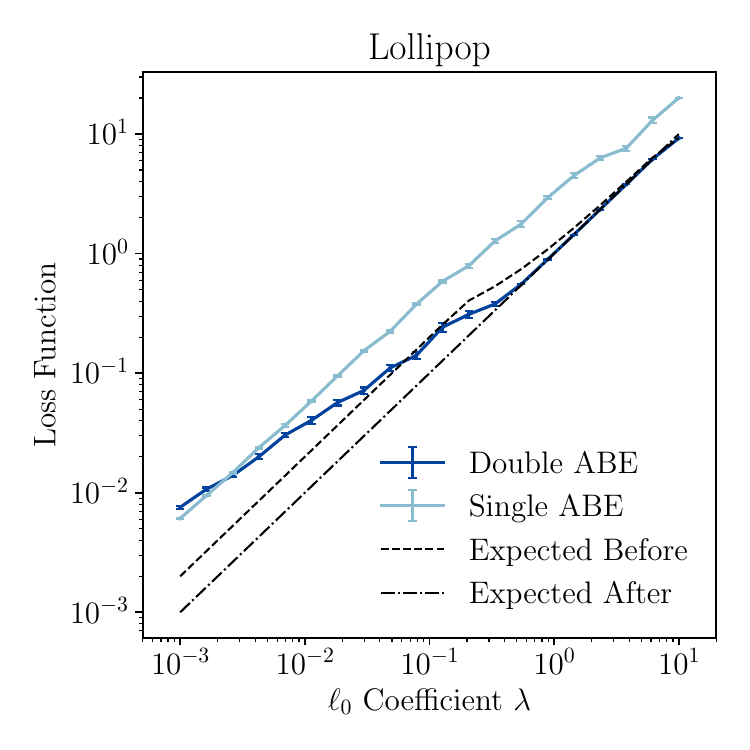}
	\label{fig:2g_Lolipop}
}
\subfloat[]{
	\includegraphics[width=0.27\textwidth]{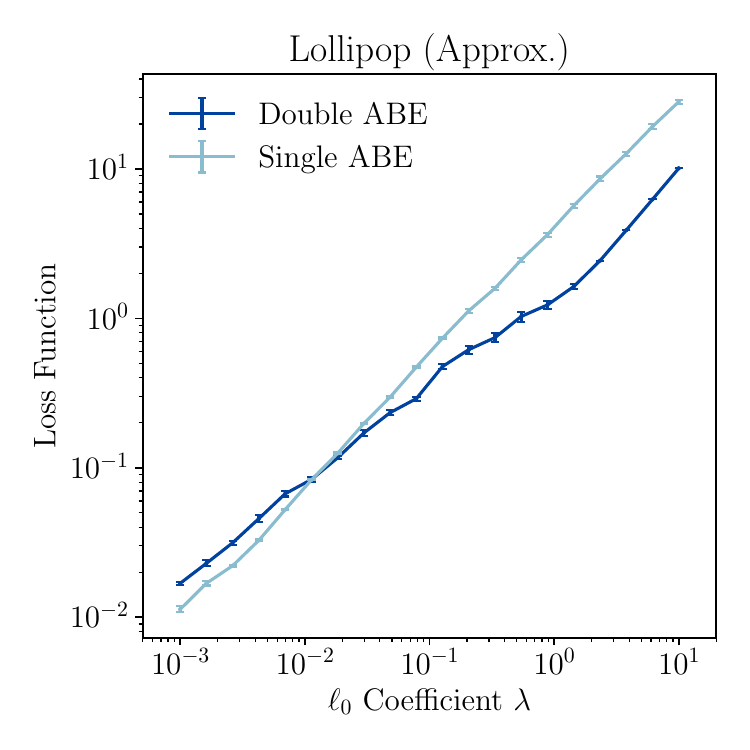}
	\label{fig:2h_LolipopApprox}
}
\subfloat[]{
	\includegraphics[width=0.27\textwidth]{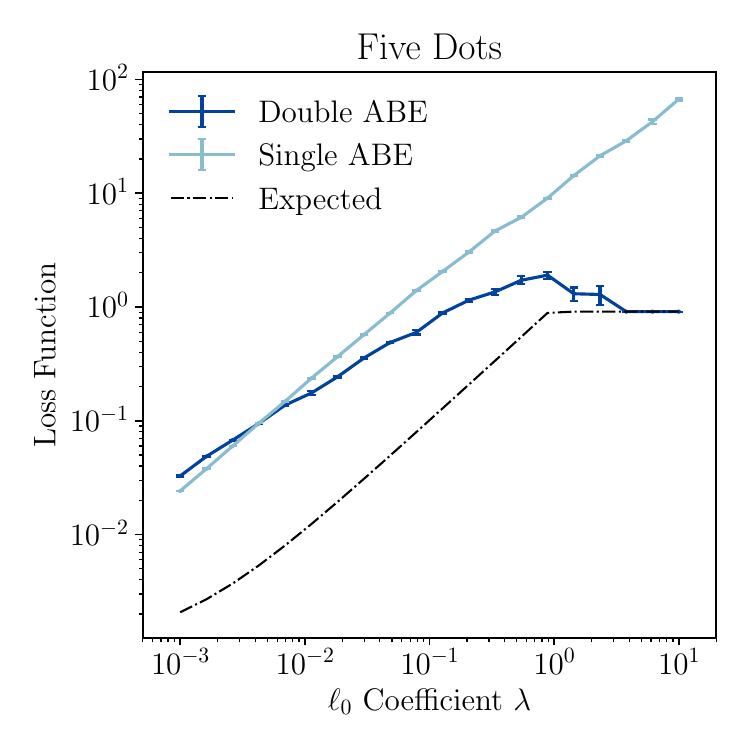}
	\label{fig:2i_FiveDots}
}\\
\subfloat[]{
	\includegraphics[width=0.27\textwidth]{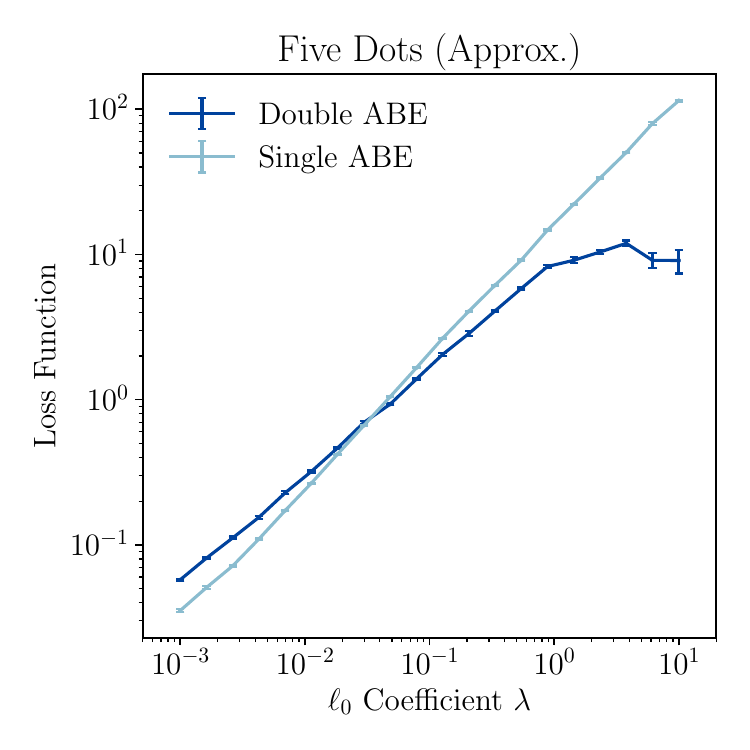}
	\label{fig:2j_FiveDots}
}
\subfloat[]{
	\includegraphics[width=0.27\textwidth]{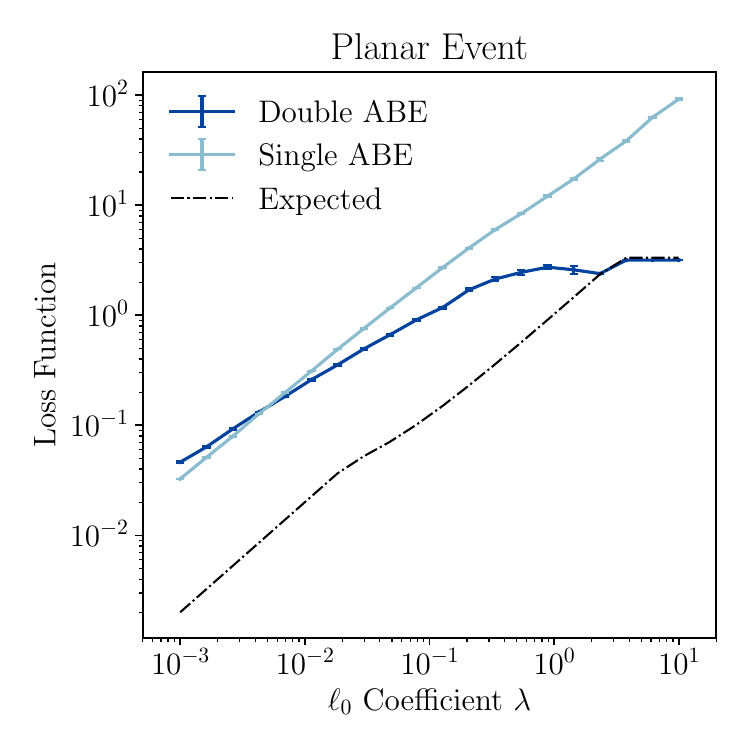}
	\label{fig:2k_PlanarEvent}
}
\subfloat[]{
	\includegraphics[width=0.27\textwidth]{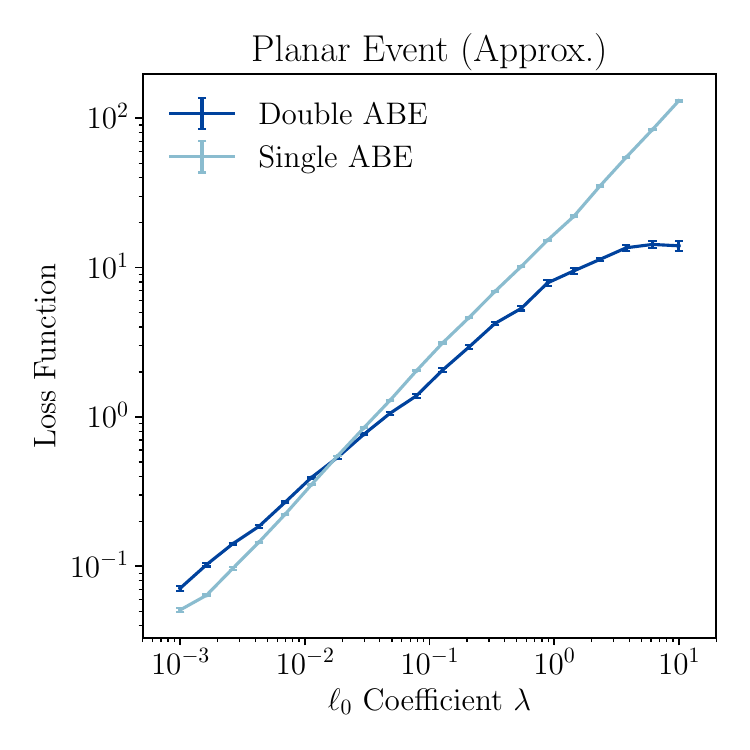}
	\label{fig:2l_PlanarEvent}
}
	\caption{
	Same as \Fig{fig:double_vs_single_nnz}, but plotting the $\ell_0$-norm regularized loss function in \Eq{eq:regularized} as a function of the $\ell_0$-norm coefficient $\lambda$. The (a)--(l) are defined in \Tab{tab:observables}.}
	\label{fig:double_vs_single_loss}
\end{figure*}

In \Fig{fig:double_vs_single_loss}, we give an alternative comparison of double ABE versus single ABE.
The advantage of using double ABE was already shown in \Fig{fig:double_vs_single_nnz} in terms of the number of identified non-zero fit coefficients as a function of the $\ell_0$-norm coefficient $\lambda$.
Here, we plot the $\ell_0$-norm regularized loss from \Eq{eq:regularized} as a function of $\lambda$, comparing the single ABE (light blue) to the double ABE (dark blue). 
For all observables, we find that the degeneracy-engineered version with double ABE performs equally well or better in terms of lowering the loss function.

\begin{figure*}
\centering
\subfloat[]{
	\includegraphics[width=0.27\textwidth]{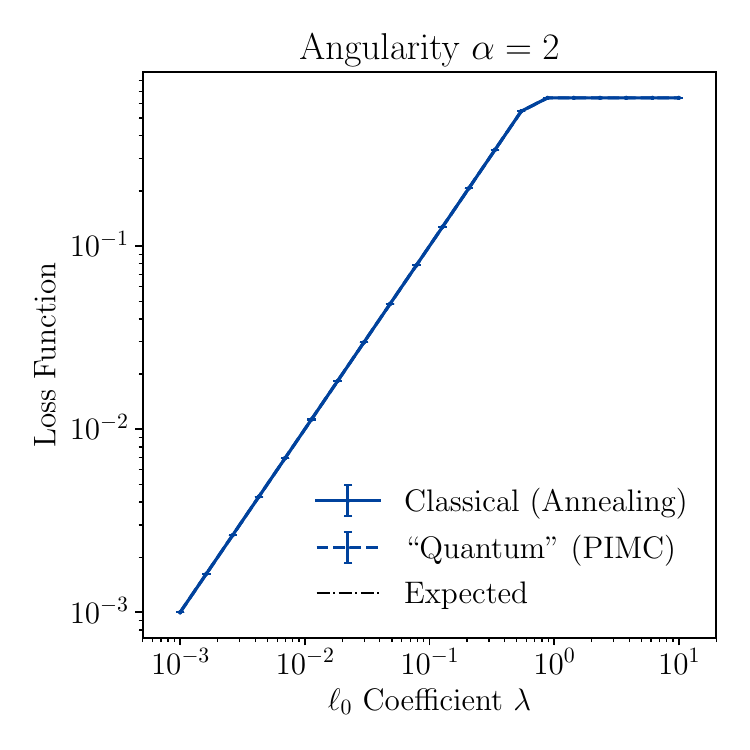}
}
\subfloat[]{
	\includegraphics[width=0.27\textwidth]{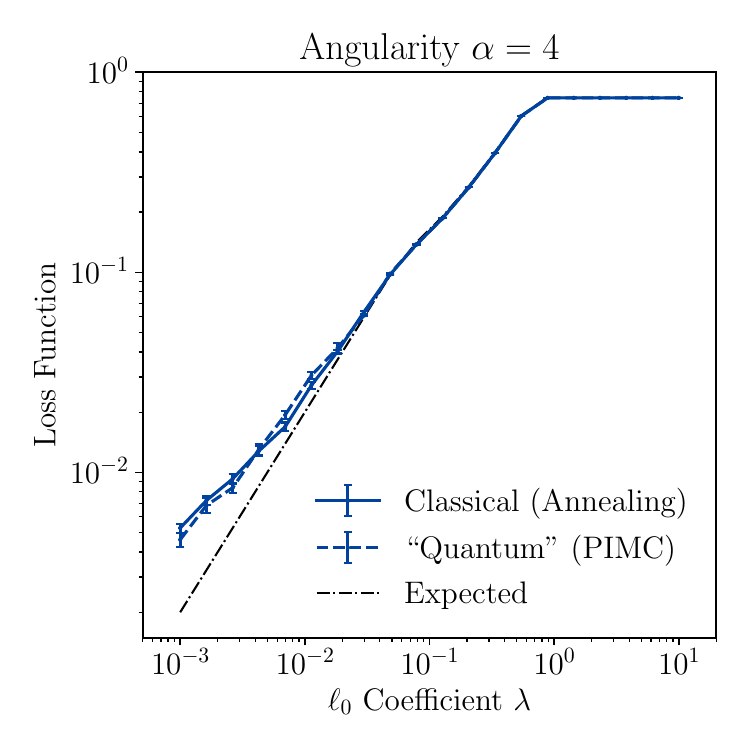}
}
\subfloat[]{
	\includegraphics[width=0.27\textwidth]{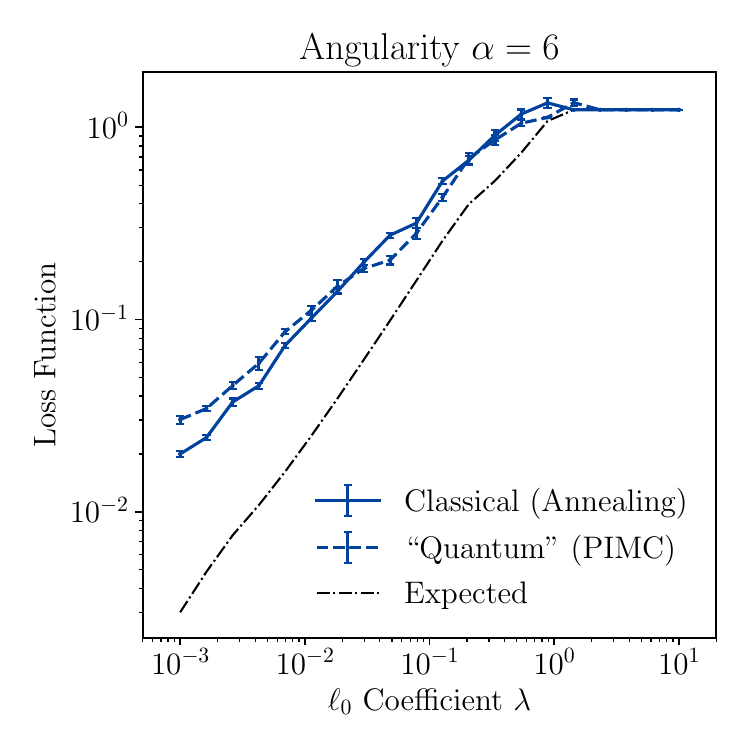}
}\\
\subfloat[]{
	\includegraphics[width=0.27\textwidth]{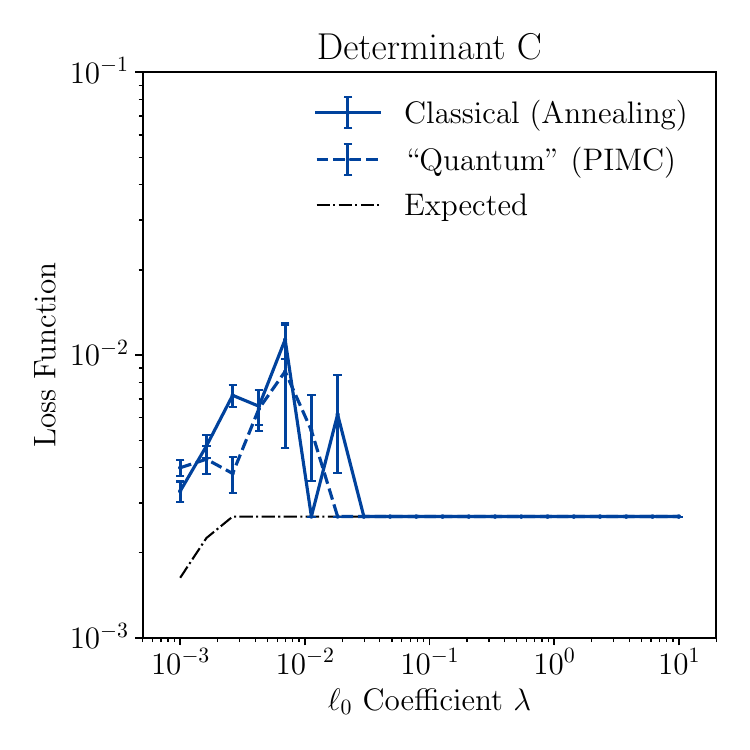}
}
\subfloat[]{
	\includegraphics[width=0.27\textwidth]{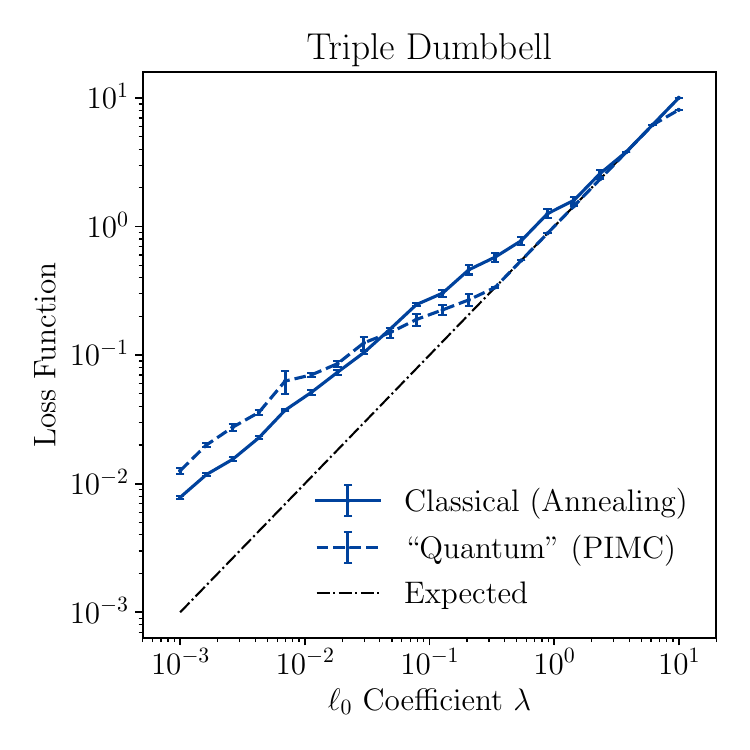}
}
\subfloat[]{
	\includegraphics[width=0.27\textwidth]{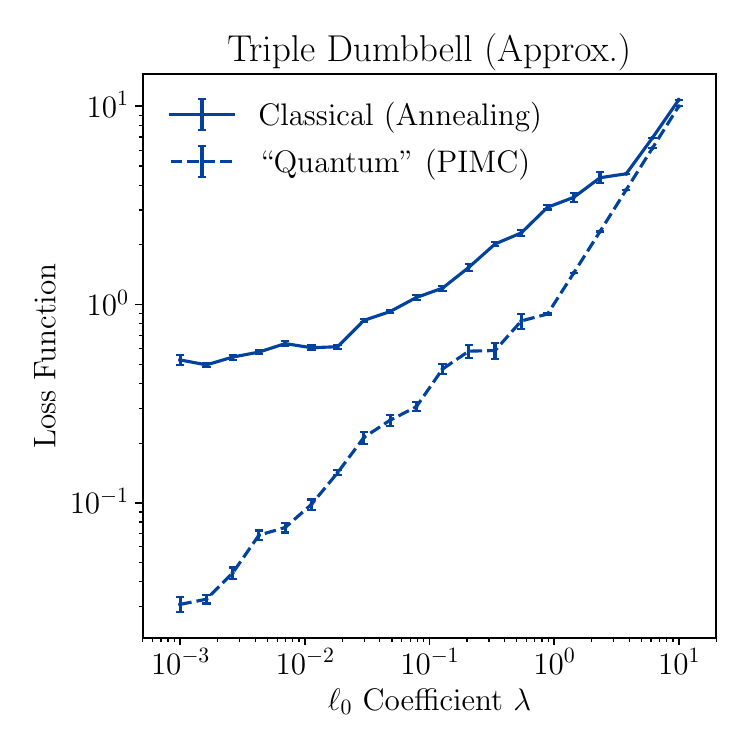}
}\\
\subfloat[]{
	\includegraphics[width=0.27\textwidth]{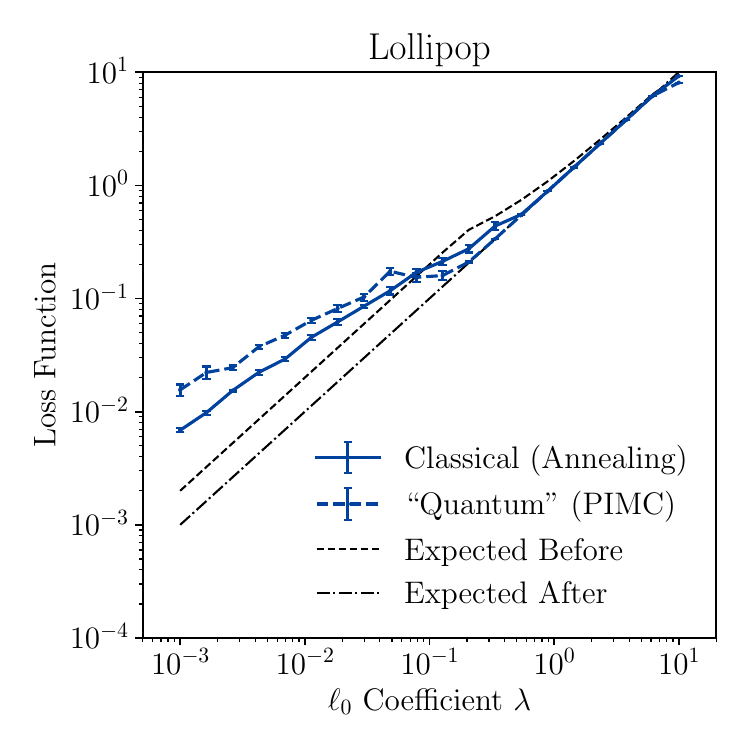}
	\label{fig:7g_Lollipop}
}
\subfloat[]{
	\includegraphics[width=0.27\textwidth]{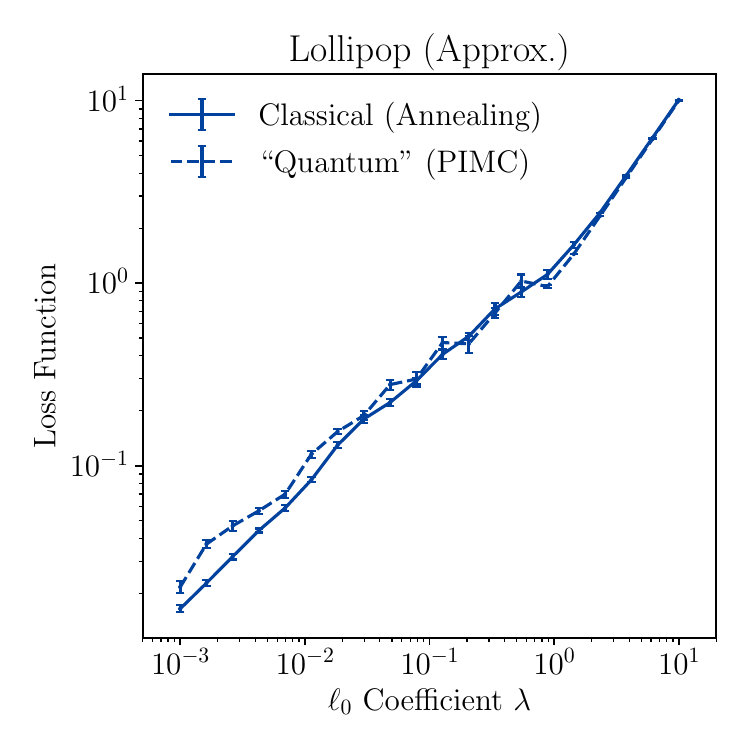}
}
\subfloat[]{
	\includegraphics[width=0.27\textwidth]{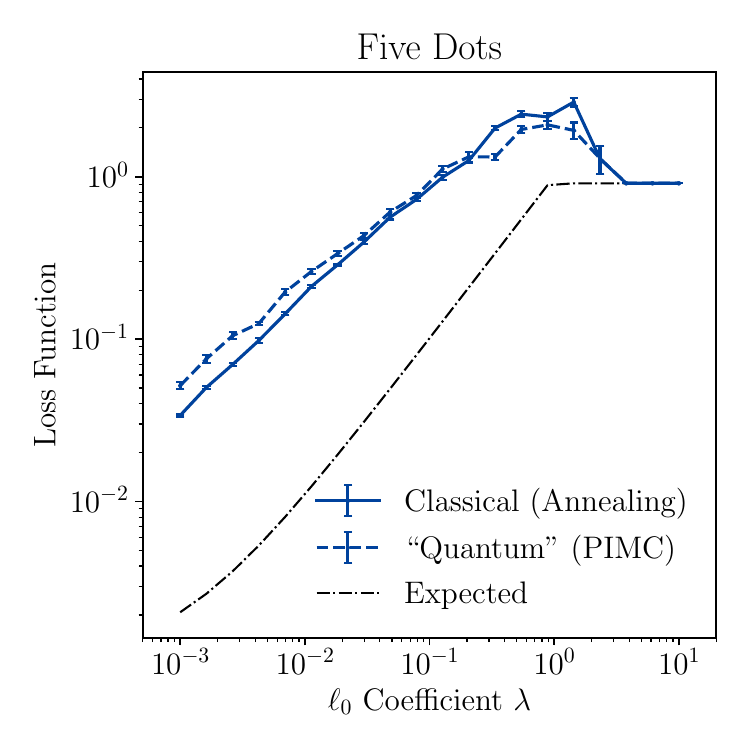}
}\\
\subfloat[]{
	\includegraphics[width=0.27\textwidth]{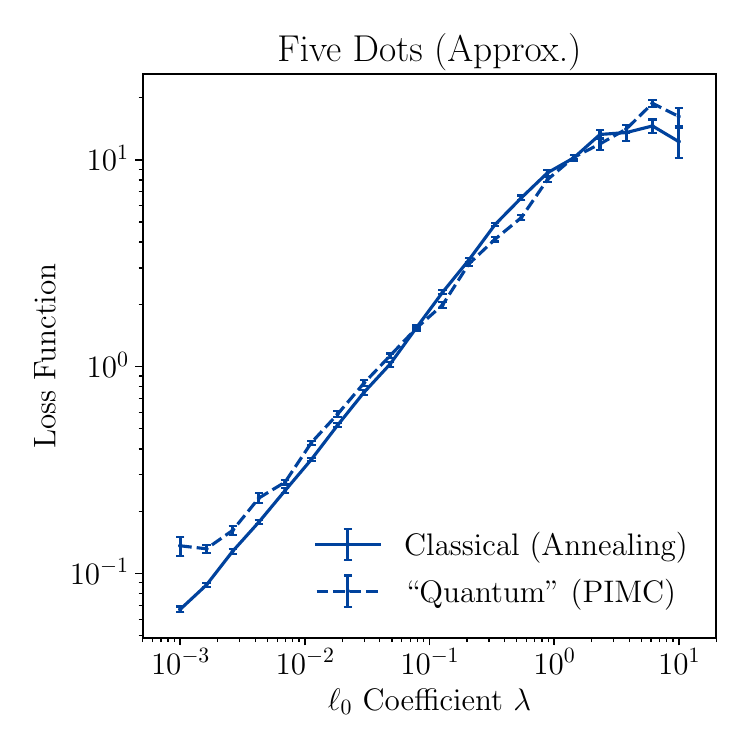}
}
\subfloat[]{
	\includegraphics[width=0.27\textwidth]{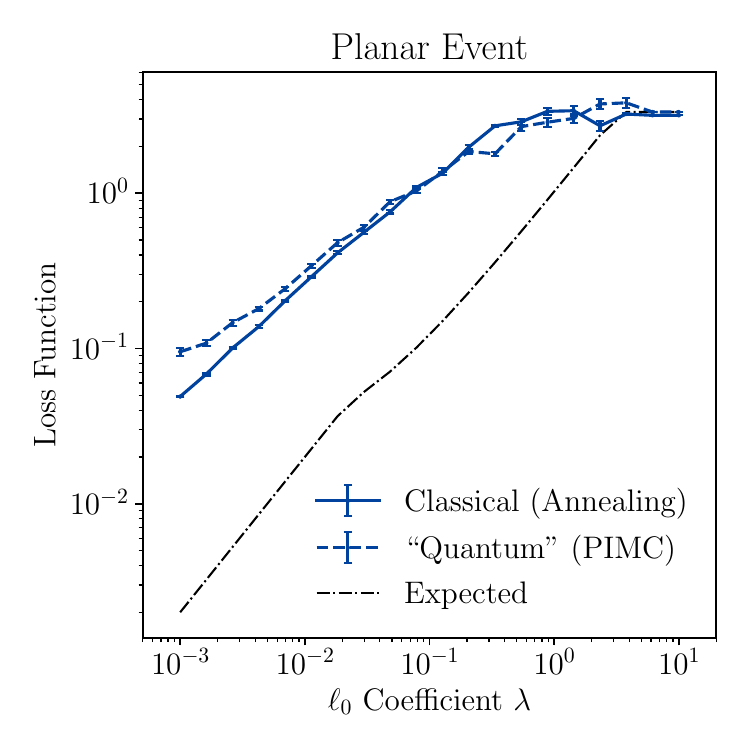}
}
\subfloat[]{
	\includegraphics[width=0.27\textwidth]{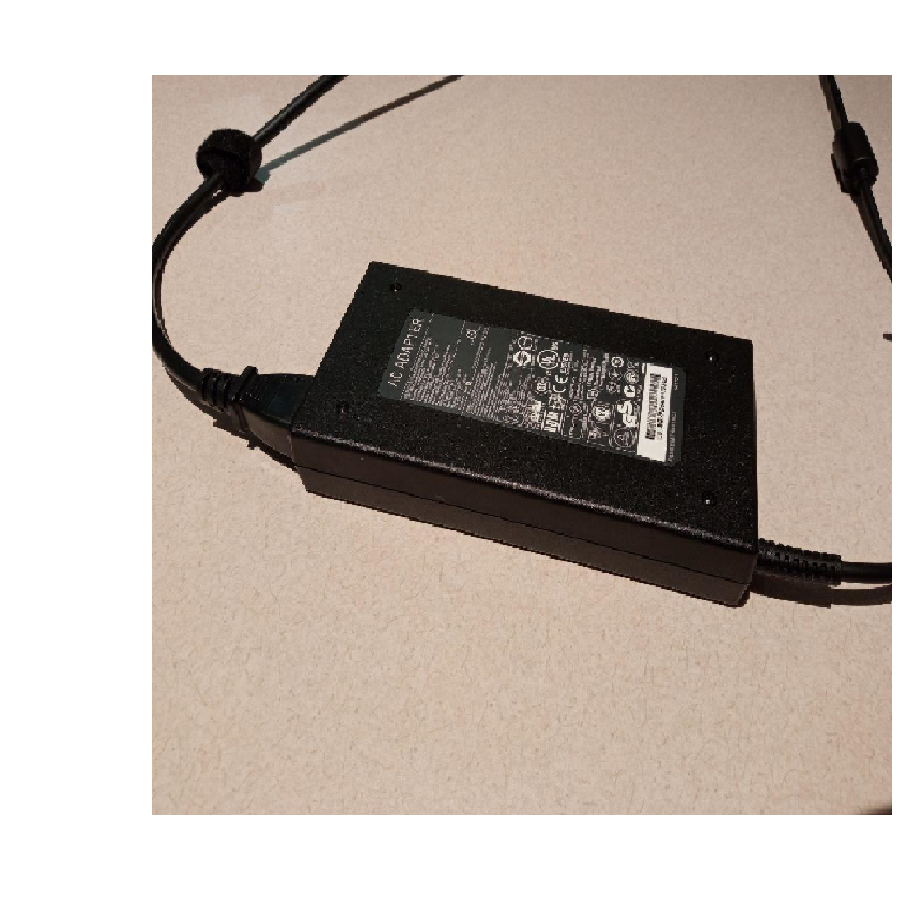}
	\label{fig:charger1}
}
	\caption{
	Same as \Fig{fig:classic_vs_pimc_nnz}, but plotting the $\ell_0$-norm regularized loss function in \Eq{eq:regularized} as a function of the $\ell_0$-norm coefficient $\lambda$. 
	\Fig{fig:charger1} is identical to \Fig{fig:charger2}. The (a)--(l) are defined in \Tab{tab:observables}.
	}
	\label{fig:classic_vs_pimc_loss}
\end{figure*}

In \Fig{fig:classic_vs_pimc_loss}, we give an alternative comparison of classical annealing and PIMC.
Like for \Fig{fig:classic_vs_pimc_nnz}, we use the degeneracy-engineered encoding with double ABE, but now plotting the $\ell_0$-norm regularized loss as a function of $\lambda$.
Comparing classical annealing (solid blue) to PIMC (dashed blue), we find similar performance across the twelve relations.

\bibliography{qubo_lzero}

\end{document}